\documentclass[12pt]{article}
\usepackage{amsfonts}
\newlength{\abstractwidth}
\usepackage{epsf}

\renewcommand{\thefootnote}{\fnsymbol{footnote}}
\renewcommand{\thanks}[1]{\footnote{#1}}
\newcommand{\starttext}{
\setcounter{footnote}{0}
\renewcommand{\thefootnote}{\arabic{footnote}}}

\newcommand{\bea}{\begin{eqnarray}}
\newcommand{\eea}{\end{eqnarray}}
\newcommand{\ee}{\end{equation}}
\newcommand{\be}{\begin{equation}}

\def\cC{{\cal C}}

\def\cG{{\cal G}}

\def\cN{{\cal N}}
\def\cO{{\cal O}}

\def\bC{{\bf C}}

\def\bR{{\bf R}}

\def\Re{{\rm Re}}
\def\Im{{\rm Im}}

\def\det{{\rm det}}

\def\half{ {1\over 2}}
\def\p{\partial}

\def\a{\alpha}
\def\b{\beta}
\def\ba{\bar \alpha}
\def\bb{\bar \beta}

\def\tet{\vartheta}
\def\ep{\varepsilon}

\def\g{\gamma}
\def\l{\lambda}
\def\o{\omega}

\def\f{\varphi}
\def\G{\Gamma}

\def\s{\sigma}

\def\ch{{\rm ch }}
\def\sh{{\rm sh }}
\def\th{{\rm th }}
\def\tg{{\rm tg  }}

\def\no{\nonumber}
\def\sm{\smallskip}

\begin{document}
\starttext
\setcounter{footnote}{0}

\begin{flushright}
UCLA/08/TEP/16 \\
3 June 2008
\end{flushright}

\vskip 0.3in

\begin{center}

{\Large \bf Exact Half-BPS Flux Solutions in M-theory I

\medskip

Local Solutions}\footnote{This work was supported in
part by National Science Foundation (NSF) grants PHY-04-56200\\
and PHY-07-57702.}

\vskip 0.7in

{\large Eric D'Hoker, John Estes, Michael Gutperle, Darya Krym}

\vskip .2in

 \sl Department of Physics and Astronomy \\
\sl University of California, Los Angeles, CA 90095, USA

\end{center}

\vskip .5in

\begin{abstract}

The complete eleven-dimensional supergravity solutions with 16 supersymmetries
on  manifolds of the form $AdS_3 \times S^3 \times S^3 \times \Sigma$, with
isometry $SO(2,2) \times SO(4) \times SO(4)$, and with either $AdS_4 \times S^7$
or $AdS_7 \times S^4$ boundary behavior, are obtained in exact form.
The two-dimensional parameter space $\Sigma$ is a Riemann surface with boundary,
over which the product space $AdS_3 \times S^3 \times S^3$ is warped.
By mapping the reduced BPS equations to an integrable system  of the
sine-Gordon/Liouville type, and then mapping this integrable system  onto a
linear equation, the general local solutions are constructed explicitly in terms of
one harmonic function on $\Sigma$, and an integral transform of two further 
harmonic functions on $\Sigma$.  The solutions to the BPS equations are
shown to automatically solve the Bianchi identities and field equations
for the 4-form field, as well as Einstein's equations. The solutions we obtain
have non-vanishing 4-form field strength on each of the three factors of
$AdS_3 \times S^3 \times S^3$, and include fully back-reacted M2-branes
in $AdS_7 \times S^4$ and M5-branes in $AdS_4 \times S^7$.
No interpolating solutions exist with mixed  $AdS_4 \times S^7$ and 
$AdS_7 \times S^4$ boundary behavior. Global regularity of these local solutions,
as well as the existence of further solutions with neither $AdS_4 \times S^7$ nor 
$AdS_7 \times S^4$ boundary behavior will be studied elsewhere.

\end{abstract}

\newpage

%\tableofcontents
\vfill\eject

\baselineskip=15pt
\setcounter{equation}{0}
\setcounter{footnote}{0}

%%%%%%%%%%%%%%%%%%%%%%%%%%%%%%%%%%%%%%%%%%%%%%%
%%%%%%%%%%%%%%%%%%%%%%%%%%%%%%%%%%%%%%%%%%%%%%%
\section{Introduction}
\setcounter{equation}{0}
%%%%%%%%%%%%%%%%%%%%%%%%%%%%%%%%%%%%%%%%%%%%%%%
%%%%%%%%%%%%%%%%%%%%%%%%%%%%%%%%%%%%%%%%%%%%%%%

The AdS/CFT correspondence 
\cite{Maldacena:1997re, Gubser:1998bc, Witten:1998qj} 
(for reviews, see \cite{Aharony:1999ti,D'Hoker:2002aw})
maps local and non-local gauge invariant operators on the CFT side onto 
solutions to supergravity on the AdS side. In the `t~Hooft limit of large gauge 
group, and for large `t~Hooft coupling, the CFT  operators map to solutions of 
classical supergravity.
As a result, the knowledge of certain classical supergravity solutions can
benefit our understanding of the dynamics on the CFT side.

\sm

Substantial progress has been made over the past few years in spelling out
this correspondence for the special case where 16 supersymmetries
are preserved by both the operators on the CFT side, and the supergravity
solutions on the AdS side. In 10-dimensional Type IIB supergravity, infinite families
of solutions which are invariant under 16 supersymmetries have been obtained
in exact form. Each one of these solutions is dual to a particular  gauge invariant
operator in  4-dimensional $\cN=4$ super Yang-Mills theory.

\sm

A first family consists of exact solutions \cite{Lin:2004nb} dual to local gauge invariant 
half-BPS operators \cite{Berenstein:2004kk}. 
A second family consists of exact solutions \cite{D'Hoker:2007xy,D'Hoker:2007xz} 
which generalize the Janus solution with no supersymmetry of \cite{Bak:2003jk}, 
and the Janus solution with 4 supersymmetries of \cite{Clark:2005te,D'Hoker:2006uu}.
The solutions in this second family are dual to half-BPS planar interface operators 
\cite{DeWolfe:2001pq}, generalized to include varying coupling constant 
in \cite{Clark:2004sb,D'Hoker:2006uv,Gomis:2006cu}, and further
generalized to include also varying instanton $\theta$-angle in 
\cite{Gaiotto:2008sd,Gaiotto:2008sa}.
A third family consists of exact solutions  \cite{D'Hoker:2007fq} 
dual to half-BPS Wilson loops \cite{Drukker:1999zq}. (Earlier work 
\cite{Yamaguchi:2006te,Gomis:2006sb,Lunin:2006xr} includes a derivation of
the reduced BPS equations, and a study of  boundary conditions.) 

\sm

The knowledge of these exact solutions should provide a powerful starting point for the 
study of the spectrum of small fluctuations and of correlation functions, either exactly
or numerically, and thus for the spectrum and correlation functions  of the dual CFT.

\sm

In the present paper we shall extend the construction of exact solutions with
16 supersymmetries to the case of 11-dimensional supergravity, or M-theory.
The AdS/CFT correspondence for M-theory actually produces two
dualities. The first duality maps $AdS_4 \times S^7$ to a 3-dimensional
CFT with 32 supersymmetries, whose canonical bosonic fields are scalars.
The second duality maps $AdS_7 \times S^4$ to a 6-dimensional CFT
with 32 supersymmetries, whose canonical bosonic fields are 2 forms
with self-dual field strength.
Our present understanding of both of these CFTs is much weaker than
our understanding of 4-dimensional $\cN=4$ super Yang-Mills theory.

\sm

The difficulty of obtaining classical supergravity solutions with 16 supersymmetries 
is, however, roughly comparable to the difficulty of obtaining the corresponding 
solutions in Type IIB supergravity. One may hope that by improving our knowledge 
of exact solutions in M-theory we will be able to deepen our understanding of the 
dynamics of the corresponding more elusive CFTs. Currently, progress is being made 
towards constructing an effective field theory for multiple M2 branes 
\cite{Bagger:2007jr,Bagger:2007vi,Gustavsson:2008dy}. It will be 
interesting to see precisely how the half-BPS supergravity solutions we find in the present 
paper are dual to half-BPS planar interface operators  in the M-brane theory.

\sm

Specifically, we shall construct, in exact form, the complete local solution invariant
under 16 supersymmetries in 11-dimensional supergravity with the geometry
$AdS_3 \times S^3 \times S^3 \times \Sigma$ invariant under $SO(2,2)\times
SO(4) \times SO(4)$, and boundary asymptotics of either $AdS_4 \times S^7$
or $AdS_7 \times S^4$. Solutions with this boundary behavior are the most
immediately relevant in the context of the AdS/CFT correspondence. Their
construction is also technically simpler than that of solutions with more general
boundary conditions. The manifold $\Sigma$ is a 2-dimensional parameter
space over which the space $AdS_3 \times S^3 \times S^3$ is, generally,  warped.
The special case of solutions with space-time $AdS_3 \times S^3 \times S^3 \times E_2$,
where $E_2$ is flat Euclidean, and the product is not warped, was 
analyzed in \cite{Boonstra:1998yu} (see also \cite{Gauntlett:1998kc,de Boer:1999rh}).
Other types of solutions on various space-times with an $AdS_3$ factor,
and various degrees of supersymmetry, have been constructed in 
\cite{Gauntlett:2006ns,Gauntlett:2006qw,Gauntlett:2007ts}.

\sm

The Ansatz $AdS_3 \times S^3 \times S^3 \times \Sigma$ encompasses
as special cases both the solutions $AdS_4 \times S^7$ and $AdS_7 \times S^4$
with the maximum of 32 supersymmetries. For these special cases,
the supergravity 4-form field strength $F$ vanishes on two out of the three
factors $AdS_3 \times S^3 \times S^3 $, and is non-vanishing
and constant  on the third factor.

\sm

The solutions we shall obtain here will have a non-vanishing
4-form field strength $F$ on all three factors of $AdS_3 \times S^3 \times S^3 $,
thus allowing for non-zero M2- and M5-brane charges. Such supergravity field
configurations include fully back-reacted solutions of M2-branes in
$AdS_7 \times S^4$ and M5-branes in $AdS_4 \times S^7$. In principle, such solutions
could also be viewed as Maldacena limits of brane configurations with 8 supersymmetries.
Indeed, the intersection of a stack of coincident M2-branes with a stack
of coincident M5-branes over a 2-dimensonal string worldsheet is 1/4-BPS,
and thus leaves 8 supersymmetries. The Maldacena limit, either nearing the M2
horizon or nearing the M5 horizon, will produce 8 additional (conformal)
supersymmetries, bringing the total to 16 supersymmetries
\cite{Boonstra:1998yu,Gauntlett:1998kc,de Boer:1999rh}. 
The problem is that the corresponding 
fully localized intersecting M2/M5-brane solution is not (yet) known.

\sm

The search for such solutions invariant under 16 supersymmetries was initiated by 
Yamaguchi \cite{Yamaguchi:2006te}, and by Lunin in \cite{Lunin:2007ab}
where the BPS equations were reduced to an
$AdS_3 \times S^3 \times S^3 \times \Sigma$ Ansatz, a harmonic
function on $\Sigma $ was identified,  a semi-quantitative investigation
into the boundary conditions was carried out, and arguments for the existence 
of solutions were presented, based on a perturbative expansion. Obtaining 
complete solutions there, however, would still require solving non-linear partial 
differential equations, which was not done in \cite{Yamaguchi:2006te,Lunin:2007ab}. 
The main novelty of the present paper lies in the following results:
(1) The reduced BPS equations are completely solved, in exact form;
(2) It is shown that any solution to the BPS equations automatically solves
the M-theory Bianchi and field equations. (3) Two further harmonic functions 
are identified in the construction of the local solution.

\sm

Examination of the reduced BPS equations reveals (a fact also observed in 
\cite{Yamaguchi:2006te,Lunin:2007ab}) that the space of solutions is naturally 
foliated by three 
constant real parameters $c_i$, with $i=1,2,3$, which are subject to the
relation, $c_1+c_2+c_3=0$, but are  otherwise free. The absolute values
$|c_i|$ correspond to the inverse radii of each of the three factors in 
$AdS_3 \times S^3 \times S^3$. To make this correspondence precise, 
it will be useful to label the factors as follows, $AdS_3 \times S^3_2 \times S^3_3$,
where the $AdS_3$ factor corresponds to label 1, but this label will not
be exhibited. Thus, $|c_1|$  is the inverse radius of $AdS_3$, while 
$|c_i|$ is the radius of $S_i^3$ with $i=2,3$. A simultaneous rescaling 
of all three $c_i$ corresponds to an overall rescaling of the solution size.
The space of the constant parameters $c_i$ modulo their overall rescaling
consists of just a single free real parameter $c$ which plays the role of
an {\sl aspect ratio} of the solution. The dependence on $c$
of the warping of $AdS_3 \times S^3_2 \times S^3_3$ over $\Sigma$ is highly 
non-trivial.  

\sm

In the limit where one of the parameters $c_i$ tends to zero, the radius
of  the factor space with label $i$ in $AdS_3 \times S^3_2 \times S^3_3 $ tends
to infinity with the effect of decompactifying that space. Due to the
relation $c_1+c_2+c_3=0$, it is not possible for two factors to decompactify
simultaneously, while the third factor is left at finite radius. But it is possible
for all three $c_i$ to tend to zero, in which case the entire solution decompactifies.
A schematic representation of the 2-dimensional parameter space generated
by $c_1, c_2, c_3$ is depicted in Figure 1 below.

\begin{figure}[tbph]
\begin{center}
\epsfxsize=4in
\epsfysize=3in
\epsffile{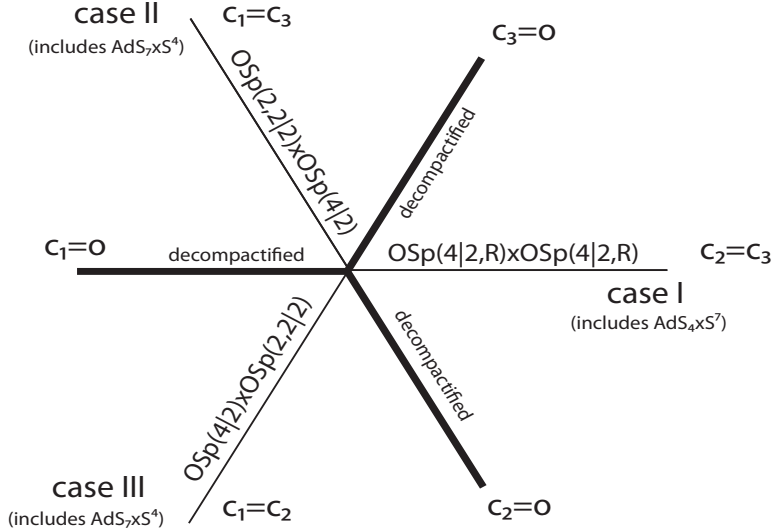}
\label{figure1}
\caption{The space of parameters $c_1,c_2,c_3$. 
Assignments differing only by an overall sign have been identified
in this representation.}
\end{center}
\end{figure}

The special solutions $AdS_4 \times S^7$ and $AdS_7 \times S^4$,
with the maximal number of 32 supersymmetries, correspond
to two of the parameters $c_i$ being coincident ($c_2=c_3$ for the first
case, and $c_1 = c_2$ or $c_1 = c_3$ for the second case). The constants
$c_i$ are determined completely by the boundary behavior of the solution.
It follows that any solution with $AdS_4 \times S^7$ boundary behavior on
all or on part of its boundary must have $c_2=c_3$. Any solution with
$AdS_7 \times S^4$ boundary behavior must have either $c_1=c_2$ or $c_1 = c_3$.
It also follows that solutions for which no two $c_i$ coincide with one another
have neither $AdS_4 \times S^7$ nor $AdS_7 \times S^4$ boundary behavior.
Finally, it also follows that no solution can interpolate between $AdS_4 \times S^7$
boundary behavior and $AdS_7 \times S^4$ boundary behavior. (Note
that this result does not assume global regularity of the solution.)

\sm

The space of solutions we obtain may also be considered from the point of
view of the supergroups under which the solutions are invariant. The
$SO(2,2) \times SO(4) \times SO(4)$ isometry group of our Ansatz
$AdS_3 \times S^3_2 \times S^3_3 \times \Sigma$ is the maximal bosonic
subgroup of the supergroup of the solutions. The cases $AdS_4 \times S^7$
and $AdS_7 \times S^4$ with maximal supersymmetry are invariant under
the supergroups $OSp(8|4,\bR)$ and $OSp(2,6|4)$ respectively.
General arguments \cite{DEGKS} show  that the supergroup under which any
solution is invariant must be a subgroup of either $OSp(8|4,\bR)$ or $OSp(2,6|4)$.
The possible supergroups with 16 supersymmetries available 
for the geometry $AdS_3 \times S^3_2 \times S^3_3 \times \Sigma$ 
are listed in Table 1 below.

\begin{table}[htdp]
\begin{center}
\begin{tabular}{|c||c|c|c|c |} \hline
case & relation & includes & supergroup of solution & maximal supergroup
\\ \hline \hline
I & $c_2=c_3$ & $AdS_4 \times S^7$ & $OSp(4|2,\bR) \times OSp(4|2,\bR) $
&  $OSp(8|4,\bR)$
\\ \hline
II & $c_3=c_1$ & $AdS_7 \times S^4$ &  $OSp(2,2|2) \times OSp(4|2) $
& $ OSp(2,6|4)$
\\ \hline
III & $c_1=c_2$ & $AdS_7 \times S^4$ &  $OSp(4|2) \times OSp(2,2|2) $
& $ OSp(2,6|4)$
\\ \hline
\end{tabular}
\end{center}
\label{table5}
\caption{Cases  with either $AdS_4 \times S^7$ or $AdS_7 \times S^4$
boundary behavior.}
\end{table}

In Table 1, only the cases where two of the $c_i$ coincide are listed,
because these are the only cases that allow for either $AdS_4 \times S^7$ or 
$AdS_7 \times S^4$ asymptotic boundary behavior. Their designations 
as cases I, II, or III will be used throughout. In Figure 1, 
the cases I, II, and III are indicated with thin lines. They are of most interest
to AdS/CFT and it is for these cases that the complete exact solutions
will be obtained in this paper.

\sm

The regions in Figure 1 where no two $c_i$ coincide with one another
correspond to solutions whose boundary behavior is neither that
of $AdS_4 \times S^7$ nor that of $AdS_7 \times S^4$. These solution spaces
are foliated by the aspect ratio parameter $c$. As a function of $c$,
the supergroup which leaves the solution invariant will change.
For example, across the decompactification lines $c_i=0$,
the boundary conditions change, and so do the corresponding supergroups,
by passing through a common Wigner-Inonu. There may also be, however,
a more intricate dependence of the supergroups on $c$, since it 
is possible to deform $OSp(4|2)$ into the exceptional supergroup $D(2,1,a)$.
Thus, in the region where no two $c_i$ coincide, the supergroup of the solution 
may actually be a non-compact real form of $D(2,1,a) \times D(2,1,b)$ with $a,b$ 
depending on the aspect ratio $c$. This possibility would be consistent 
with the M-brane analysis of \cite{Boonstra:1998yu,Gauntlett:1998kc}.

\sm

A final result concerns the integrable system onto which the reduced BPS
equations are mapped, and from which the exact solutions will be 
constructed. Its field is a single real scalar function $\tet$ on $\Sigma$.
The field equation is given by,
\bea
\label{intro1}
4  \p_{\bar w} \p_w \tet 
- \p_{\bar w} \bigg ( i e^{2 i \tet} \p_w \ln h \bigg )
+ \p_w \bigg (  i e^{- 2i \tet} \p_{\bar w} \ln h \bigg )
 =0
\eea
where $h$ is an arbitrary harmonic function on $\Sigma$.
This integrable system is of the sine-Gordon/Liouville type,
with translation invariance in the direction perpendicular to the
coordinate $h$, but broken translation invariance along the direction $h$.
This field theory is akin to the Liouville theory with broken translation
invariance of \cite{Liouville} and shares with it the remarkable 
property that its full solution may be 
obtained in explicit form. A similar integrable system was encountered
in the solution of Type IIB theory in \cite{D'Hoker:2007xy,D'Hoker:2007xz}.
Yet, the two systems are different from one another.

\sm

The remainder of this paper is organized as follows. 

\sm

In section 2, we briefly review 11-dimensional supergravity, spell out the
$AdS_3 \times S^3 _2 \times S^3 _3 \times \Sigma$ Ansatz for the supergravity
fields and supersymmetry spinor, and produce the reduced BPS equations,
which are the starting point of our exact solution. 

\sm

In section 3, we solve for the metric factors of the $AdS_3 \times S^3_2 \times S^3_3$ 
spaces, express these metric factors in terms of bilinears in the supersymmetry 
spinors, and reproduce the maximally supersymmetric solutions $AdS_4 \times S^7$
and $AdS_7 \times S^4$. 

\sm

In section 4, we derive from the reduced BPS equations, for all values of the 
constants $c_i$, a holomorphic 1-form $\kappa$ on $\Sigma$. 

\sm

In section 5, this holomorphic 1-form $\kappa$ is used to produce the complete 
solution to 
the BPS equations for case III, namely when $c_1=c_2$. This is achieved by 
mapping the reduced BPS  equations to the integrable system (\ref{intro1}), 
and then mapping this system onto a linear equation. The metric factors  of the 
solutions are computed explicitly in terms of the solutions to this linear equation.
In section 6, case II is solved by showing that it is related to case III by a 
simple discrete symmetry of the reduced BPS equations. In section 7, 
case I is solved by methods identical to the ones used to solve for case III.

\sm

In section 8, the remaining linear differential equation is solved exactly in terms 
of an integral transform of two holomorphic functions on $\Sigma$. 
It is shown that, alternatively, the solution may be obtained by
projection from a single 3-dimensional harmonic function.

\sm

In section 9, the flux fields are computed as well. It is shown that
the Bianchi identities, the field equations for the 4-form field strength, and
Einstein's equations all hold automatically for any solution to the BPS equations.

\sm

In Appendix A, we give a Clifford algebra representation adapted to
the $AdS_3 \times S^3 \times S^3 \times \Sigma$ geometry.
In Appendix B, we summarize the geometry of Killing spinors
for odd-dimensional spheres and odd-dimensional Minkowski AdS space-times.

\newpage

%%%%%%%%%%%%%%%%%%%%%%%%%%%%%%%%%%%%%%%%%%%%%%%
%%%%%%%%%%%%%%%%%%%%%%%%%%%%%%%%%%%%%%%%%%%%%%%
\section{M-theory Ansatz and reduced BPS equations}
\setcounter{equation}{0}
%%%%%%%%%%%%%%%%%%%%%%%%%%%%%%%%%%%%%%%%%%%%%%%
%%%%%%%%%%%%%%%%%%%%%%%%%%%%%%%%%%%%%%%%%%%%%%%

In this section, we shall construct the most general $SO(2,2) \times SO(4) \times SO(4)$
invariant Ansatz for the bosonic fields of eleven-dimensional supergravity on the 
manifold\footnote{Throughout, we shall use the notation in which the $AdS_3$ 
space corresponds to the label $a=1$ (which will not be exhibited on $AdS_3$), 
and the spheres $S^3 _a$  correspond to the  labels $a=2,3$.}
$AdS_3 \times S^3_2 \times S^3_3$ warped over a two-dimensional parameter 
space $\Sigma$. 
Since $\Sigma$ inherits an orientation and a metric from supergravity,
$\Sigma$ is automatically a Riemann surface, endowed with a complex structure.
The BPS equations will be reduced to this Ansatz and simplified through the use
of a Killing spinor basis for the spinors on $AdS_3 \times S^3_2 \times S^3_3$.

\subsection{Eleven-dimensional supergravity}

Eleven dimensional supergravity  \cite{Cremmer:1978km} contains two bosonic fields,
the metric $g_{MN}$, and a four form field strength $F_{PQRS}$,
with $M,N,P,Q,R,S=0,1, \cdots , 9,\natural \equiv 10$. 
It will be convenient to recast the tensor
$F_{PQRS}$ in terms of a 4-form
$F= F_{PQRS} dx^P\wedge dx^Q \wedge dx^R \wedge dx^S/24$, which
is given in terms of a 3-form potential $C$ by $F= dC$.
Supergravity also contains a fermion field, the gravitino $\Psi _M$, but here
we shall be interested in purely bosonic solutions, and thus set $\Psi_M=0$
throughout. The action is then given by
\bea
S={1\over 2 \kappa_{11}^2} \int d^{11}x \sqrt{-g} \Big(R-{1\over 48} F_{MNPQ}F^{MNPQ}\Big) -{1\over 12 \kappa_{11}^2} \int C \wedge F \wedge F
\eea
The field equation for the metric $g_{MN}$ is,
\bea
\label{Einstein}
R_{MN}-{1\over 12} F_{MPQR}F_N^{\;\; PQR} +{1\over 144} g_{MN} F_{PQRS}F^{PQRS}=0
\eea
The Bianchi identity, and field equation  for the 4-form $F$ are respectively,
\bea
\label{Maxwell}
dF=0 \hskip 1in
d*F + {1\over 2} F \wedge F=0
\eea
Supersymmetry of a purely bosonic field configuration under a transformation with
supersymmetry parameter $\ep$ requires that the supersymmetry variation of
$\Psi _M$ with respect to $\ep$ must vanish for vanishing $\Psi _M$. This requirement
yields the BPS equations,
\bea
\label{fullgravitino1}
\nabla_M \ep +{1\over 288} \Big(\Gamma_M^{\;\; NPQR}
- 8 \delta_M{}^N \Gamma^{PQR} \Big) F_{NPQR} \, \ep=0
\eea
Here, $\ep$ is an eleven dimensional Majorana spinor, and $\nabla _M$ is
the covariant derivative with respect to the Levi-Civita connection for $g_{MN}$.
It will be convenient to use the identity
\bea
 \Gamma ^ {MNPQR} - 8 \delta ^{M[N} \Gamma^{PQR]}
 = -\half \G^M \G^{NPQR} + {3 \over 2} \G^{NPQR} \G^M
\eea
to recast the BPS equation in the form,\footnote{Throughout, we shall use the
notation $\G \cdot T = \G^{M_1 \cdots M_p} T_{M_1 \cdots M_p}$ for the
contraction of an antisymmetric tensor field of rank $p$ with the $\G$-matrix
of the same rank.}
\bea
\label{fullgravitino}
\nabla_M \ep +{1\over 24^2} \Big( - \Gamma_M (\G \cdot F)
+ 3 (\G \cdot F) \G_M \Big) \ep=0
\eea
The advantage of this form of the BPS equations is that only the
combination $\G \cdot F$ appears.

\subsection{Invariant Ansatz for supergravity fields}

The $SO(2,2) \times SO(4) \times SO(4)$ invariant Ansatz for the supergravity
fields on $AdS_3 \times S^3_2 \times S^3_3$ warped over $\Sigma$ are as follows.
The Ansatz for metric is given by,
\bea
ds^2= f_1^2 \;ds_{AdS_3}^2+ f_2^2 \;ds_{S^3_2}^2
+ f_3^2 \; ds_{S^3_3}^2+ ds_{\Sigma}^2
\eea
Here, $ds_{AdS_3}^2$,  $ds_{S^3_2}^2$, and $ds_{S^3_3}^2$, are the unit radius
metrics on the corresponding spaces, respectively invariant under $SO(2,2)$, $SO(4)$
and $SO(4)$. The metric $ds^2 _\Sigma$ is an unspecified Riemannian metric on $\Sigma$,
and the {\sl metric factors} $f_1, f_2$, and $f_3$ are unspecified real (but not
necessarily positive) functions on $\Sigma$,
the expressions for all of which will be determined by imposing the BPS equations.
It will be convenient to cast the supergravity metric (and other fields) in terms of
an eleven-dimensional Lorentz frame $e^A \equiv dx^M \, e_M {}^A$,
with $A=0,1, \cdots , 9, \natural$,  chosen as follows,
\bea
e^{i_1} = f_1 \hat e^{i_1} & \hskip 1in &  i_1=0,1,2
\no \\
e^{i_2} = f_2 \hat e^{i_2} &&  i_{2}=3,4,5
\no \\
e^{i_3} = f_3 \hat e^{i_3} &&  i_{3}=6,7,8
\no \\
e^a \hskip 0.57in &&  a=9,10
\eea
The frames $\hat e^{i_1}, \hat e^{i_2} $, and $ \hat e^{i_3} $ correspond to the unit
radius metrics on  $AdS_3, S^3_2$, and $S^3_3$ respectively.
The Ansatz for the 3-form potential $C$ is conveniently expressed as follows,
\bea
C= b_1 \hat e^{012}+ b_2 \hat e^{345}+ b_3 \hat e^{678}
\eea
where $b_1, b_2$, and $b_3$ are real functions on $\Sigma$.
The corresponding field strength  takes the form,
\bea
F =  g_{1 a} e^{012a} + g_{2 a} e^{345a} + g_{3 a} e^{678a}
\eea
with
\bea
\label{bianchipre}
g_{i a} = - {1 \over f_i^3} D_a b_i \hskip 1in i=1,2,3
\eea
Throughout, the covariant derivative on $\Sigma$ acting on scalar functions on
$\Sigma$ will be denoted by $D_a \equiv e_a {}^M \p_M$ where $e_A {}^ M$ is
the inverse of the Lorentz frame $e_M {}^A$.

\subsection{Invariant Ansatz for the supersymmetry parameters}

The supersymmetry parameter $\ep$ must be compatible with the
$SO(2,2) \times SO(4) \times SO(4)$ symmetry of the problem, and
globally well-defined on the symmetric spaces $AdS_3$, $S_2^3$, and $S_3^3$.
Killing spinors on $AdS_3$, $S_2^3$, and $S_3^3$ provide a convenient basis for
the invariant and globally well-defined spinors $\ep$ on these symmetric
spaces.\footnote{Conventions for $\G$ matrices are in Appendix A, and basic results
on Killing spinors for odd dimensional  symmetric spaces are collected in  Appendix B.}
The factors $AdS_3$, $S_2^3$, and $S_3^3$ each carry two-dimensional irreducible
spinors. It will be convenient to organize the invariant spinors on
$AdS_3 \times S_2^3 \times S_3^3$ directly in terms of 8-dimensional spinors $\chi$.
The Killing spinor equations on $AdS_3 \times S_2^3 \times S_3^3$ take the following
form,
\bea
\label{symmkilling}
0 &=& \bigg( \hat \nabla_{i_1}
- {\eta_1 \over 2 } (\gamma_{i_1} \otimes I_2 \otimes I_2)
\bigg) \chi^{\eta_1, \eta_2, \eta_3}
\no\\
0 &=& \bigg( \hat \nabla_{i_2}
- i {\eta_2 \over 2 } (I_2 \otimes \gamma_{i_2} \otimes I_2)
\bigg) \chi^{\eta_1, \eta_2, \eta_3}
\no\\
0 &=& \bigg( \hat \nabla_{i_3}
- i {\eta_3 \over 2 } (I_2 \otimes I_2 \otimes \gamma_{i_3})
\bigg) \chi^{\eta_1, \eta_2, \eta_3}
\eea
The covariant derivatives $\hat \nabla _{i_1}, \hat \nabla _{i_2}$, and $\hat \nabla _{i_3}$
are with respect to the unit radius metrics of the corresponding spaces.
The parameters $\eta_1$, $\eta_2$ and $\eta_3$ are restricted to take the
values $\pm 1$ as a result of requiring that the integrability conditions be satisfied.
The equations (\ref{symmkilling}) then have solutions of maximal rank, and
a basis of Killing spinors is provided by the 8 linearly independent solutions
corresponding to $\eta _1 =\pm 1, \eta _2 = \pm 1$, and $\eta _3 = \pm 1$.

\sm

The supersymmetry parameter $\ep$ may be decomposed in the basis of Killing
spinors $\chi ^{\eta_1, \eta _2, \eta _3}$, and we shall denote the corresponding
coefficients by $\zeta_{\eta_1, \eta_2, \eta_3}$. For a fixed assignment of
$\eta_1$, $\eta_2$, and $\eta_3$, the object $\zeta_{\eta_1, \eta_2, \eta_3}$ is
a four component spinor, and we have
\bea
\ep = \sum_{\eta_1, \eta_2, \eta_3}
\chi^{\eta_1, \eta_2, \eta_3} \otimes \zeta_{\eta_1, \eta_2, \eta_3}
\eea
Without  loss of generality, we may impose a reality condition on the basis of
Killing spinors $\chi^{\eta_1 ,\eta_2 ,\eta_3}$,  since the coefficients
$\zeta_{\eta_1, \eta_2, \eta_3}$ are, in general, allowed to be complex spinors.
The proper reality condition is
\bea
(\chi^{\eta_1, \eta_2, \eta_3})^*
= (I_2 \otimes \sigma^2 \otimes \sigma^2) \chi^{\eta_1, \eta_2, \eta_3}
\eea
This reality condition on the Killing spinors $\chi^{\eta_1 ,\eta_2 ,\eta_3}$,
together with the Majorana condition $\ep ^* = B \ep$ on the full 32-component
spinor $\ep$ implies a corresponding reality condition on the coefficients
$\zeta_{\eta_1, \eta_2, \eta_3}$, which is found to be,
\bea
\label{zetareality}
(\zeta_{\eta_1, \eta_2, \eta_3})^* = (\sigma^3 \otimes \sigma^1) \zeta_{\eta_1, \eta_2, \eta_3}
\eea
Finally, we may use the Killing spinor equations (\ref{symmkilling}) to express the
eleven-dimensional covariant derivatives along the symmetric spaces in terms of
an algebraic action by the eleven-dimensional $\G$-matrices as follows,
\bea
\label{elevenkilling}
\nabla_{i_1} \epsilon &=&
\sum_{\eta_1, \eta_2, \eta_3} \bigg( + {\eta_1 \over 2 f_1} \Gamma_{i_1} \Gamma^{012}
+ \half \omega_{i_1 j_1 a} \Gamma^{j_1 a} \bigg)
\chi^{\eta_1, \eta_2, \eta_3} \otimes \zeta_{\eta_1, \eta_2, \eta_3}
\no\\
\nabla_{i_2} \epsilon &=&
\sum_{\eta_1, \eta_2, \eta_3} \bigg( - {\eta_2 \over 2 f_2} \Gamma_{i_2} \Gamma^{345}
+ \half \omega_{i_2 j_2 a} \Gamma^{j_2 a} \bigg)
\chi^{\eta_1, \eta_2, \eta_3} \otimes \zeta_{\eta_1, \eta_2, \eta_3}
\no\\
\nabla_{i_3} \epsilon &=&
\sum_{\eta_1, \eta_2, \eta_3} \bigg( - {\eta_3 \over 2 f_3} \Gamma_{i_3} \Gamma^{567}
+ \half \omega_{i_3 j_3 a} \Gamma^{j_3 a} \bigg)
\chi^{\eta_1, \eta_2, \eta_3} \otimes \zeta_{\eta_1, \eta_2, \eta_3}
\eea
These equations will guarantee that the 9 space-time vector components of the
BPS equations along the directions of $AdS_3 \times S_2^3 \times S_3^3$
will result in algebraic conditions on $\zeta _{\eta_1, \eta _2, \eta _3}$.

\subsection{The reduced BPS equations}

To reduce the BPS equations to the $SO(2,2) \times SO(4) \times SO(4)$
invariant Ansatz, we need
the components of the 11-dimensional Lorentz connection $\o^A {}_B$. 
They are given as follows,
\bea
\label{spincon}
\omega^{i_1} {}_{j_1} = \hat \omega^{i_1} {}_{j_1}
&\qquad&
\omega^{i_1} {}_a = e^{i_1} {D_a f_1 \over f_1}
\no\\
\omega^{i_2} {}_{j_2} = \hat \omega^{i_2} {}_{j_2}
&\qquad&
\omega^{i_2} {}_a = e^{i_2} {D_a f_2 \over f_2}
\no\\
\omega^{i_3} {}_{j_3} = \hat \omega^{i_3} {}_{j_3}
&\qquad&
\omega^{i_3} {}_a = e^{i_3} {D_a f_3 \over f_3}
\eea
together with the connection on $\Sigma$, given by $\o^a {}_b = \hat \o \epsilon ^a {}_b$
for which we use the convention that $\epsilon ^{9\natural }= \epsilon ^9 {}_\natural= +1$.
All other components vanish.
The connections with hats refer to the connections on the unit radius manifolds
$AdS_3$, $S_2^3$, and $S_3^3$ respectively. 

\sm

The Killing spinor equations on the basis spinors $\chi ^{\eta _1, \eta _2, \eta _3}$  
allow us to recast the action of the covariant derivatives $\nabla _{i_1}$, $\nabla _{i_2}$,
and $\nabla _{i_3}$ on the supersymmetry parameter $\ep$ in terms of a purely
algebraic linear action on $\ep$. To simplify the action of the covariant derivatives
$\nabla _a$ on $\ep$ along $\Sigma$, we make use of
\bea
\nabla_a \chi^{\eta_1, \eta_2, \eta_3} \otimes \zeta_{\eta_1, \eta_2, \eta_3}
= \chi^{\eta_1, \eta_2, \eta_3}
\otimes \bigg( D_a \zeta_{\eta_1, \eta_2, \eta_3}
+ {i \over 2} \hat \omega_a (1 \otimes \sigma^3) \zeta_{\eta_1, \eta_2, \eta_3} \bigg)
\eea
To simplify the terms in the BPS equation that involve $F$, we use the
following relation,
\bea
{1 \over 24} \G \cdot F = g_{1a} \G^{012a} + g_{2a} \G^{345a} + g_{3a} \G^{678a}
\eea
where the multiple $\G$-matrices take the form, (see Appendix A for $\G$-matrix conventions),
\bea
\G^{0123} & = & + I_8 \otimes \s^1 \otimes \s^3 \s^a
\no \\
\G^{345a} & = & -i I_8 \otimes \s^2 \otimes \s^3 \s^a
\no \\
\G^{678a} & = & -i I_8 \otimes \s^3 \otimes \s^3 \s^a
\eea
Here, $I_8$ is the identity matrix in the 8-component spinor space
generated by the Killing spinors $\chi ^{\eta _1, \eta _2, \eta _3}$,
while the last two factors in the tensor product act on the
4-component spinors $\zeta _{\eta _1, \eta _2, \eta _3}$.
The resulting reduced BPS equations take the following form.

\sm

On the $AdS_3$ space, we have,
\bea
\label{grav1}
0 &=&
{\eta_1 \over 2 f_1} (\sigma^1 \otimes \sigma^3) \zeta_{\eta_1, \eta_2, \eta_3}
+ \half {D_a f_1 \over f_1} (1 \otimes \sigma^a) \zeta_{\eta_1, \eta_2, \eta_3}
\\&&
- {i \over 12} g_{2 a} (\sigma^2 \otimes \sigma^3 \sigma^a) \zeta_{\eta_1, \eta_2, \eta_3}
- {i \over 12} g_{3 a} (\sigma^3 \otimes \sigma^3 \sigma^a) \zeta_{\eta_1, \eta_2, \eta_3}
- {1 \over 6} g_{1 a} (\sigma^1 \otimes \sigma^3 \sigma^a) \zeta_{\eta_1, \eta_2, \eta_3}
\no \eea
On the $S_2^3$ space, we have
\bea
\label{grav2}
0 &=&
i {\eta_2 \over 2 f_2} (\sigma^2 \otimes \sigma^3) \zeta_{\eta_1, \eta_2, \eta_3}
+ \half {D_a f_2 \over f_2} (1 \otimes \sigma^a) \zeta_{\eta_1, \eta_2, \eta_3}
\\&&
+ {1 \over 12} g_{1 a} (\sigma^1 \otimes \sigma^3 \sigma^a) \zeta_{\eta_1, \eta_2, \eta_3}
- {i \over 12} g_{3 a} (\sigma^3 \otimes \sigma^3 \sigma^a) \zeta_{\eta_1, \eta_2, \eta_3}
+ {i \over 6} g_{2 a} (\sigma^2 \otimes \sigma^3 \sigma^a) \zeta_{\eta_1, \eta_2, \eta_3}
\no
\eea
On the $S_3^3$ space, we have,
\bea
\label{grav3}
0 &=&
i {\eta_3 \over 2 f_3} (\sigma^3 \otimes \sigma^3) \zeta_{\eta_1, \eta_2, \eta_3}
+ \half {D_a f_3 \over f_3} (1 \otimes \sigma^a) \zeta_{\eta_1, \eta_2, \eta_3}
\\&&
+ {1 \over 12} g_{1 a} (\sigma^1 \otimes \sigma^3 \sigma^a) \zeta_{\eta_1, \eta_2, \eta_3}
- {i \over 12} g_{2 a} (\sigma^2 \otimes \sigma^3 \sigma^a) \zeta_{\eta_1, \eta_2, \eta_3}
+ {i \over 6} g_{3 a} (\sigma^3 \otimes \sigma^3 \sigma^a) \zeta_{\eta_1, \eta_2, \eta_3}
\no \eea
Finally, on $\Sigma$ the BPS equations reduce to differential equations in $\zeta$,
\bea
\label{grav4}
0 &=&
D_a \zeta_{\eta_1, \eta_2, \eta_3}
+ {i \over 2} \hat \omega_a (1 \otimes \sigma^3) \zeta_{\eta_1, \eta_2, \eta_3}
- {i \over 12}  \epsilon_a{}^b (\sigma^1 \otimes 1) g_{1 b} \zeta_{\eta_1, \eta_2, \eta_3}
\\&&
- {1 \over 12} \, \epsilon_a{}^b (\sigma^2 \otimes 1) g_{2 b} \zeta_{\eta_1, \eta_2, \eta_3}
- {1 \over 12} \, \epsilon_a{}^b (\sigma^3 \otimes 1) g_{3 b} \zeta_{\eta_1, \eta_2, \eta_3}
\no\\&&
+ {1 \over 6} \, (\sigma^1 \otimes \sigma^3) g_{1 a} \zeta_{\eta_1, \eta_2, \eta_3}
- {i \over 6} \, (\sigma^2 \otimes \sigma^3) g_{2 a} \zeta_{\eta_1, \eta_2, \eta_3}
- {i \over 6} \, (\sigma^3 \otimes \sigma^3) g_{3 a} \zeta_{\eta_1, \eta_2, \eta_3}
\no
\eea
The task at hand is to simplify these equations and the goal is to solve
them for $\zeta$, as well as for the metric factors $f_1,f_2,f_3$, the metric 
$ds^2 _\Sigma$, and the fluxes $g_1,g_2,g_3$.
To do so, we begin by exhibiting discrete symmetries that will allow us to reduce
the above equations for all 8 possible assignments of $\eta_1, \eta_2$, $\eta _3$ 
to the equation for just a single value $\eta _1 = \eta _2 = \eta _3 =+1$.

\subsection{Discrete symmetries}

The reduced BPS equations (\ref{grav1}), (\ref{grav2}), (\ref{grav3}), and (\ref{grav4})
are invariant under the following involutions,
\bea
S_0 : \zeta \, \to \, i (I \otimes \sigma ^3) \, \zeta
& \hskip 0.7in &
\eta _i \to - \eta _i
\\ && \no \\
S_j : \zeta \, \to \, s_j (\sigma ^j  \otimes I) \, \zeta
& \hskip 0.7in &
\eta _i \to - (-)^{\delta _{i,j}} \eta _i \hskip 1in g_{ia} \to - (-)^{\delta _{i.j}}  g_{ia}
\no \eea
where $j=1,2,3$, with $s_1=i$, and $ s_2=s_3=1$.
All other fields are being left invariant.  The factors of $i$ in the definitions of these
transformations have been chosen so that these discrete symmetries also leave
the reality condition  (\ref{zetareality}) on $\zeta$ invariant. Taking $S_0,\, S_1$,
and $S_3$ as the functionally independent generators of these commuting involutions,
the full symmetry group consists of the eight elements
$\{ I, \, S_0, \, S_1, \, S_0S_1, \, S_2, \, S_0S_2, \, S_3, \, S_0 S_3 \}$.
These 8 generators map $\zeta _{+++}$ to all eight components.
Thus, it suffices to solve for $\zeta _{+++}$.

\subsection{Reducing the BPS equations to 2-component spinors
\label{compconv}}

The discrete symmetries, spelled out in the preceding subsection,
allow us to reduce the BPS equations for all 8 components
$\zeta _{\eta _1, \eta _2, \eta_3}$ to a reduced BPS equation for just
any single one of these 8 components, which we choose to be $\zeta _{+++}$.
The other seven components are then recovered by applying the involutions
found above. The reality condition (\ref{zetareality}) on $\zeta _{+++}$ reads
$(\zeta_{+++})^* = (\sigma^3 \otimes \sigma^1) \zeta_{+++}$. In a basis
where $I_2 \otimes \s^3$ is block-diagonal, we  have
\bea
\s^3 \otimes \s^1 = \left ( \matrix{ 0 & \s^3 \cr \s^3 & 0 \cr} \right )
\eea
and the reality condition on the 4-component spinor $\zeta _{+++}$
may be solved in terms of a 2-component complex spinor $\xi$
and its complex conjugate $\xi^*$, as follows,
\bea
\zeta _{+++}= \left ( \matrix{ \s^3 \xi ^* \cr \xi \cr} \right )
\eea
To recast the reduced BPS equations in terms of the 2-component
spinors $\xi$ and $\xi^*$, it will be convenient to make the complex structure
of the tangent space of $\Sigma$ explicit. This is because the
complex structure that separates $\xi$ from $\xi^*$ is intimately
intertwined with the complex structure on $\Sigma$ by the reduced BPS
equations.

\sm

We shall use the following conventions for the frame metric and 
anti-symmetric tensor, 
\bea
\delta ^{z \bar z} = 2 \hskip 0.6in \delta _{z \bar z} = \half
\hskip 1in 
\epsilon _z {}^z = - \epsilon _{\bar z} {} ^{\bar z} = i
\eea
We introduce the following complex frame components $e^z$ and $e ^{\bar z}$ 
for $\Sigma$,
\bea
e^z = (e^9 + i e^\natural )/2  & \hskip 1in & e_z = e^9 - i e^\natural
\no \\
e^{\bar z} = (e^9 - i e^\natural )/2 && e_{\bar z} = e^9 + i e^\natural
\eea
so that $ds^2 _\Sigma = 4 e^z e^{\bar z}$, and extend this pattern for
any vector or tensor, such as,
\bea
g_{iz} = g_{i9} - i g_{i \natural} \hskip 1in \s _z = \s^9 - i \s^\natural = \s^1 - i \s^2
\eea
It will also be convenient to introduce conformal coordinates $w$, $\bar w$ defined by
\bea
e^z = \rho(w,\bar w) \, dw
\qquad \qquad
e^{\bar z} = \rho(w, \bar w) \, d \bar w
\eea
where $\rho$ is the scale factor appearing in the metric, 
$ds^2_\Sigma = 4 \rho^2 |dw|^2$.  In these coordinates, 
the spin-connection 1-form $\hat \o$ on $\Sigma$ is given by
\bea
\hat \o = -i \rho^{-1} \p_{\bar w} \rho \, d \bar w +i \rho^{-1} \p_w \rho \, dw
\eea
Expressing the reduced BPS equations (\ref{grav1}), (\ref{grav2}), (\ref{grav3}), 
and (\ref{grav4}), in terms of $\xi$, $\xi^*$, and conformal coordinates $w, \bar w$ 
further reduces the BPS equations. We find three purely algebraic equations  
in $\xi$ and $\xi^*$, (recall that we have set $\eta_1=\eta _2 = \eta _3 =+1$), 
\bea
\label{sum1}
0 & = &
- {i \over 2 f_1} \sigma^2 \xi^*
+  {D_z f_1 \over 2f_1} \xi
- {1 \over 6} g_{1 z} \sigma^1 \xi
- {i \over 12} g_{2 z} \sigma^2 \xi
- {i \over 12} g_{3 z} \sigma^3 \xi
\no\\
0 & = &
- {1 \over 2 f_2} \sigma^1 \xi^*
+  {D_z f_2 \over 2f_2} \xi
+ {1 \over 12} g_{1 z} \sigma^1 \xi
+ {i \over 6} g_{2 z} \sigma^2 \xi
- {i \over 12} g_{3 z} \sigma^3 \xi
\no\\
0 & = &  + {i \over 2 f_3} \xi^*
+  {D_{z} f_3 \over 2f_3} \xi
+ {1 \over 12} g_{1 z} \sigma^1 \xi
- {i \over 12} g_{2 z} \sigma^2 \xi
+ {i \over 6} g_{3 z} \sigma^3 \xi
\eea
and two differential equations on $\xi$,
\bea
\label{sum2}
0 &=& D_z \xi
- {i \over 2} \hat \omega_z  \xi
- {1 \over 12} \,  g_{1 z}  \s^1  \xi
+ {i \over 12} \, g_{2 z} \sigma^2   \xi
+ {i \over 12} \, g_{3 z} \sigma^3 \xi
\no \\
0 & =  & D_{z} \xi^\dag + {i \over 2} \hat \omega_{z} \xi^\dag
- {1 \over 4} \,  g_{1 z} \xi^\dag \s^1
- {i \over 4} \, g_{2 z} \xi^\dag \s^2
- {i \over 4} \, g_{3 z} \xi^\dag \s^3
\eea
The  complex conjugates of the above differential equations  are given by,
\bea
0 & =  & D_{\bar z} \xi - {i \over 2} \hat \omega_{\bar z} \xi
- {1 \over 4} \, g_{1 \bar z} \s^1  \xi
+ {i \over 4} \, g_{2 \bar z}  \s^2 \xi
+ {i \over 4} \, g_{3 \bar z}  \s^3  \xi
\no\\
0&=&
D_{\bar z} \xi^\dag
+ {i \over 2} \hat \omega_{\bar z}  \xi^\dag
- {1 \over 12} \,  g_{1 \bar z}    \xi^\dag \s^1
- {i \over 12} \, g_{2 \bar z}    \xi^\dag \sigma^2
- {i \over 12} \, g_{3 \bar z}  \xi^\dag \sigma^3
\eea
The reduced BPS equations (\ref{sum1}) and (\ref{sum2}) will constitute
the basic point of departure for the construction of our solutions. In the subsequent
sections, we shall systematically progress towards obtaining their complete
local solutions for the cases I, II, and III defined in the Introduction, namely
for boundary behavior given by $AdS_4 \times S^7$ or $AdS_7 \times S^4$.

\newpage

%%%%%%%%%%%%%%%%%%%%%%%%%%%%%%%%%%%%%%%%%%%
%%%%%%%%%%%%%%%%%%%%%%%%%%%%%%%%%%%%%%%%%%%
\section{Metric factors of the $AdS_3$ and $S^3$ spaces}
\setcounter{equation}{0}
%%%%%%%%%%%%%%%%%%%%%%%%%%%%%%%%%%%%%%%%%%%
%%%%%%%%%%%%%%%%%%%%%%%%%%%%%%%%%%%%%%%%%%%

In this section, we shall  solve for the metric factors $f_i$ as  functions
of bilinears in the supersymmetry parameters $\xi$ and $\xi^*$. In doing so,
three real integration constants $c_1, c_2, c_3$ will emerge.  Their significance
was announced in the Introduction: their absolute values $|c_1|, |c_2|$, 
and $ |c_3|$ correspond to the inverse radii of the factor spaces $AdS_3$, 
$S_2^3$, and $S_3^3$ respectively.
The constants $c_1,c_2$, and $c_3$ will obey a single relation, $c_1+c_2+c_3=0$,
which implies a harmonic-like relation between the radii of the $AdS_3$, $S_2^3$, and
$S_3^3$ factors. The solutions with 32 supersymmetries,
namely $AdS_4 \times S^7$ and $AdS_7 \times S^4$, correspond to a different
assignment of these constants. It will follow that for any solution with
$AdS_4 \times S^7$ (respectively $AdS_7 \times S^4$) boundary asymptotics, the
constants $c_1, c_2, c_3$ will be completely fixed as well, and that no solutions
can have mixed $AdS_4 \times S^7$ and $AdS_7 \times S^4$ boundary asymptotics.

\subsection{Solving for the metric factors $f_i$}

The algebraic reduced BPS equations (\ref{sum1}) involve the metric factors
$f_i$, while the differential reduced BPS equations (\ref{sum2}) do not. 
The metric factors $f_i$ 
may be solved as bilinear functions of $\xi$ and $\xi^*$. To derive this result, 
it will be convenient to introduce a shorthand notation for bilinears in $\xi$ and $\xi^*$.
A convenient basis is given by,
\bea
\lambda _i = \xi ^\dag \s_i \xi  \hskip 1in
\mu _i = \xi ^t \s_i \xi
\eea
where $i=0,1,2,3$, and we have defined $\s_0 \equiv I$, as usual.
As  a result, the combinations $\l_i$ are real, but $\mu_i$ are generally complex.
We have $\mu _2=0$, as well as the relations,
\bea
\label{lam1}
0 & = & \l_0^2 - \l_1^2 - \l_2^2 - \l_3^2
\no \\
0 & = & \mu_0^2 - \mu_1^2 - \mu _3^2
\eea
From the differential equations (\ref{sum2}) for $\xi$ and $\xi^\dagger$, 
we deduce the following
differential equations for $\l_i$,
\bea
\label{Dl}
D_{z} \l_0 & = &
+ {1 \over 3} g_{1 z} \l_1 + {i \over 6} g_{2 z} \l_2 + {i \over 6} g_{3 z} \l_3
\no \\
D_{z} \l_1 & = &
+ {1 \over 3} g_{1 z} \l_0 + {1 \over 3} g_{2 z} \l_3 - {1 \over 3} g_{3 z} \l_2
\no \\
D_{z} \l_2 & = &
+ {i \over 6} g_{1 z} \l_3 + {i \over 6} g_{2 z} \l_0 + {1 \over 3} g_{3 z} \l_1
\no \\
D_{z} \l_3 & = &
- {i \over 6} g_{1 z} \l_2 - {1 \over 3} g_{2 z} \l_1 + {i \over 6} g_{3 z} \l_0
\eea
These equations are functionally dependent, in view of the quadratic relation
(\ref{lam1}). Differential equations for  $\mu_i$ can also
be derived, but those will not be needed here.

\sm

To solve for the metric factors $f_i$, we begin by recasting the three algebraic
reduced BPS equations (\ref{sum1}) in terms of an equivalent set of two groups
of equations for the bilinears of $\xi$. The first set is obtained by multiplying
each equation of (\ref{sum1}) on the left by $\xi ^t \s^2$. This procedure has the
effect of eliminating all the derivative terms in $f_1,f_2,f_3$, and we find,
\bea
\label{alg1}
0 & = & - {i  \over 2 f_1} \l_0 + {i \over 6} g_{1 z} \mu_3
- {i \over 12} g_{2 z} \mu_0 +  {1 \over 12} g_{3 z} \mu_1
\no \\
0 & = & + {i  \over 2 f_2} \l_3 - {i \over 12} g_{1 z} \mu_3
+ {i \over 6} g_{2 z} \mu_0 +  {1 \over 12} g_{3 z} \mu_1
\no \\
0 & = & - {i  \over 2 f_3} \l_2 - {i \over 12} g_{1 z} \mu_3
- {i \over 12} g_{2 z} \mu_0 -  {1 \over 6} g_{3 z} \mu_1
\eea
The second set is obtained by multiplying the first equation of (\ref{sum1})
by $\xi^\dag$, the second by $\xi^\dag \s^3$ and the third by $\xi ^\dag \s^2$.
This procedure produces combinations in terms of bilinears in $\xi$
that are linearly independent of the ones produced in (\ref{alg1}).
This will guarantee that all information contained in (\ref{sum1}) will
have been included in the set (\ref{alg1}) and the set (\ref{difrad}) below.
The effect of this procedure is to eliminate the first term in each of the
equations of (\ref{sum1}). The resulting equations are,
\bea
\label{difrad}
0 & = &  {D_{z} f_1 \over 2 f_1} \l_0
- {1 \over 6} g_{1 z} \l_1 - {i \over 12} g_{2 z} \l_2 - {i \over 12} g_{3 z} \l_3
\no \\
0 & = &  {D_{z} f_2 \over 2 f_2} \l_3
+ {i \over 12} g_{1 z} \l_2 + {1 \over 6} g_{2 z} \l_1 - {i \over 12} g_{3 z} \l_0
\no \\
0 & = &  {D_{z} f_3 \over 2 f_3} \l_2
- {i \over 12} g_{1 z} \l_3 - {i \over 12} g_{2 z} \l_0 - {1 \over 6} g_{3 z} \l_1
\eea
By taking suitable linear combinations of the three equations in (\ref{difrad}) 
with those for
$D_z \l_i$ for $i=0,2,3$ in (\ref{Dl}), we obtain,
\bea
\label{grad}
f_1 D_{z} \left ( {\l_0 \over f_1} \right ) =
f_2 D_{z} \left ( {\l_3 \over f_2} \right ) =
f_3 D_{z} \left ( {\l_2 \over f_3} \right ) = 0
\eea
These equations are easily integrated to give expressions for the metric factors in
terms of the spinor bilinears $\lambda_i$,
\bea
\label{rads1}
f_1 = {\l_0 \over c_1}
\qquad
\qquad
f_2 = - {\l_3 \over  c_2}
\qquad
\qquad
f_3 = {\l_2 \over  c_3}
\eea
where the $c_i$ are the corresponding real integration constants.  The minus
sign in the definition of $c_2$ has been inserted in order to exhibit a higher
degree of symmetry between the constants $c_1,c_2,c_3$, as will become clear later. 
Note that
the absolute value $|c_i|$ may be viewed as the inverse radius of the
$AdS_3$ or $S^3$ factor multiplying the corresponding bilinear $\l_i$.

\subsection{Summary of the remaining reduced BPS equations}

Using the results of (\ref{rads1}), the remaining algebraic equations (\ref{alg1})
may be considerably simplified. They become, after multiplication by an overall factor of 2,
\bea
\label{alg2}
0 & = &
c_1 - { 1 \over 3} g_{1z} \mu _3 + { 1 \over 6} g_{2z} \mu_0 + {i \over 6} g_{3z} \mu_1
\no \\
0 & = &
 c_2 + { 1 \over 6} g_{1z} \mu _3 - { 1 \over 3} g_{2z} \mu_0 + {i \over 6} g_{3z} \mu_1
\no \\
0 & = &
c_3 + { 1 \over 6} g_{1z} \mu _3 + { 1 \over 6} g_{2z} \mu_0 - {i \over 3} g_{3z} \mu_1
\eea
The sum of the three equations yields the relation on the integration constants,
\bea
\label{c123}
0 =  c_1  + c_2  + c_3
\eea
which was announced in the Introduction. Equations (\ref{alg2}) only involve 
$\xi$, $\xi^*$, and the reduced  flux fields $g_{iz}$. 
There also remain the differential reduced BPS equations (\ref{sum2}),
\bea
\label{sum2a}
0 &=& D_z \xi
- {i \over 2} \hat \omega_z  \xi
- {1 \over 12} \,  g_{1 z}  \s^1  \xi
+ {i \over 12} \, g_{2 z} \sigma^2   \xi
+ {i \over 12} \, g_{3 z} \sigma^3 \xi
\no \\
0 & =  & D_{z} \xi^\dag + {i \over 2} \hat \omega_{z} \xi^\dag
- {1 \over 4} \,  g_{1 z} \xi^\dag \s^1
- {i \over 4} \, g_{2 z} \xi^\dag \s^2
- {i \over 4} \, g_{3 z} \xi^\dag \s^3
\eea

\subsection{Recovering the solutions with 32 supersymmetries}

To understand better the significance of the constants $c_i$, it will be
useful to recover, from the reduced BPS equations, the solutions with
32 supersymmetries, namely $AdS_7 \times S^4$, and $AdS_4 \times S^7$.
Actually, there really are 3 different cases, because $AdS_7$ can be
built up from the first $S_2^3$ or the second $S_3^3$, with corresponding
assignments of the non-vanishing 4-form flux and charge. These
possible cases are as given in the table below.
\begin{table}[htdp]
%\caption{default}
\begin{center}
\begin{tabular}{|c||c|c|c|c||c|c|c |} \hline
 space-time & belonging to & $c_1$  & $c_2$ & $c_3$  & $g_{1z}$ & $g_{2z}$  & $g_{3z}$
\\ \hline \hline
 $AdS_4 \times S^7$ &   case I & 
 	$-2  $   &  $1$ &  $1$ &  $-3$ &   $0$ & $ 0$  \\ \hline
$AdS_7 \times S^4$ &  case II & 
	$1  $   &  $-2$ &  $1$ &  $0$ &   $-3i$ &  $ 0$ \\ \hline
$AdS_7 \times S^4$  & case III & 
	$1 $   &  $1$ &  $-2$ &  $0$ & $0$ & $3i$  \\ \hline
\end{tabular}
\end{center}
\label{table1}
\caption{Assignments of $c_i$ and $g_{iz}$ for the maximally supersymmetric cases.}
\end{table}\\
For each case, we shall introduce local complex coordinates
$w,\bar w$ for which $\rho=1$, so that $ds^2 _\Sigma = 4 |dw|^2$.
The metric factors $f_i$ are then most conveniently expressed
in terms of real coordinates $x,y$, defined by $w=x+i y$.

\subsubsection{$AdS_4 \times S^7$ (belonging to case I)}

The solution  $AdS_4 \times S^7$ has the following metric factors,
\bea
f_1 = - \ch(2x) \hskip 0.7in
f_2 =  -2 \cos(y) \hskip 0.7in
f_3 =  -2 \sin(y)
\eea
The supersymmetry parameter  is given by
\bea
\xi = \sqrt{2} \left ( \matrix{
\ch({w + 3 \bar w \over  4}) \cr
\sh({w + 3 \bar w \over 4}) \cr}
\right )
=  \sqrt{2} \left ( \matrix{ \ch(x-iy/2) \cr \sh(x-iy/2) \cr} \right )
\eea
The 10-dimensional metric is given by
\bea
ds^2 = \ch(2x)^2 ds^2_{AdS_3} + 4dx^2
+ 4\cos(y)^2 ds^2_{S_2^3} +  4\sin(y)^2 ds_{S^3_3} +  4dy^2
\eea
which is the standard product metric for the range $x \in \bR$ and $0 \leq y \leq \pi/2$.

\subsubsection{$AdS_7 \times S^4$ (belonging to case II)}

The solution  $AdS_7 \times S^4$ with flux on the first $S_2^3$ has the following metric 
 factors,
\bea
f_1 =  2 \ch(x)
\hskip 0.7in
f_2 =   \sin(2y)
\hskip 0.7in
f_3 =  2 \sh(x)
\eea
The supersymmetry parameter  is given by
\bea
\xi = {1 \over \sqrt{2}} \left ( \matrix{
i \, \exp({w - 3 \bar w \over 4}) - \exp({-w + 3 \bar w \over 4}) \cr
\exp({w - 3 \bar w \over 4}) - i \, \exp({-w + 3 \bar w \over 4}) \cr}
\right )
=
{1 \over \sqrt{2}} \left ( \matrix{
i \, e^{-x/2+iy} - e^{x/2-iy} \cr
e^{-x/2+iy } - i e^{x/2-iy} \cr}
\right )
\eea
The 11-dimensional  metric is given by
\bea
ds^2 =  4\ch(x)^2 ds^2_{AdS_3}
+  4\sh(x)^2 ds^2_{S_3^3} +  4dx^2 + \sin(2y)^2 ds_{S^3_2} + 4dy^2
\eea
which is the standard metric for the range $0<x$ and $0 \leq y \leq \pi/2$.

\subsubsection{$AdS_7 \times S^4$ (belonging to case III)}

The solution describing $AdS_7 \times S^4$ with flux on the second $S_3^3$
has the following metric factors,
\bea
f_1 = 2 \ch(x)
\hskip 0.7in
f_2 = 2 \sh(x)
\hskip 0.7in
f_3 = \sin(2y)
\eea
The supersymmetry parameter  is given by
\bea
\xi = \left ( \matrix{
\exp({+w - 3 \bar w \over 4}) \cr
\exp({-w + 3 \bar w \over 4}) \cr}
\right )
=
\left ( \matrix{ e^{-x/2+iy} \cr e^{+x/2-iy } \cr} \right )
\eea
The 11-dimensional  metric is given by
\bea
ds^2 = 4 \ch(x)^2 ds^2_{AdS_3}
+ 4 \sh(x)^2 ds^2_{S_2^3} + 4 dx^2 + \sin(2y)^2 ds_{S^3_3} + 4dy^2
\eea
which is the standard metric for the range $0<x$ and $0 \leq y \leq \pi/2$.

\newpage

%%%%%%%%%%%%%%%%%%%%%%%%%%%%%%%%%%%%%%%%%%%%%%%
%%%%%%%%%%%%%%%%%%%%%%%%%%%%%%%%%%%%%%%%%%%%%%%
\section{Identifying the holomorphic form $\kappa$}
\setcounter{equation}{0}
%%%%%%%%%%%%%%%%%%%%%%%%%%%%%%%%%%%%%%%%%%%%%%%
%%%%%%%%%%%%%%%%%%%%%%%%%%%%%%%%%%%%%%%%%%%%%%%

In this section, we shall identify, for any assignment of the constants $c_1, c_2, c_3$,
a combination of the spinors $\xi$, $\xi^*$ and the metric factor $\rho$, which
is holomorphic on $\Sigma$ as a result of the BPS equations.

\sm

From the remaining differential reduced BPS equations in (\ref{sum2a}), we
observe the following. Viewing the reduced flux fields $g_{iz}$ as forming
part of a generalized connection on the spin bundles on $\Sigma$ of which 
$\xi$ and $\xi^*$ are sections, equations (\ref{sum2a}) show that the
connection acting on $\xi^*$ is {\sl minus three times} the connection acting on $\xi$.
This suggests that holomorphic objects may be found in combinations of the
type $\xi ^* \otimes \xi ^{\otimes 3}$. This tensor product space spans an 8-dimensional
vector space,
and we shall seek linear combinations in this space which are holomorphic.
The exact solutions for the $AdS_4 \times S^7$ and $AdS_7 \times S^4$ cases
confirm that such objects lead to holomorphic combinations for those cases.

\subsection{Computing the derivatives of $\xi ^* \otimes \xi ^{\otimes 3}$}

To carry out this calculation in practice, it is convenient to first recast
(\ref{sum2a}) in terms of the two complex components $\a$ and $\b$ of $\xi$,
\bea
\xi = \left ( \matrix{\a \cr \b \cr} \right )
\eea
in terms of which we have
\bea
\mu_0 & = &  \a^2 + \b^2
\no \\
\mu_1 & = & 2 \a \b
\no \\
\mu _3 & = & \a^2 - \b^2
\eea
The differential equations (\ref{sum2a}) take the following form,
\bea
\label{diffab}
D_z \a  & = & + { i \over 2} \hat \o_z \a +
{1 \over 12} g_{1z} \b - {1 \over 12} g_{2z} \b - { i \over 12} g_{3z} \a
\no \\
D_z \b & = & + { i \over 2} \hat \o_z \b +
{1 \over 12} g_{1z} \a + {1 \over 12} g_{2z} \a + { i \over 12} g_{3z} \b
\no \\
D_z \ba  & = & - { i \over 2} \hat \o_z \ba +
{1 \over 4} g_{1z} \bb - {1 \over 4} g_{2z} \bb + { i \over 4} g_{3z} \ba
\no \\
D_z \bb  & = & - { i \over 2} \hat \o_z \bb +
{1 \over 4} g_{1z} \ba + {1 \over 4} g_{2z} \ba - { i \over 4} g_{3z} \bb
\eea
We shall also need the derivatives of the 4 components of the tensor power
$\xi^{\otimes 3}$. They may be deduced from the above equations for
$D_z \a$ and $D_z \b$, by direct calculation, and are given by the following
expressions,
\bea
\label{diffabcub}
D_z ( \a^3) & = &  { 3i \over 2} \hat \o_z \a^3 +
{1 \over 4} g_{1z} \a^2 \b - {1 \over 4} g_{2z} \a^2 \b - { i \over 4} g_{3z} \a^3
\no \\
D_z (\a^2\b) & = &  { 3i \over 2} \hat \o_z \a^2\b
+ {1 \over 12} g_{1z} (\a^3 + 2 \a \b^2)
+ {1 \over 12} g_{2z} (+\a^3 - 2 \a \b^2)
- { i \over 12} g_{3z} \a^2\b
\no \\
D_z (\a\b^2) & = &  { 3i \over 2} \hat \o_z \a\b^2
+ {1 \over 12} g_{1z} (\b^3 + 2 \a^2 \b)
+ {1 \over 12} g_{2z} (- \b^3 + 2 \a^2 \b)
+ { i \over 12} g_{3z} \a\b^2
\no \\
D_z (\b^3) & = &  { 3i \over 2} \hat \o_z \b^3 +
{1 \over 4} g_{1z} \a \b^2 + {1 \over 4} g_{2z} \a \b^2 + { i \over 4} g_{3z} \b^3
\eea
It is now straightforward to calculate the derivatives of $\xi^* \otimes \xi ^{\otimes 3}$
by combining the last two equations of (\ref{diffab}) with the four
equations of (\ref{diffabcub}).

\subsection{Holomorphicity modulo the algebraic reduced BPS equations}

We now investigate the following problem: find a linear combination, which we shall
denote by  $\bar \kappa_0$, in the 8-dimensional space $\xi ^* \otimes \xi ^{\otimes 3}$,
whose $D_z$-derivative vanishes, upon the use of the algebraic remaining reduced
BPS equations of (\ref{alg2}), and up to factors of $\rho$. 
A general linear combination in the 8-dimensional
space $ \xi^* \otimes \xi ^{\otimes 3} $ takes the form,
\bea
\label{kap1}
\bar \kappa_0 & = & \ba ( A_3  \a^3 + A_2  \a \b^2 + A_1 \a \b^2 + A_0 \b^3)
\no \\ &&
+ \bb ( B_3 \a^3 + B_2 \a^2 \b + B_1 \a \b^2 + B_0 \b^3)
\eea
where $A_i,B_i$ are complex constants, which remain to be determined.

\sm

From the structure of $D_z \bar \kappa_0$ as a function of $\a,\b, \bar \a, \bar \b$,
it is clear that only a combination of  (\ref{alg2})
which is homogeneous in $\a$ and $\b$ can enter. This combination
is unique, and is obtained from (\ref{alg2}) by eliminating the inhomogeneous
terms. To find it, we use the relation $c_1+c_2+c_3=0$ of (\ref{c123})
to recast the two remaining algebraic equations in the form,
\bea
\label{gg1}
g_{1z} (\a^2 - \b^2) & = & 2(c_1 - c_3)  + 2i g_{3z} \a \b
\no \\
g_{2z} (\a^2 + \b^2) & = & 2(c_2 - c_3)  + 2i g_{3z} \a \b
\eea
The unique homogeneous combination may be cast in the form of
the equation  $\cC=0$, where the combination $\cC$ is defined by,
\bea
\label{cC}
\cC \equiv
(c_2 - c_3) g_{1z} (\a^2 - \b^2) + (c_3 - c_1) g_{2z} (\a^2 + \b^2) - 2i (c_1-c_2) g_{3z}  \a \b
\eea
Either one of the equations of (\ref{gg1}) may be retained as the
combination which is linearly independent of $\cC$.

\sm

Returning to the search for $\bar \kappa_0$, we seek to solve the equation,
\bea
\label{Dkap1}
D_z \bar \kappa_0 = i \o _z \bar \kappa_0
+ {1 \over 24} (\ell_0 \ba \a + \ell_1 \ba \b + \ell_2 \b \a + \ell_3 \bb \b ) \, \cC
\eea
{\sl for all values of $g_{1z}, g_{2z}, g_{3z}$ and $\a, \b, \ba, \bb$}, (and no longer
constrained by the algebraic equations (\ref{gg1})), for some
complex constants $\ell_i$, $i=0,1,2,3$.
The meaning of (\ref{Dkap1}) is that a non-vanishing solution $\bar \kappa_0$ will be
anti-holomorphic (up to a multiplicative factor of $\rho$) when the
homogeneous algebraic equation $\cC=0$ is satisfied.
It is straightforward to solve (\ref{Dkap1}) for $\bar \kappa_0$ given
by (\ref{kap1}). We readily find that, for any assignment of $c_1, c_2, c_3$,
we must have,
\bea
A_0 = A_2 = \ell_0 = 0 & \hskip 1in & \ell_2=\ell_1
\no \\
B_1= B_3 = \ell_3 = 0 &&
\eea
Furthermore, the non-vanishing entries of the solution are
\bea
A_1 = - B_2 & = & -  \ell_1 (c_1-c_2)
\no \\
A_3 = - B_0 & = & +  \ell_1 c_3
\eea
Clearly, the constant $\ell_1$ is just an overall multiple of the solution, which
may be chosen at will. The anti-holomorphic combination $\bar \kappa$ is then given by
\bea
\label{kap2}
\bar \kappa = \rho \bar \kappa _0 = c_3 \rho ( \ba \a^3 - \bb \b^3 )
- (c_1 - c_2) \rho \a \b (\ba  \b - \bb  \a)
\eea
up to any convenient multiplicative constant.

\subsection{The differential form $\kappa$ is of type $(1,0)$}

It will be useful to know that $\kappa$ is a differential form on $\Sigma$ 
of weight $(1,0)$. To see this, it suffices to examine the coefficients of the connection 
$\hat \o_z$ in the differentials of $\a$ and $\b$; this allows us to identify forms 
of pure $(0,n)$ type, namely combinations which have no $\hat \o_z$ contribution. 
Thus, the combinations $\sqrt{\rho}\, \a$, and $ \sqrt{\rho} \, \b$
must be of type $(0,n)$, for some real number $n$. Similarly, the combinations
$\ba / \sqrt{\rho}$, and $ \bb / \sqrt{\rho}$ must be of type $(0,n')$, for some real
number $n'$ (which may be different from $n$).
As a result, the combinations $\sqrt{\rho} \, \ba$, and $ \sqrt{\rho} \, \bb$ are of type $(n,0)$,
while the combinations $\a / \sqrt{\rho}$, and $ \b / \sqrt{\rho}$ are of type $(n',0)$.
Finally, we know that $\rho^2$ must be of type $(1,1)$ since it is the metric.
From the ratio of $\sqrt{\rho}\, \a$ and $ \a / \sqrt{\rho}$, we find that
$\rho$ is of type $(-n',n)$ so that $n=-n'= 1/2$. Similarly we find,
$\a$ and $\b$ are of type $ ( - 1/4 ,  + 1/4 )$, so that
 $\kappa $ is a holomorphic form of type $ (1,0)$.

\sm

The holomorphic form $\kappa$ and its complex conjugate $\bar \kappa$
may be regarded as given, on the same footing as initial value conditions.
For given $\kappa, \bar \kappa$, one could  then proceed to obtain the
unknowns $\beta, \bar \beta$ as a function of the unknowns $\a$ and $\ba$,
thereby further reducing the number of unknowns in the differential
BPS equations.  Since the equations for $\b$, $\bb$ as a function of
$\a$, $\ba$, and $\kappa, \bar \kappa$ are quartic, however, this way of proceeding 
does not appear to be useful.

\newpage

%%%%%%%%%%%%%%%%%%%%%%%%%%%%%%%%%%%%%%%%%%%%
%%%%%%%%%%%%%%%%%%%%%%%%%%%%%%%%%%%%%%%%%%%%
\section{Exact local solution of the BPS equations,  case III}
\setcounter{equation}{0}
%%%%%%%%%%%%%%%%%%%%%%%%%%%%%%%%%%%%%%%%%%%%
%%%%%%%%%%%%%%%%%%%%%%%%%%%%%%%%%%%%%%%%%%%%

In this section, we shall give the complete exact local solution for the
BPS equations when the constants $c_i$ satisfy the relation $c_1=c_2$. 
In view of the relation (\ref{c123}), we thus have $c_3=-2 c_1$. 
Without loss of generality, an overall scaling may be applied to the solution 
to set $c_1=c_2=1$, and $c_3=-2$. This assignment of $c_i$ values
includes  the $AdS_7 \times S^4$ solution with 32 supersymmetries.
It also includes all solutions that behave asymptotically as  $AdS_7 \times S^4$
near any part of its boundary.

\sm

The complete general solution is obtained by using a parametrization
of the reduced flux field $g_{iz}$ adapted to the special relation $c_1=c_2$,
changing variables from $(\rho, \a, \b, \ba, \bb)$ to
$(\rho, \kappa, \bar \kappa, \sigma, \bar \sigma)$,
where $\kappa$ is the holomorphic 1-form identified in the preceding section,
and $\sigma$ is a suitably chosen dual combination. A number of further successive
changes of variables allows us to map the BPS equations onto a
 linear system, which can be solved exactly.

\subsection{Variables adapted to $c_1=c_2$}

The algebraic reduced BPS equations, derived in (\ref{gg1}), may be solved 
for $c_1=c_2$ by parametrizing the fluxes $g_{1z}$, $g_{2z}$, and $g_{3z}$  in 
terms of a single complex function $\psi$, as follows,
\bea
\label{gs}
g_{1z} + g_{2z} & = & 4 \a^2 \psi
\no \\
g_{1z} - g_{2z} & = & 4 \b^2 \psi
\no \\
i g_{3z} & = & - \, { 3  \over  \a \b} +  {\a^4 - \b^4 \over \a \b} \, \psi
\eea
The holomorphic form $\kappa$ of (\ref{kap2}) also simplifies in this case, and we have,
\bea
\kappa = \rho (\a \ba^3 - \b \bb^3)
\eea
We now think of $\kappa$ as a {\sl given} holomorphic form; its $D_z$ differential
is thus also given (up to knowledge of $\rho$), and not vanishing. This gives a
new {\sl algebraic} equation, which we shall now determine. To this end, we
first compute,
\bea
\p_w ( \rho \ba \a^3 ) & = & \rho^2 \psi \a^2 \b^2 (\a \bb + \ba \b)
\no \\
\p_w ( \rho \bb \b^3 ) & = & \rho^2 \psi \a^2 \b^2 (\a \bb + \ba \b)
\no \\
\p_w (\rho^{-1}  \a \ba^3) & = & {1 \over 3} \psi \ba^2 \b^2 (\ba \b + 9 \a \bb)
+ { 2 i \over 3} g_{3z} \a \ba^3
\no \\
\p_w (\rho^{-1}  \b \bb^3) & = & {1 \over 3} \psi \a^2 \bb^2 (\a \bb + 9 \ba \b)
- { 2 i \over 3} g_{3z} \b \bb^3
\eea
Subtracting the first two equations, we recover $\p_w \bar \kappa=0$.
Thus, we may retain from the first two equations only their sum, and
introduce a new variable for this combination,
\bea
\sigma = \rho (\a \ba^3 +  \b \bb^3)
\eea
The key to the complete solution of the BPS equations lies in a first change
of variables from $(\rho, \a, \b, \ba, \bb)$ to the new variables,
$(\rho, \kappa, \bar \kappa, \sigma, \bar \sigma)$. To recover the
original $\a, \b, \ba, \bb$, we form the combinations,
\bea
\s + \kappa & = & 2\rho \a \ba ^3
\no \\
\s - \kappa & = & 2\rho \b \bb^3
\eea
The following quantities will be required during the course of the
planned change of variables,
\bea
2 \rho  \a^4 = { (\bar \s + \bar \kappa)^{3/2} \over (\s + \kappa)^{1/2}}
& \hskip 1in &
{\ba  \over \a} = {(\s + \kappa)^{1/2} \over (\bar \s + \bar \kappa)^{1/2}}
\no \\
2 \rho  \b^4 = { (\bar \s - \bar \kappa)^{3/2} \over (\s - \kappa)^{1/2}}
& \hskip 1in &
{\bb  \over \b} = {(\s - \kappa)^{1/2} \over (\bar \s - \bar \kappa)^{1/2}}
\eea
The square roots prove to be inconvenient, and may be remedied
by uniformizing the problem in terms of hyperbolic functions. We
define a new complex function $\f$ instead of $\sigma$, by
\bea
\label{ff}
\s = \kappa \, \ch (2 \f) \hskip 0.1in 
& \hskip 1in & 
\s + \kappa = 2 \kappa (\ch \f)^2
\no \\
\s^2 - \kappa^2 =  \kappa ^2 (\sh 2 \f)^2
& \hskip 1in & 
\s - \kappa = 2 \kappa (\sh \f)^2
\eea
In terms of $\f$, we have
\bea
\label{phich}
\rho  \a^4 =
 { \bar \kappa ^{3/2} \over \kappa ^{1/2}} ~ {\ch^3  \bar \f \over \ch \f}
 & \hskip 0.5in &
{\ba  \over \a} = { \kappa ^{1/2} \over \bar \kappa ^{1/2}} ~ {\ch   \f \over \ch \bar \f}
\no \\
\rho  \b^4 = { \bar \kappa ^{3/2} \over \kappa ^{1/2}} ~ {\sh ^3  \bar \f \over \sh \f}
 & \hskip 0.5in &
{\bb  \over \b} = { \kappa ^{1/2} \over \bar \kappa ^{1/2}} ~
{\sh \f \over \sh \bar \f }
\eea
The variables $\kappa, \bar \kappa, \f, \bar \f$ will prove to be very
well-adapted to the resolution of the reduced BPS equations (\ref{alg2})
and (\ref{sum2a}), and we now perform this change of variables there.

\subsection{Changing variables in the homogeneous BPS equation}

Although $\f$ and $\bar \f$ will be our ultimate variables, it will be
convenient to first change variables to $\sigma$.
The $\p_w$ derivative of its complex conjugate is  given by
\bea
\label{psi1}
\p_w \bar \sigma  = 2 \rho^2 \psi \a^2 \b^2 (\a \bb + \ba \b )
\eea
In addition, the $D_z$-derivative equations for $\kappa$ and $\sigma$
may be cast in the following form,
\bea
\p_w \left ( \rho ^{-2} \kappa \right )
& = &
{1 \over 3} \psi \ba^2 \b^2 (\ba \b + 9 \a \bb)
- {1 \over 3} \psi \a^2 \bb^2 (\a \bb + 9 \ba \b) + {2i \over 3\rho } g_{3z}  \sigma
\no \\
\p_w \left ( \rho ^{-2} \sigma \right )
& = &
{1 \over 3} \psi \ba^2 \b^2 (\ba \b + 9 \a \bb)
+ {1 \over 3} \psi \a^2 \bb^2 (\a \bb + 9 \ba \b) + {2i \over 3\rho } g_{3z}  \kappa
\eea
Eliminating $g_{3z}$ between these  equations, and further
eliminating $\psi$ using (\ref{psi1}) gives,
\bea
\p_w \left ( {\sigma ^2 - \kappa ^2 \over  \rho ^4} \right )
& = &
{\p_w \bar \sigma  \over 3 \rho^4}    \left \{
(\sigma + \kappa) { \bb^2 \over \b^2}  + ( \sigma - \kappa) {\ba ^2 \over \a^2}
+ 8 \sigma {\ba  \bb \over \a \b}   \right \}
\no \\ && 
+ {8 \kappa \p_w \bar \sigma \over 3 \rho^4}  \left (
{\ba  \bb \over \a \b} \right ) \left ( {\a \bb - \ba \b \over  \a \bb + \ba \b} \right )
\eea
Expressing all $\a, \b, \ba, \bb$ in terms of the known $\kappa, \bar \kappa$,
and the unknown $\sigma, \bar \sigma$ gives a single $\p_w$ differential
equation with unknowns $\sigma $ and $\bar \sigma$,
\bea
\p_w \ln \left ( {\sigma ^2 - \kappa ^2 \over  \rho ^4} \right )
& = &
{\p_w \bar \sigma  \over 3 }   \left \{
{1 \over \bar \sigma - \bar \kappa}  + {1 \over \bar \sigma +\bar \kappa}
+ 8 {\sigma \over |\sigma ^2 - \kappa ^2|}   \right \}
\no \\
&& + {8 \kappa \p_w \bar \sigma \over 3 |\sigma ^2 - \kappa ^2| }
\left ( { \sigma \bar \sigma - \kappa \bar \kappa - |\sigma ^2 - \kappa ^2| 
\over \sigma \bar \kappa - \kappa \bar \sigma} \right )
\eea
Note that all $\rho$-dependence has been grouped on the left hand side.
The first two terms of the first line on the right hand side may be integrated,
using $\p_w \bar \kappa=0$, and the remaining terms may be regrouped,
\bea
\p_w \ln \left ( {\sigma ^2 - \kappa ^2 \over  \rho ^4 
(\bar \sigma ^2 - \bar \kappa^2)^{1/3}} \right )
=
{8 \p_w \bar \sigma \over 3 |\sigma ^2 - \kappa ^2| }
\left (  { \bar \kappa (\sigma^2 - \kappa ^2)  - \kappa  |\sigma ^2 - \kappa ^2|
  \over \sigma \bar \kappa - \kappa \bar \sigma} \right )
\eea
We redefine $\rho$ as follows,
\bea
\rho ^4 = \tilde \rho ^4 |\sigma ^2 - \kappa^2 |^2
\eea
in terms of which we have
\bea
\label{bh1}
\p_w \ln \bigg (  \tilde \rho ^3 (\bar \sigma ^2 - \bar \kappa^2) \bigg )
=
- {2 \p_w \bar \sigma \over  |\sigma ^2 - \kappa ^2| }
\left (  { \bar \kappa (\sigma^2 - \kappa ^2)  - \kappa  |\sigma ^2 - \kappa ^2|
  \over \sigma \bar \kappa - \kappa \bar \sigma} \right )
\eea
Changing variables from $\sigma $ to $\f$ further simplifies the equation, and we get,
\bea
\label{eq1}
\p_w \ln \bigg (  \tilde \rho ^{3/2} \sh ( 2 \bar \f ) \bigg )
= - 2 \p_w \bar \f  \left ( {\ch (\f + \bar \f) \over
\sh (\f + \bar \f) } \right )
\eea

\subsection{Changing variables in the  inhomogeneous BPS equation}

The starting point is the expressions for the reduced flux fields in (\ref{gs}),
and the differential equation for the ratio $\a/\b$, and for $\bar \sigma$,
\bea
\label{psi4}
D_z \ln {\a \over \b} = {1 \over 12} (g_{1z} - g_{2z}) {\beta \over \a}
- {1 \over 12} (g_{1z} + g_{2z}) {\a \over \b} - { i \over 6} g_{3z}
\eea
Eliminating the reduced flux field $g_{iz}$  in favor of $\psi$ using (\ref{gs}),
eliminating $\psi $ in favor of $\p_w \bar \sigma$, using (\ref{psi1}),
and recasting the derivatives $D_z$ in terms of derivatives with
respect to the local complex coordinate $w$, we find,
\bea
\p_w  \ln {\a^4  \over \b^4 } =
 { 2 \rho  \over \a \b}
+   \left ( {1 \over \rho \a^4}  - { 1 \over \rho \b^4} \right )
\left ( {\bar \a \over \a} + { \bb \over \b} \right )^{-1}  \p_w \bar \sigma
\eea
All terms, except the first one on the right hand side of this equation,
are readily expressed in terms of $\f$, using (\ref{phich}). The missing
term is handled by forming  the combination,
\bea
{\rho ^8 \over \a^8 \b^8} =  { 16 \rho ^{12} \over 16 \rho^4 \a^8 \b^8}
=  16 \rho ^{12}  {\s^2 - \kappa ^2 \over (\bar \s^2 - \bar \kappa ^2)^3}
= 16 \tilde \rho ^{12}  (\s^2 - \kappa ^2)^4
\eea
To obtain the last equality above, we have made use of the definition
of $\tilde \rho$, given by $\rho ^4 = \tilde \rho ^4 |\s^2 - \kappa ^2|^2$.
The 8-th root of this equation yields the quantity needed in the
inhomogeneous reduced BPS equations. Expressing the result
in terms of $\f$, we find,
\bea
\label{rhotildedef}
{\rho \over \a \b} = \sqrt{2} \kappa \tilde \rho ^{3/2} \sh (2 \f)
\eea
Putting all together, we obtain,
\bea
\p_w \ln \left ( { (\ch \bar \f)^3 \sh \f \over \ch \f (\sh  \bar \f)^3} \right )
=
2 \sqrt{2}  \kappa \tilde \rho ^{3/2} \sh (2 \f)
+ 16 \p_w \bar \f {  \ch \f  (\sh  \bar \f)^3  -  \sh \f  (\ch  \bar \f)^3 \over
 \sh ( \f + \bar \f) \sh ( 2 \bar \f) }
\eea
A further simplified version of this equation will be presented in the summary
below. It is obtained by addition to both sides the quantity $4\p_w \ln (\sh \bar \f/\ch \bar \f)$.

\subsection{Summary of reduced BPS equations in $\tilde \rho, \f$ variables}

The homogeneous and inhomogeneous BPS equations produce respectively,
\bea
\label{sum5}
\p_w \ln \bigg (  \tilde \rho ^{3/2} \sh ( 2 \bar \f ) \bigg )
& = & - 2 \p_w \bar \f  \left ( {\ch (\f + \bar \f) \over \sh (\f + \bar \f) } \right )
\no \\
\p_w \ln \Big | \th (\f) \Big |^2
& = &
2 \sqrt{2} \kappa  \, \tilde \rho ^{3/2}  \sh (2 \f)
+ 4  \p_w \bar \f  \left ( {\ch (\f - \bar \f)   \over \sh (\f + \bar \f)} \right )
\eea
For completeness, we recall the relation between $\rho$ and $\tilde \rho$
with these fields,
\bea
\rho^2 = \tilde \rho ^2 \kappa \bar \kappa | \sh (2 \f)  |^2
\eea

\subsection{Further change of variables}

The system of equations (\ref{sum5}) exhibits some degree of resemblance with
the corresponding Type IIB equations (7.13-14) of \cite{D'Hoker:2007xy}. 
The key difference is that, here, only a single holomorphic object $\kappa$
has been exposed thus far, while in the Type IIB case, there were two. 
While the corresponding Type IIB equations (7.13-14) of 
\cite{D'Hoker:2007xy} had 2 unknown real functions, the
system (\ref{sum5}) exhibits three unknown real functions, $\tilde \rho,
\Re (\f), \Im (\f)$. Although this situation presents a further complication,
we shall be able to solve for $\tilde \rho$, and map the system (\ref{sum5})
to a linear partial differential equation which may be solved completely.

\sm

Given the similarity of system (\ref{sum5}) with the corresponding system
(7.13-14) of \cite{D'Hoker:2007xy} in Type IIB, we now perform a further 
change of variables analogous to the one we carried out in equation (9.1) of 
\cite{D'Hoker:2007xy} for the Type IB problem. We define the real fields $\mu$
and $\tet$ as follows,
\bea
\label{ab1}
\f - \bar \f = i \mu \hskip 1in { \sh ( 2 \f ) \over \sh ( 2 \bar \f)} = e^{ 2 i \tet}
\eea
To convert the formulas (\ref{sum5}) into these variables, we also need to
have $\f + \bar \f$, which may be deduced from the above relations, and
is given by,
\bea
\label{ab2}
\th (\f + \bar \f) = { \tg (\mu) \over \tg (\tet)}
\eea
Other useful formulas are
\bea
\label{ab3}
|\sh (2 \f)|^2 & = &  { \cos (\mu)^2 \sin (\mu)^2 \over \sin (\tet)^2 - \sin (\mu)^2}
\no \\
\p (\f + \bar \f) & = & \half \, { \sin (2 \tet) \p \mu -  \sin (2 \mu) \p \tet
\over \sin (\tet)^2 - \sin (\mu)^2}
\no \\
{\sh (\f + \bar \f) \over |\sh (2 \f) |} & = & {\cos (\tet) \over \cos (\mu)}
\eea
The range of the hyperbolic tangent function $|\th (\f + \bar \f)|\leq 1$
places a restriction on the ranges of the variables $\mu$ and $\tet$
in view of relation (\ref{ab2}), namely $|\tg (\mu) | \leq |\tg (\tet)|$.

\subsection{The reduced BPS equations in the new variables}

The first reduced BPS equation in (\ref{sum5}) may be expressed as
\bea
\p_w \ln \bigg (  \tilde \rho ^{3/2} \sh ( 2 \bar \f ) \bigg )
& = &
-  \p_w (\f + \bar \f)   \left ( {\ch (\f + \bar \f) \over
\sh (\f + \bar \f) } \right )
+  \p_w (\f - \bar \f)   \left ( {\ch (\f + \bar \f) \over
\sh (\f + \bar \f) } \right )
\no \\
& = &
- \p_w \ln \sh (\f + \bar \f) + { i \p_w \mu \over \th (\f + \bar \f)}
\eea
Moving the first term on the rhs to the lhs, and using (\ref{ab2}) on the second term,
we find,
\bea
\p_w \ln \bigg (  \tilde \rho ^{3/2} \sh ( 2 \bar \f ) \sh (\f + \bar \f) \bigg )
=
 +  i \, \tg (\tet) \, \p_w \ln \sin (\mu)
\eea
Finally, making use of (\ref{ab1}), the expression inside
the ln may be recast as follows,
\bea
\p_w \ln \left (  \tilde \rho ^{3/2} |\sh ( 2 \f )|^2
e^{- i \tet} { \cos (\tet) \over \cos (\mu)}  \right )
=
 +  i \, \tg (\tet) \, \p_w \ln \sin (\mu)
\eea
Here, we have left the expression $|\sh (2 \f)|^2$ unconverted, because it
will be combined with the function $\tilde \rho ^{3/2}$ at a later stage.
The second reduced BPS equation of (\ref{sum5}) may be handled analogously,
and we find, 
\bea
\p_w \tet + i \p_w \ln \sin (\mu) =
2 \sqrt{2}   \, \kappa \, \tilde \rho ^{3/2} |\sh (2 \f ) |^2 e^{ i \tet} {\cos (\tet)
\over \sin (2 \mu)}
\eea
Clearly, there is a particular combination of the metric factor $\tilde \rho$
which enters in both reduced BPS equations. Therefore, we shall define
a new variable $\hat \rho$ instead of $\tilde \rho$, by
\bea
\label{rhohatdef}
\hat \rho ^{3/2} \equiv  2 \sqrt{2}  \, \tilde \rho ^{3/2} { |\sh ( 2 \f)|^2
\over \sin (2 \mu)}
\eea
Note that there is some arbitrariness in defining this combination. We do not
include the extra factor of $\cos \tet$ because leaving this factor in will
facilitate integrating the equations later on.
In terms of the fields $\tet, \mu, \hat \rho$, the reduced BPS equations
now take the form,
\bea
\label{eq9}
\p_w \ln \Big (  \hat \rho ^{3/2}  e^{- i \tet}  \cos (\tet)  \Big )
& = &
  \Big ( i \, \tg (\tet) -1 \Big ) \, \p_w \ln \sin (\mu)
 \no \\
\p_w \tet + i \p_w \ln \sin (\mu) & = &
 \kappa \, \hat \rho ^{3/2} e^{ i \tet} \cos (\tet)
\eea
where in the first equation, we have moved a term $- \p_w \ln \sin (\mu)$ to the
right hand side.

\subsection{Integrating the equation for $\hat \rho$}

To integrate  the reduced BPS equations of (\ref{eq9}) further,
we note that,
\bea
i \, \tg (\tet) -1 & = &  - {1 \over e^{i \tet} \cos (\tet)}
\no \\
\p_w \ln \Big ( e^{- i \tet} \cos (\tet) \Big ) & = & { - i \p_w \tet \over e^{i \tet} \cos (\tet)}
\eea
Hence the two BPS equations of (\ref{eq9}) become,
\bea
\p_w \ln   \hat \rho ^{3/2}
& = &
  {  i \p_w \tet - \p_w \ln \sin (\mu)  \over e^{i \tet} \cos (\tet)}
 \no \\
 i \p_w \tet - \p_w \ln \sin (\mu)  & = &
 i  \kappa \, \hat \rho ^{3/2}  e^{ i \tet} \cos (\tet)
\eea
Remarkably, it is now manifest that all dependence on $\tet$ and $\mu$
may be eliminated between these equations. Doing so, we obtain,
\bea
\label{rho3}
\p_w  \ln   \hat \rho ^{3/2} =  i \, \kappa \, \hat \rho ^{3/2}
\eea
The holomorphic differential $i \kappa$ may be written as the $\p$-differential of a
real harmonic function which we shall denote by $h$,
\bea
-i  \kappa = \p_w h
\eea
Recasting (\ref{rho3}) in terms of $h$ gives
an equation which may be integrated, and we have
\bea
\label{rho5}
{1 \over \hat \rho ^{3/2}} = h
\eea
As given by $-i  \kappa = \p_w h$, the harmonic function
$h$ is determined only up to an additive constant, which one may view as fixed by
the final equation (\ref{rho5}). Note that the positivity of $\hat \rho ^{3/2}$
requires that $h \geq 0$ throughout $\Sigma$.

\subsection{Mapping the remaining BPS equation onto a linear equation}

Substituting the result (\ref{rho5}) for $\hat \rho^{3/2}$ into either one
of the BPS equations of (\ref{sum5}), produces our final equation to be solved,
\bea
\label{bps2}
i\p_w \tet - \p_w \ln \sin (\mu) =  - e^{ i \tet} \cos (\tet) \p_w \ln h
\eea
This is a complex first order differential equation  for two
real variables $\tet$ and $\mu$, with the harmonic function $h$
considered as given.
The integrability condition in $\mu$ requires a second order partial
differential equation for $\tet$ alone, 
\bea
2  \p_{\bar w} \p_w \tet 
- \p_{\bar w} \bigg ( i e^{ i \tet} (\cos \tet ) \, \p_w \ln h \bigg )
+ \p_w \bigg (  i e^{- i \tet} (\cos \tet )\, \p_{\bar w} \ln h \bigg )
 =0
\eea
This equation is of the sine-Gordon/Liouville type and is equivalent to the 
equation (\ref{intro1}) given in the Introduction.
Our task is to solve (\ref{bps2}). To this end, we combine the
real variables $\tet$ and $\mu$ into a single complex field $G$, defined by
\bea
\label{Gdef}
G \equiv  \sin (\mu ) e^{- i \tet}
\eea
It follows that $\bar G  / G=  e^{2 i \tet}$, which allows us to recover $\tet$ from  $G$.
Similarly, we have $\sin (\mu)^2 = G \bar G$, which allows us to recover $\mu$ from $G$. 
In terms of  $G $, equation (\ref{bps2}) becomes,
\bea
\label{bps3}
\p_w G = \half (G  + \bar G) \, \p_w \ln h
\eea
{\sl Remarkably, we have succeeded in mapping the non-linear BPS equation
(\ref{sum5}) into a linear differential equation in $G$, and its complex conjugate equation. }

\subsection{Metric factors}

The metric factors $f_i$ may be readily obtained in terms of  $G, \bar G$, and $ h$,
with the help of their expressions  in terms of $\a$ and $\b$ of (\ref{rads1}).
The relevant quantities we need to compute are then $\a \bar \a$, $\b \bar \b$
and $\bar \a \b$.  Putting together the formulas from (\ref{phich}) we get the
following expressions in terms of $\kappa, \bar \kappa, \f,\bar\f$, and $\rho$,
\bea
\alpha\bar\alpha&=&
{|\kappa^{1/2}|\over\rho^{1/2}}|\ch \f|
\no\\
\beta\bar\beta&=&|\kappa^{1/2}|\rho^{-1/2}|\sh \f|
\no\\
\bar\alpha\beta&=&
2^{1/2}{|\kappa^{1/2}|\over\rho^{1/2}}{\sh \bar\f\, \ch \f \over |\sh (2\f)|^{1/2}}
\eea
Next we express these quantities in terms of $G$, $\bar G$, and $h$.
To do so, we first change variables to $\tet$, $\mu$, $\tilde\rho$, using (\ref{ab1}) 
and the identities (\ref{ab2}) and (\ref{ab3}) for $\tet$, $\mu$, and $(\ref{rhotildedef})$ 
for $\tilde \rho$.
Then we change variables  to $G$, $\bar G$, and $h$ using the definition of $G$ (\ref{Gdef}),
the definition of $\hat \rho$ (\ref{rhohatdef}) and the fact $\hat \rho^{-3/2} = h$.
An ubiquitous combination entering the results is defined by
\bea
\label{Wdef}
W^2 & = & - 4 |G|^4 - (G-\bar G)^2
\no \\ &&
= - 4 \sin (\mu)^2 \Big ( \sin(\mu)^2 - \sin (\tet)^2 \Big )
\eea
In view of the inequality $|\tg (\mu) | \leq |\tg (\tet)|$, which was part of the definition
of the variables $\mu$ and $\tet$ in section 5.5, it follows that for this
range of parameters $\mu, \tet$, one always has $W^2 \geq 0$.
The boundary is given by $W=0$. The corresponding range for $G$   in the 
complex plane such that $W^2 \geq 0$ is  depicted in Figure 2 of section 8.
For this range of parameters adapted to the solution, $W$ is real, 
and we may take it to be positive without loss of generality. 

\sm

In terms of $G$, $h$, and $W$, we derive the following expressions for 
the metric factor.
Using the definitions of $\tilde \rho$ in (\ref{rhotildedef}) and $\hat \rho$ in 
(\ref{rhohatdef}), we derive  $\rho$ in terms of $G$, $\bar G$ and $h$,
\bea
\label{rhosolution}
\rho^6={|\p_w h|^6 \over 16 h^4 }(1- |G|^2)W^2
\eea
The metric factors $f_1, f_2, f_3$ are given by the following relations,
\bea
\label{metricsolution2}
(f_1^2-f_2^2)^3&=& 128 h^2 {(1- |G |^2 ) |G |^6 \over W^4 }
\no\\
(f_1 f_2)^3&=& - 4 h^2  {(1 - |G|^2 ) \over W} 
\no\\
f_3^3 &=& -  {h W \over 4 (1- |G|^2) }
\eea
These relations for $f_1$ and $f_2$ may be solved more explicitly, and we have,
\bea
\label{metricsolution}
f_1^6 &=&  4 h^2   {(1 - |G|^2) \over W^4} \bigg( |G - \bar G| + 2 |G|^2 \bigg)^3
\no\\
f_2^6 &=&  4 h^2 {(1 - |G|^2) \over W^4} \bigg(  |G - \bar G| - 2 | G|^2 \bigg)^3
\eea
Using $W^2\geq 0$, we automatically have $ |G - \bar G| \pm 2 |G|^2 \geq 0$. 
The corresponding sign has been chosen so that $f_1 ^6 \geq f_2^6$, as 
required by the solution (\ref{rads1}), the fact that $c_1=c_2$, and the 
inequality $|\l_3|\leq \l_0$. Note that the product 
$f_1f_2f_3$ of  the metric factors yields a simple expression solely
in terms of the harmonic function $h$, 
\bea
f_1 f_2 f_3 = \pm h
\eea
This relation was used in \cite{Lunin:2007ab} as the definition of the 
harmonic coordinate $h$.
The flux fields $g_{iz}$ may be computed analogously, and will be given in section 9.

\subsection{Calculating $G$ and $h$  for the $AdS_7 \times S^4$ solution}

We use the expressions for the supersymmetry parameter $\xi$,
obtained section 3.3 for the $AdS_7 \times S^4$ case, to evaluate 
the forms $\kappa$, $\sigma$, and  $\f$, and we find, 
\bea
\ch (2 \f) = - { \ch (2w) \over \sh (2 w)}
\hskip 1in
\sh (2 \f) = - { 1 \over \sh (2 w)}
\eea
These relations allow us to compute the variables $\mu$ and $\tet$, 
\bea
\sin \mu = - i \, {\sh ( w - \bar w) \over |\sh (2 w)|}
\hskip 1in
e^{- i \tet} = {|\sh (2 w)| \over \sh ( 2 \bar w)}
\eea
and hence $G$ and $h$, 
\bea
G & = & -i \, {\sh (w-\bar w) \over \sh ( 2 \bar w) }
\no \\
h & = & {1 \over \hat \rho ^{3/2}} =  - i \Big ( \ch (2w) - \ch (2 \bar w ) \Big )
\eea
which is remarkably simple. For $w=x+iy$, and for the range given by $x \in \bR$, and 
$0 \leq y \le \pi/2$, we have $\Im (G) \geq 0$, so that $G$ belongs to the 
range of the upper sphere in Figure 2 of section 8.

\newpage

%%%%%%%%%%%%%%%%%%%%%%%%%%%%%%%%%%%%%%%%%%%%%%%
%%%%%%%%%%%%%%%%%%%%%%%%%%%%%%%%%%%%%%%%%%%%%%%
\section{Exact local solution of  BPS equations, case II}
\label{cisymm}
\setcounter{equation}{0}
%%%%%%%%%%%%%%%%%%%%%%%%%%%%%%%%%%%%%%%%%%%%%%%
%%%%%%%%%%%%%%%%%%%%%%%%%%%%%%%%%%%%%%%%%%%%%%%

From a physical point of view, cases II and III are mirror images of one another,
obtained by interchanging the spheres $S_2^3$ and $S_3^3$. Thus, we expect 
the solution of case II to be substantially the same as that of case III, and to be
obtained by interchanging the constants $c_2$ and $c_3$, the metric factors  
$f_2$ and $f_3$,  the fluxes $g_{2z} $ and $g_{3z}$, all while leaving the constant 
$c_1$, the metric factor $f_1$, and the flux  $g_{1z}$ unchanged. We shall see 
that this natural recipe is indeed correct, up to sign factors produced under the 
effects of the interchanges.

\sm

We shall establish this result as an answer to a more general question,
namely, {\sl is it possible to find a set of transformations of the reduced
BPS equations, under which the BPS equations associated with one
assignment of constants $c_1, c_2, c_3$ is mapped onto the BPS
equations associated with another assignment of constants 
$\tilde c_1, \tilde c_2, \tilde c_3$ ?} It is assumed, of course, that 
the relations $c_1+ c_2+ c_3=0$, and $\tilde c_1+ \tilde c_2+ \tilde c_3=0$
hold, since these equations form an integral part of the BPS equations. 
The map between case II and case III is a simple special case of this question. 

\sm

It would be difficult to try and address this question in all generality, 
since it is difficult to investigate the behavior of the BPS equations under a
completely general transformation. Thus, in the present  paper, 
we shall restrict attention to {\sl the most general constant transformation which acts 
 linearly on the reduced supersymmetry parameters $\xi$}.

\sm

In the preceding sections, the BPS equations have been reduced to
a set of first order differential equations (\ref{sum2a}) on the supersymmetry parameter,
\bea
\label{sum8}
0 &=& D_z \xi
- {i \over 2} \hat \omega_z  \xi
- {1 \over 12} \,  g_{1 z}  \s^1  \xi
+ {i \over 12} \, g_{2 z} \sigma^2   \xi
+ {i \over 12} \, g_{3 z} \sigma^3 \xi
\no \\
0 & =  & D_{\bar z} \xi - {i \over 2} \hat \omega_{\bar z} \xi
- {1 \over 4} \, g_{1 \bar z} \s^1  \xi
+ {i \over 4} \, g_{2 \bar z}  \s^2 \xi
+ {i \over 4} \, g_{3 \bar z}  \s^3  \xi
\eea
and a subsidiary set of algebraic equations (\ref{alg2}), equivalent to,
\bea
\label{gg6}
g_{1z} \mu_3 & = & 2c_1  - 2 c_3 + i g_{3z} \mu_1
\no \\
g_{2z} \mu_0 & = & 2 c_2 - 2 c_3 + i g_{3z} \mu_1
\eea
The $g_{iz}$ are components of  real tensors $g_{ia}$, and  satisfy
$g_{i\bar z} = \left ( g_{iz} \right )^*$ for $ i=1,2,3$.
The bilinears $\mu_i$ are defined by $\mu_i = \xi ^t \sigma _i \xi $ with $i=0,1,2,3$.
The $\mu_i$  automatically satisfy $\mu_0 ^2 - \mu_1^2 - \mu_3^2=0$.
The real constants $c_i$ must be non-zero and are related by $c_1 +c_2 + c_3=0$.
The differential equations may be recast in a more compact form by
introducing the following combination,
\bea
\cG_z & = &  g_{1 z}  \s^1 -i  g_{2 z} \sigma^2  -i  g_{3 z} \sigma^3
\no \\
-\s^3 (\cG_z)^* \s^3 & = &
 g_{1 \bar z}  \s^1 - i  g_{2 \bar z} \sigma^2  -  i  g_{3 \bar z} \sigma^3
\eea
As a result, the differential BPS equations (\ref{sum8}) become,
\bea
\label{sum8a}
0 &=& D_z \xi
- {i \over 2} \hat \omega_z  \xi
- {1 \over 12} \cG_z  \xi
\no \\
0 & =  & D_{\bar z} \xi - {i \over 2} \hat \omega_{\bar z} \xi
+ {1 \over 4} \s^3 (\cG_z)^* \s^3  \xi
\eea
The combinations $(\cG_z, \cG_{\bar z})$ may be thought of as the components 
of a generalized connection 1-form, coupling to $\xi$ in a left-right asymmetric way.

\subsection{Symmetries of the differential BPS equations}

We begin by seeking constant (i.e. $\Sigma$-independent) linear transformations 
on $\xi$, and suitable associated transformations on $g_i$ (which will not be linear) 
such that the differential BPS equations are left invariant, irrespective of the algebraic 
equations. The next step will then be to require that also the algebraic reduced 
BPS equations  are invariant.

\sm

A general constant complex transformation $S$, which acts on $\xi$ and $\cG_z$ by
\bea
\label{Sinv}
\xi & = & S \tilde \xi
\no \\
\cG_z & = & S \tilde \cG_z S^{-1}
\eea
will leave the first differential equation of (\ref{sum8a})  invariant. 
The overall multiplicative factor in $S$ acts as a constant $U(1)$ phase and a 
real scale factor on $\xi$. The phase symmetry
guarantees that $\xi$ and $\xi ^*$ do not mix. The second differential equation will 
also be invariant under (\ref{Sinv}) provided that we  have
\bea
\s^3 (\cG_z)^* \s^3 = S \s^3 (\tilde \cG_z)^* \s^3 S^{-1}
\eea
Taking the complex conjugate of this equation, and eliminating $\tilde \cG_z$ 
between the resulting equation and the second equation of (\ref{Sinv})
gives the following requirement, 
\bea
\label{curlyg}
 \cG_z  = \s^3 S^* \s^3 S^{-1}  \cG_z \left ( \s^3 S^* \s^3 S^{-1} \right )^{-1}
\eea
In other words,  the matrix $\s^3 S^* \s^3 S^{-1}$ must commute with $\cG_z$. 
For a general solution to the BPS equations with 16 supersymmetries,  
the flux field $\cG_z$ will  take on generic values.\footnote{Note that for the 
solutions with 32 supersymmetries, the fluxes {\sl do not take on generic values}, 
since only a single one of the fluxes $g_{1z}, g_{2z}, g_{3z}$ will be non-vanishing.}
By Shur's lemma,  we are led to require $\s^3 S^* \s^3 S^{-1} = t^2  I$,
where $t$ is a constant complex parameter.
Taking the determinant on both sides of (\ref{curlyg}) informs us that $|t|=1$. Since
the phase transformation was a symmetry of the differential BPS
equations, we may choose $t=1$ without loss of generality. This gives
rise to our final equation for $S$,
\bea
S =  \s^3 S^* \s^3
\eea
The complete solution for $S$ is given as follows,
\bea
S =  \left ( \matrix{a & i b \cr ic & d \cr} \right ) \hskip 1in a,b,c, d \in {\bf R}
\eea
The determinant of $S$ does not act on $\cG_z$ and acts on $\xi$ as a scale
factor under which the differential equations are invariant. In the algebraic
equations, the scale factor will scale all $c_i$ in the same manner, and
so is also a symmetry. As a result, the scale factor in $S$ is irrelevant,
and we set $\det (S)=1$ without loss of generality. Thus, we have $ad+bc=1$.
The set of these matrices forms a group isomorphic to $SL(2,\bR)$,
under which the differential BPS equations (\ref{sum8}) and (\ref{sum8a}) are invariant.

\subsection{No continuous symmetries of the full set of BPS equations}

To investigate the behavior under $S$ of the algebraic equations,
we need  the behavior of $g_{iz}$ and of $\mu_i$ under $S$.
They may be obtained from (\ref{Sinv}), and are given by
\bea
\tilde \cG_z & = & S^{-1} \cG_z S
\no \\
\tilde \mu_i & = & \tilde \xi ^t \s_i \tilde \xi = \xi ^t (S^t)^{-1} \s_i S^{-1} \xi
\eea
It is convenient to write out these transformations in components, and we find, 
\bea
\left ( \matrix{ \tilde g_{1z} \cr \tilde g_{2z} \cr \tilde g_{3z} \cr } \right )
=
\left ( \matrix{ u_{11} & u_{12} & u_{13} \cr
u_{21} & u_{22} & u_{23} \cr
u_{31} & u_{32} & u_{33} \cr } \right )
\left ( \matrix{ g_{1z} \cr g_{2z} \cr g_{3z} \cr } \right )
\eea
where the entries $u_{ij}$ are given in terms of the entries $a,b,c,d$ of $S$ by, 
\bea
\label{uss}
2u_{11} = a^2 + b^2 + c^2+d^2
\qquad & u_{13} =  -ac + bd \qquad & u_{31} = -ab + cd
\no \\
2u_{12} = a^2 + b^2 - c^2 -d^2
\qquad & u_{23} =  -ac - bd \qquad & u_{32} = ab+cd
\no \\
2u_{21} =  a^2 - b^2 + c^2 - d^2
\qquad &  &  u_{33} = ad-bc
\no \\
2u_{22} =  a^2 - b^2 - c^2 + d^2 \qquad &&
\eea
Note that, although $S$ is complex, the action of $S$ on the flux
fields $g_{1z}, g_{2z}, g_{3z}$ is real. Similarly, the bilinears $\mu_i$
transform in a related manner, 
\bea
\left ( \matrix{ \tilde \mu_3 \cr \tilde \mu_0 \cr \tilde \mu_1 \cr } \right )
=
\left ( \matrix{ u_{11} & - u_{12} & -i u_{13} \cr
-u_{21} &  u_{22} & i u_{23} \cr
-i u_{31} & -i u_{32} & u_{33} \cr } \right )
\left ( \matrix{ \mu_3  \cr \mu_0 \cr \mu_1 \cr } \right )
\eea
It remains to analyze the algebraic equations. Assuming that the original
configuration satisfies the algebraic equations,
\bea
\label{gg7}
g_{1z} \mu_3 - i g_{3z} \mu_1 & = & 2c_1  - 2 c_3
\no \\
g_{2z} \mu_0 - i g_{3z} \mu_1 & = & 2 c_2 - 2 c_3
\eea
for some assignment of $c_i$ satisfying $c_1+c_2+c_3=0$, we shall
require that the transformed configuration also satisfy the algebraic
equations, but possibly for a different set of $\tilde c_i$,
\bea
\label{gg8}
\tilde g_{1z} \tilde \mu_3 - i \tilde g_{3z} \tilde \mu_1
& = & 2\tilde c_1  - 2 \tilde c_3
\no \\
\tilde g_{2z} \tilde \mu_0 - i \tilde g_{3z} \tilde \mu_1
& = & 2 \tilde c_2 - 2 \tilde c_3
\eea
Computing the left hand side of the first line, we get
\bea
\tilde g_{1z} \tilde \mu_3 - i \tilde g_{3z} \tilde \mu_1
& = &
g_{1z} \mu_3 (u_{11}^2 - u_{31}^2)  +
g_{2z} \mu_0 (-u_{12}^2 - u_{32}^2)
-i g_{3z} \mu_1 ( u_{13}^2 + u_{33}^2)
\no \\ &&
- g_{1z} \mu_0 ( u_{11} u_{12}  + u_{31} u_{32})
-i g_{1z} \mu_1 ( u_{11} u_{13}  + u_{31} u_{33})
\no \\ && +
g_{2z} \mu_3 (u_{11} u_{12}  - u_{31} u_{32})
-i g_{2z} \mu_1 ( u_{12} u_{13}  + u_{32} u_{33})
\no \\ && +
g_{3z} \mu_3 ( u_{11} u_{13}  - u_{31} u_{33})
- g_{3z} \mu_0 ( u_{12} u_{13}  + u_{32} u_{33})
\eea
The last six terms do not fit into the patterns of the algebraic equations,
and thus should vanish. It is immediate that this requires
\bea
\label{alg9}
u_{11} u_{12}= u_{31}u_{32}= u_{11} u_{13}=u_{31} u_{33}
=u_{12} u_{13} + u_{32} u_{33}=0
\eea
Since from its very expression in (\ref{uss}), we have $u_{11} \not=0$, it must be that
$u_{12}=u_{13}=0$. These relations imply the following relations on the 
parameters $a,b,c,d$ of $S$, 
\bea
a^2 + b^2 - c^2 - d^2 = ac-bd=0
\eea
Given that also $ad+bc=1$, these conditions are immediately solved by
the following parametrization in terms of a single real angle $\theta$,
\bea
a=d & = & \cos \theta
\no \\
b=c & = & \sin \theta
\eea
for any $0 \leq \theta < 2 \pi$.  As a result, we have $u_{11}=1$, 
and $u_{12}=u_{13} = u_{21} = u_{31}=0$, as well as,
\bea
u_{22} = u_{33} & = & \cos 2 \theta
\no \\
- u_{23}= u_{32} & = & \sin 2 \theta
\eea
Condition (\ref{alg9}), however,  implies that we must have $\sin 4 \theta=0$.
Thus, the parameter $\theta$ can  take only discrete values, and we conclude
that the reduced BPS equations admit no continuous linear symmetries.  

\subsection{Discrete symmetries of the full BPS equations}

The only nontrivial transformations left to investigate correspond to values 
of $\theta$ such that $\sin 4 \theta=0$. Letting $\theta \to \theta + \pi$
reverses the sign of $a,b,c,d$, and thus leaves all $u_{ij}$ invariant;
this transformation simply reverses the sign of $\xi$, leaving the fluxes
$g_{1z}, g_{2z}, g_{3z}$ unchanged. This transformation is trivial.
The case $\theta =0$ is simply the identity, and is also trivial, leaving 
only the following possible values, $\theta = \pi/4, \pi/2, 3 \pi/4$
as non-trivial cases.

\sm

The case $\theta = \pi/2$ yields $u_{11}=1$, and $ u_{22}=u_{33}=-1$,
so that $g_{2z}, g_{3z}, \mu _0, \mu _1$ change sign, but 
$g_{1z}$ and $\mu _3$ are unchanged. This transformation 
leaves the values of $c_1, c_2, c_3$ unchanged as well, and is
thus also essentially trivial.

\sm

The only non-trivial case left is  $\theta = \pi / 4$. (The case $\theta = 3 \pi/4$
is equivalent to the case $\theta = \pi/4$ combined with the sign changes
on the fluxes and bilinears produced by $\theta=\pi/2$. Therefore, 
the case $\theta = 3 \pi/4$ need not be considered separately.) 
The case $\theta = \pi/4$ corresponds to 
interchanging the two $S^3$ spheres as follows,
\bea
S = { 1 \over \sqrt{2}} \left( \matrix{1 & i \cr i & 1} \right)
\hskip1in
u_{ij} =   \left( \matrix{1 & 0 & 0 \cr 0 & 0 & -1 \cr 0 & 1 & 0} \right)_{ij}
\eea
This produces the following transformation of the $g_{i z}$
\bea
\tilde g_{1 z} = g_{1 z}
\qquad \qquad
\tilde g_{2 z} = - g_{3 z}
\qquad \qquad
\tilde g_{3 z} =  g_{2 z}
\eea
In order for $(\ref{gg7})$ to remain invariant, the metric constants must transform as
\bea
\tilde c_1 = c_1
\qquad \qquad
\tilde c_2 = c_3
\qquad \qquad
\tilde c_3 = c_2
\eea
Finally using the expression for the metric factors in terms of the spinor bi-linears (\ref{rads1}), one may also work out the transformation of the metric factors
\bea
\tilde f_1 = f_1
\qquad \qquad
\tilde f_2 = -f_3
\qquad \qquad
\tilde f_3 = f_2
\eea
We see that the effect of swapping $c_2$ with $c_3$ amounts to swapping 
the two $S^3$ spheres and physically corresponds to the same geometry.

\newpage

%%%%%%%%%%%%%%%%%%%%%%%%%%%%%%%%%%%%%%%%%%%%%%%
%%%%%%%%%%%%%%%%%%%%%%%%%%%%%%%%%%%%%%%%%%%%%%%
\section{Exact local solution of  BPS equations, case I}
\setcounter{equation}{0}
%%%%%%%%%%%%%%%%%%%%%%%%%%%%%%%%%%%%%%%%%%%%%%%
%%%%%%%%%%%%%%%%%%%%%%%%%%%%%%%%%%%%%%%%%%%%%%%

Case I is physically different from cases II and III. In principle, the geometry
of case I could be obtained from the geometry of cases II or III by a double
analytic continuation in which the $AdS_3$ of case III is mapped into one
of the $S^3$ of case I and one of the $S^3$ of case III is mapped into
the $AdS_3$ of case I. This analytic continuation is very delicate, however, and 
we have found it difficult to carry it out with absolute confidence. For this reason, 
we shall derive the solution for case I again from first principles, just as we did 
for case III.

\sm

For case I, we have $c_2=c_3$, and thus $c_1=-2c_2$. By an overall rescaling 
of the case I geometry, we may choose  $c_2=c_3=1$ and thus $c_1=-2$.
The holomorphic 1-form $\kappa$, given by (\ref{kap2}) up to an overall
multiplicative constant, will be normalized as follows, 
\bea
\kappa = \rho (\a + \b) (\ba - \bb)^3 + \rho ( \a - \b) (\ba + \bb)^3
\eea
The structure of $\kappa$ in terms of $\a$ and $\b$ suggests carrying out 
the following change of variables, 
\bea
a & \equiv & \a + \b
\no \\
b & \equiv & \a - \b
\eea
Expressing the reduced BPS equations (\ref{diffab})  in terms of $a,b,\bar a, \bar b$, 
the corresponding differential  equations are given as  follows,
\bea
\label{diffab3}
D_z a  & = & + { i \over 2} \hat \o_z a +
{1 \over 12} g_{1z} a + {1 \over 12} g_{2z} b - { i \over 12} g_{3z} b
\no \\
D_z b & = & + { i \over 2} \hat \o_z b -
{1 \over 12} g_{1z} b - {1 \over 12} g_{2z} a - { i \over 12} g_{3z} a
\no \\
D_z \bar a  & = & - { i \over 2} \hat \o_z \bar a +
{1 \over 4} g_{1z} \bar a + {1 \over 4} g_{2z} \bar b + { i \over 4} g_{3z} \bar b
\no \\
D_z \bar b   & = & - { i \over 2} \hat \o_z \bar b -
{1 \over 4} g_{1z} \bar b - {1 \over 4} g_{2z} \bar a + { i \over 4} g_{3z} \bar a
\eea
while the corresponding algebraic equations are
\bea
\label{Ialg}
 g_{2z} (a^2 + b^2) - i  g_{3z} (a^2 - b^2) & = & 0
\no \\
2  g_{1z}ab  - i  g_{3z} (a^2 - b^2) & = & 12
\eea

\subsection{Variables adapted to $c_2=c_3$}

The first equation of (\ref{Ialg}) is homogeneous, and allows for a convenient
parametrization of the flux field in terms of a single complex form $\psi$,
\bea
g_{2z} + i g_{3z} & = & + 4 a^2 \psi
\no \\
g_{2z} - i g_{3z} & = & - 4 b^2 \psi
\eea
In terms of $a$ and $b$, the holomorphic 1-form $\kappa$ and
its conjugate $\sigma$, take on simple forms,
\bea
\kappa = \rho ( a \bar b^3 + b \bar a^3) & \hskip 1in &
\bar \kappa = \rho ( \bar a b^3 + \bar b a^3)
\no \\
\sigma = \rho ( a \bar b^3 - b \bar a^3) & \hskip 1in &
\bar \sigma = \rho ( \bar a b^3 - \bar b a^3)
\eea
The four differential reduced BPS equations of (\ref{diffab3}) are equivalent to the
following two equations for $ \bar \kappa, \bar \sigma$,
\bea
\p_w \bar \kappa & = & 0
\no \\
\p_w \bar \sigma & = & -2 \rho ^2 \psi a^2 b^2 (a\bar a - b \bar b)
\eea
and the following two equations for $\kappa , \sigma$,
\bea
\p_w \left ( { \kappa \over \rho^2} \right )
& = &
- {2 \over 3} g_{1z} { \sigma \over \rho}
- {1 \over 3} \psi a^2 \bar a^2  ( a \bar a - 9 b \bar b)
+ {1 \over 3} \psi b^2 \bar b^2  ( 9 a \bar a -  b \bar b )
\no \\
\p_w \left ( { \sigma \over \rho^2} \right )
& = &
- {2 \over 3} g_{1z} { \kappa \over \rho}
+ {1 \over 3} \psi a^2 \bar a^2  ( a \bar a - 9 b \bar b)
+ {1 \over 3} \psi b^2 \bar b^2  ( 9 a \bar a -  b \bar b )
\eea
The advantage of the variables $\kappa, \bar \kappa, \sigma, \bar \sigma$ 
is that $\kappa$ and $\bar \kappa$ should now be viewed as a
{\sl given} holomorphic forms, thus reducing the number of unknowns to
only two, namely $\sigma, \bar \sigma$. In analogy with the steps
followed in the solution of case III, we introduce the combination,
\bea
\p_w \left ( { \sigma ^2 - \kappa ^2 \over 2 \rho^4} \right )
=
{\psi \over 3 \rho^2} \bigg [ a^2 \bar a^2  ( a \bar a - 9 b \bar b) (\sigma + \kappa)
+  b^2 \bar b^2  ( 9 a \bar a -  b \bar b ) (\sigma - \kappa) \bigg ]
\eea
Next, we eliminate $\psi$, $a,b,\bar a, \bar b$ in favor of $\kappa, \bar \kappa,
\sigma , \bar \sigma$ throughout. To do so, it is useful to have the
following formulas,
\bea
\label{formab}
{ \bar a ^2 \over b^2} = - { \sigma - \kappa \over \bar \sigma + \bar \kappa}
\hskip 0.7in
{ a \bar a \over b \bar b} = \left | {\sigma - \kappa \over \sigma + \kappa} \right |
\hskip 0.7in
{  \bar a \bar b \over a b} = - {\sigma^2  - \kappa^2 \over |\sigma ^2 -  \kappa^2 |}
\eea
As for case III, it is advantageous to introduce a rescaled
metric factor $\tilde \rho ^2$ on $\Sigma$, defined by
\bea
\rho^4 = \tilde \rho ^4 |\sigma ^2 - \kappa ^2|^2
\eea
Upon carrying out the above elimination, we find,
\bea
\p_w \ln \bigg ( \tilde \rho ^3 (\bar \sigma ^2 - \bar \kappa ^2)  \bigg )
=
 { 2 \p_w \bar \sigma  \over |\sigma ^2 - \kappa ^2|} \,
\left ( { \bar \kappa (\sigma ^2 - \kappa ^2) - \kappa |\sigma ^2 - \kappa ^2|
\over \sigma \bar \kappa + \kappa \bar \sigma} \right )
\eea
Finally, using the further change of variables of (\ref{ff}) to $\f, \bar \f$,
we find,
\bea
\p_w \ln \bigg ( \tilde \rho ^{3/2} \sh ( 2 \bar \f) \bigg )
= 2 \p_w \bar \f \, { \sh ( \f - \bar \f) \over \ch (\f - \bar \f)}
\eea
We now perform the corresponding changes of variables in the
inhomogeneous algebraic reduced BPS equation of (\ref{Ialg}) as well. 
The starting point is the second algebraic equation in (\ref{Ialg}), 
expressed in terms of $\psi$,
\bea
g_{1z} = { 6 \over ab} + { a^4 - b^4 \over ab} \, \psi
\eea
as well as the differential relation for the ratio $a/b$,
\bea
\p_w \ln { a \over b} = { 1 \over ab} + { a^4 - b^4 \over 2 ab} \, \psi
\eea
Eliminating $\psi $ and $a,b, \bar a, \bar b$ in favor of $\f, \bar\f$ in
all terms requires the computation of $\rho /(ab)$.  As for case III, this
term is first computed to the 8-th power, and we find,
\bea
{ \rho ^8 \over a^8 b^8}= 16 \kappa ^8 \tilde \rho ^{12}  \sh (2 \f) ^8
\eea
Taking its 8-th root will generally introduce an 8-th root of unity $\nu$,
\bea
\label{rhoab}
{ \rho  \over a b}= \sqrt{2} \nu  \kappa  \tilde \rho ^{3/2} \sh (2 \f)
\eea
This yields the following equation,
\bea
\p_w \ln \left ( { \th \f \over \th \bar \f} \right )
=
4 \sqrt{2} \nu \kappa \tilde \rho ^{3/2} \sh ( 2 \f)
+ 4 \p_w \bar \f \, { \sh ( \f + \bar \f) \over \ch (\f - \bar \f) }
\eea
The 8-th root of unity may be further constrained by the following
considerations. Taking the square of (\ref{rhoab}), and dividing the
result by the first formula in (\ref{formab}), we find a positive
combination, and this requires that $\nu ^2 =-1$,
so that $\nu$ is actually a 4-th root of unity only, and equal to $\nu = \pm i$.

\subsection{Further Change of Variables}

The natural variables $\tet$, $\mu$ for this case are
\bea
\f + \bar \f = \mu \hskip 1in
{\sh (2 \f) \over \sh (2 \bar \f)} = e^{2 i \tet}
\eea
In case I, (in contrast with case III) there are no restrictions on the ranges
of $\mu$ and $\tet$, which can take any real values. 
The combination $\f-\bar \f$ is then given by
\bea
\th ( \f - \bar \f) = i \, \tg (\tet) \th (\mu)
\eea
The following formulas will be useful,
\bea
\label{ab13}
|\sh (2 \f)|^2 & = &  { \ch (\mu)^2 \sh (\mu)^2 \over \cos (\tet)^2 + \sh (\mu)^2}
\no \\
\p (\f - \bar \f) & = & {i \over 2}  \, { \sin (2 \tet) \p \mu +  \sh (2 \mu) \p \tet
\over \cos (\tet)^2 + \sh (\mu)^2}
\no \\
{\sh (\f - \bar \f) \over |\sh (2 \f) |} & = & \pm i \, {\sin (\tet) \over \ch (\mu)}
\eea
As in case III, it will be useful to introduce a rescaled metric factor 
$\hat \rho ^{3/2}$, defined by
\bea
\hat \rho ^{3/2} = { 4 \sqrt{2} |\sh (2 \f)|^2  \over \sh (2 \mu)} \, \tilde \rho ^{3/2}
\eea
In terms of $\kappa$, $\hat \rho, \mu$ and $\tet$, the reduced BPS equations become,
\bea
\label{red5}
\p_w \tet + i \p_w \ln \ch (\mu) & = & -i \nu \hat \rho ^{3/2} e^{i \tet} \cos (\tet)
\no \\
\p_w \ln \Big ( \hat \rho ^{3/2} e^{-i \tet } \cos ( \tet) \Big ) & = &
\Big ( i \, \tg (\tet) -1 \Big ) \p_w \ln \ch (\mu)
\eea
Expressing the holomorphic 1-form $\kappa$ as the $(1,0)$-differential
of a real harmonic function $h$,
\bea
\kappa =  \nu \p_w h
\eea
the equation for $\hat \rho$ may be integrated,
\bea
{1 \over \hat \rho ^{3/2}} = h
\eea
To integrate the remaining equation, we introduce yet one more
change of variables,
\bea
G \equiv \ch (\mu) e^{-i \tet}
\eea
By its very construction, and the fact that $\mu$ and $\tet$ are real, we must 
have $|G|\geq 1$.
In terms of $G$, the remaining equation of (\ref{red5}) takes the form,
\bea
\p_w G = \half (G + \bar G) \p_w \ln h
\eea
This equation is identical to the one encountered for case III
and may be solved by the same methods. We shall do so 
explicitly for both cases in section 8.

\subsection{Metric factors}

For case I, the metric factors $f_1, f_2, f_3$ are given in terms of 
$a$ and $b$ by
\bea
f_1 & = & - \half \l_0 = - {1 \over 4} (a \bar a+ b \bar b)
\no \\
f_2 & = & - \l_3 = - {1 \over 2} ( a \bar b + \bar a b)
\no \\
f_3 & = & + \l_2 = { i \over 2} (\bar a b - a \bar b)
\eea
It is convenient to first work out the form of the $\Sigma$-metric factor $\rho$,
by converting $\rho$ to $\hat \rho$, the latter being known directly
in terms of the harmonic function $h$. The following combination
\bea
W^2 \equiv 4 |G|^4 + (G- \bar G)^2
\eea
enters ubiquitously. As a result of the defining range $|G|\geq 1$ of $G$,
it readily follows that within this range, we automatically have $W^2 \geq 0$,
so that $W$ is real, and we shall take it to be positive throughout.
To calculate $\rho$, a helpful formula is as follows, 
\bea
\cos (\tet)^2 + \sh (\mu)^2 = {W^2 \over 4 |G|^2}
\eea
The result is most easily expressed as a formula for $\rho^6$, given by,
\bea
\rho^6 = { |\p_w h|^6 \over (4h)^4}  ( |G|^2 -1 ) W^2
\eea
The metric factors $f_1, f_2, f_3$ are given by the following expressions,
\bea
f_1 ^3 & = & { h W \over 16 (|G|^2-1)}
\no \\
(f_2^2+ f_3^2)^3 & = &  {16 h^2 |G|^6 (|G|^2 -1) \over  W^4}
\no \\
f_2^3 f_3^3 & = & {h^2 (|G|^2-1) \over 4W}  
\eea
Solving the above relations for $f_2^6$ and $f_3^6$, we find,
\bea
f_2 ^6 & = & { 2 h^2 (|G|^2-1) \over W^4} \Big ( |G|^2 \pm  \half |G-\bar G| \Big )^3
\no \\
f_3 ^6 & = & { 2 h^2 (|G|^2-1) \over W^4} \Big ( |G|^2 \mp \half |G-\bar G| \Big )^3
\eea
The correlated sign choices under the 3-rd powers in the above formulas
are reversed under the interchange of the spheres $S_2^3$ and $S_3^3$.
The product of the metric factors is again proportional to $h$, and we have 
$f_1f_2f_3 = \pm h/4$.

\subsection{Calculating $G$ and $h$ for the $AdS_4\times S^7$ solution}

As a check, using the above changes of variables, one may 
evaluate $G$ and $h$ for the $AdS_4\times S_7$ solution with 
32 supersymmetries. The starting point may be taken to be the definition
of the function $\f$ in terms of $w$, given in subsection 3.3,
\bea
\sh (2\f)={i\over\ch (2w)}
\eea
All other functions needed may be computed from this correspondence,
and we find, 
\bea
\ch (\mu)={\ch (w+\bar w)\over |\ch (2w)|}
\hskip 1in 
e^{- i \tet} = i {|\ch (2 w)| \over \ch (2 \bar w)}
\eea
Using the relation between $\hat \rho$ and $h$, and  the relations
between $\hat \rho, \tilde \rho$, with $\rho=1$, we find, 
\bea
G & = & i{\ch (w+\bar w)\over \ch (2\bar w)}
\no \\
h & = & 4i(\sh (2w)-\sh (2\bar w)) 
\eea
For the range $w=x+iy$, with $x \in \bR$, and $0 \leq y \leq \pi/2$, the 
function $G$ sweeps through the entire range $|G|\geq 1$ of Figure 2 in section 8.

\newpage

%%%%%%%%%%%%%%%%%%%%%%%%%%%%%%%%%%%%%%%%%%%%%%%
%%%%%%%%%%%%%%%%%%%%%%%%%%%%%%%%%%%%%%%%%%%%%%%
\section{General solution of the linear equation for $G$ \label{solvng BPS eqns}}
\setcounter{equation}{0}
%%%%%%%%%%%%%%%%%%%%%%%%%%%%%%%%%%%%%%%%%%%%%%%
%%%%%%%%%%%%%%%%%%%%%%%%%%%%%%%%%%%%%%%%%%%%%%%

For cases I, II, and III, the harmonic function $h$  and the complex function 
$G$ satisfy the same complex linear partial differential equation on $\Sigma$, 
\bea
\label{Geq4}
\p_w G = \half (G + \bar G) \p_w \ln h
\eea
The key difference between case I on the one hand, and cases II and III on the 
other hand, is the allowed range of the function $G$. These ranges are defined by
\bea
{\rm case ~ I }  \hskip 0.68in & \hskip 1in & |G|\geq 1
\no \\
{\rm cases ~ II ~  and ~ III} && W^2 = - 4 |G|^4 - (G-\bar G)^2 \geq 0
\eea
and  are depicted in Figure 2 below.

\begin{figure}[tbph]
\begin{center}
\epsfxsize=4.5in
\epsfysize=2in
\epsffile{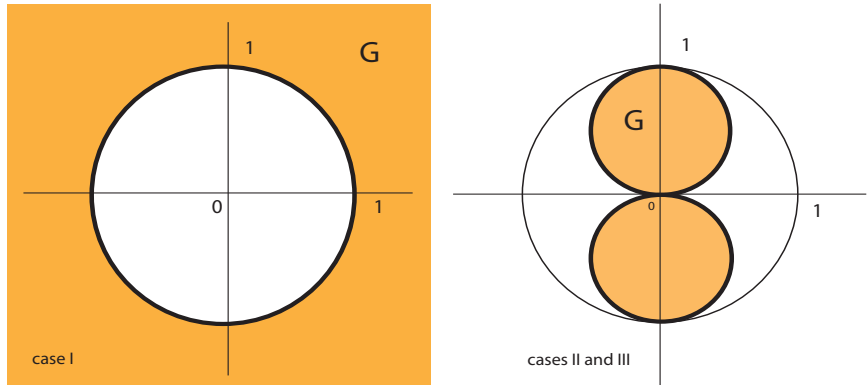}
\label{figure2}
\caption{Allowed ranges of $G$  in the complex plane 
for  case I, and cases II and III. }
\end{center}
\end{figure}

Equation (\ref{Geq4}) for $G $ is manifestly covariant under conformal reparametrizations
of the local conformal coordinate $w$. We shall take advantage of this invariance
to choose conformal coordinates $u = r + i x$ adapted to the harmonic function $h$,
and defined as follows,
\bea
h =  r & \hskip 1in & 2\p_u =  \p_r - i \p_x \qquad
\no \\
\tilde h =  x && 2 \p_{\bar u} =  \p_r + i \p_x \qquad
\eea
Here, we have introduced also the harmonic function $\tilde h$ dual to $h$, so that 
$\p_{\bar u} ( h + i \tilde h)=0$. Recall that since $r=h= \hat \rho ^{-3/2}$,
only the domain $r \geq 0$ is allowed for regular real solutions. It is for this reason that 
we have used the notation $r$, typical of a radial variable for $h$.
Next, we  decompose equation (\ref{bps3}) into its real and imaginary parts.
To do so, decompose $G$ into its real and imaginary parts,
\bea
G(x,r)  = G_r (x,r) + i G_x (x,r)
\eea
for $G_r,G_x$ real functions. The real and imaginary parts of (\ref{bps3})
are respectively given by,
\bea
\label{bps4}
\p_r G_r + \p_x G_x & = & { G_r \over r}
\no \\
\p_r G_x - \p_x G_r & = & 0
\eea
The second equation of (\ref{bps4}) is solved completely  by expressing
the components $G_r$ and $G_x$ in terms of the gradient of a single real function.
For later convenience, we shall include an extra factor of $r$ in the
definition of this function, and denote it by $r \Psi (x,r)$, so that
\bea
\label{Psidef}
G_r  & = & \p_r  ( r \Psi   )
\no \\
G_x  & = & \p_x ( r \Psi   )
\eea
The  first  equation of (\ref{bps4}) then becomes a second order partial
differential equation on $\Psi$,
\bea
\label{Psieq}
\Big ( \p_x^2 + \p_r ^2 + { 1 \over r} \p_r    - { 1 \over r^2} \Big ) \Psi (x,r)=0
\eea
The motivation for including the extra factor of $r$ in the definition
of $\Psi (x,r)$ was to assure that the $r$-part of the above differential
equation is of the form of a 2-dimensional Laplace equation in cylindrical 
coordinates of which $r$ is the radial coordinate. 

\subsection{Solving by  three-dimensional harmonic functions}

A simple geometrical interpretation of equation (\ref{Psieq}) is obtained by
relating it to the  Laplace equation in 3-dimensional Euclidean flat space. 
All our arguments are local; global regularity conditions will be studied
and imposed on the local solutions in a subsequent paper.

\sm

The solutions of (\ref{Psieq}) are in two-to-one correspondence with harmonic 
functions in 3-dimensional Euclidean space. To show this, we first note that, 
if $\Psi (x,r)$ satisfies (\ref{Psieq}), then the real 3-dimensional function $\Phi (x,y,z)$,
defined by
\bea
\Phi (x,r \cos \theta, r \sin \theta ) 
=  \Psi (x,r) e^{+i (\theta - \theta _0)} +  \Psi (x,r) e^{-i (\theta - \theta _0)}
\eea
satisfies the three-dimensional Laplace equation,
\bea
\label{lap3}
\left ( \p_x ^2 + \p_y^2 + \p_z^2 \right ) \Phi (x,y,z)=0
\eea
for any relative constant phase $\theta _0$. It is straightforward to check this
by casting the 3-dimensional Laplace equation in terms of cyclindrical coordinates 
$r,\theta$ for the directions $y,z$, using the fact that 
\bea
{\p ^2 \Phi (x,y,z) \over \p \theta ^2} = - \Phi (x,y,z)
\eea
and then using (\ref{Psieq}). Conversely, let  $\Phi (x,y,z)$ be a solution of the 
3-dimensional Laplace equation (\ref{lap3}).
The complex-valued function $\Psi _c (x,r)$, constructed by the Fourier transform,
\bea
\Psi _c(x,r) = \int _0 ^{ 2 \pi} \! d \theta \, e^{i \theta} \, \Phi (x,r \cos \theta, r \sin \theta )
\eea
automatically satisfies (\ref{Psieq}). The real and imaginary parts of $\Psi _c (x,r)$ 
provide two possible real functions $\Psi (x,r)$ satisfying (\ref{Psieq}).
Thus, all solutions to (\ref{Psieq}) may
be obtained by this projection method from 3-dimensional harmonic functions.

\subsection{Solving by Fourier transform}

A more direct method of solving (\ref{Psieq}) involves direct Fourier transformation.
Since the equation is invariant under arbitrary translations of the variable $x$,
we use Fourier analysis in $x$. Since the function $\Psi (x,r)$ is real,
we have the following general Fourier representation,
\bea
\Psi (x,r) = \int _0 ^ \infty { dk \over 2 \pi} \,
\Big ( \Psi _k (r) \, e^{-ikx} + \Psi _k (r)^* \, e^{+ikx} \Big )
\eea
The individual Fourier modes $\Psi _k (r)$ satisfy the modified Bessel equation,
for $r >0$,
\bea
\left ( - k^2 + \p_r ^2 + { 1 \over r} \p_r   - { 1 \over r^2} \right )  \Psi _k (r)  =0
\eea
The general solution of this equation for $k\geq 0$ is given by
\bea
\label{psi5}
\Psi _k (r) = - \pi \psi_1 (k) I_1 (kr) + \psi_2 (k) K_1 (kr)
\eea
where $I_1(kr)$ and $K_1(kr)$ are  modified Bessel functions,
and $\psi _1(k)$ and $\psi _2(k)$ are arbitrary complex functions of $k\geq 0 $.
(The extra factor of $- \pi$ has been introduced for later convenience.)
The modified Bessel functions admit the following useful integral
representations,
\bea
I_1 (kr) & = &  - { 1 \over \pi} \int _{-1} ^1 { t \, dt \over  \sqrt{1-t^2}}  \, e^{- tkr}
\no \\
K_1 (kr) & = & \int _1 ^\infty { t \, dt \over  \sqrt{t^2 -1 }}  \, e^{- tkr}
\eea
These representations are absolutely convergent for all real $kr>0$, which
is indeed the case here.
Next recast the solution $\Psi (x,r)$ in terms of these integral representations,
\bea
\Psi (x,r) & = &
+ \int _0 ^\infty { dk \over 2 \pi } \, \psi_1 (k) e^{-ikx}
\int _{-1} ^1 { t \, dt \over  \sqrt{1-t^2}}  \, e^{- t kr } + {\rm c.c.}
\no \\ &&
+ \int _0 ^\infty { dk \over 2 \pi } \, \psi_2 (k) e^{-ikx}
\int _1 ^\infty { t \, dt \over  \sqrt{t^2-1}}  \, e^{- t kr } + {\rm c.c.}
\eea
Let us now define the following functions,
\bea
C_1 (tr+ix) & \equiv &  \int _0 ^\infty { dk \over 2 \pi } \, \psi_1 (k) \,  e^{-  k(t r+ ix) }
\no \\
C_2 (tr+ix) & \equiv &  \int _0 ^\infty { dk \over 2 \pi } \, \psi_2 (k) \, e^{-  k(t r+ ix) }
\eea
and their complex conjugates.
Since $\psi_1(k)$ and $\psi_2(k)$ were arbitrary complex functions of $k$,
the functions $C_1$ and $C_2$ are arbitrary functions of their argument.
The argument is a complex variable $tr+ix$; the functions $C_1$ and $C_2$
depend on this variable, but not on its complex conjugate. (The complex
conjugated functions $C_1(tr+ix)^*$ and $C_2 (tr+ix)^*$ depend on the
complex conjugate variable $tr-ix$ but not on $tr+ix$.) Thus, it is appropriate to
interpret the functions $C_1$ and $C_2$ as {\sl holomorphic functions
of their argument}. The functions entering the real integrals are thus
harmonic functions, but not of $w$, but instead of $t h + i \tilde h$.
Then the solution $\Psi (x,r)$ may be expressed as follows,
\bea
\label{Psi3}
\Psi (x,r)& = &
\int _{-1} ^1 { t \, dt \over  \sqrt{1-t^2}}  \bigg ( C_1 (tr+ix) + C_1 (tr+ix)^* \bigg )
\no \\ &&
+ \int _1 ^\infty { t \, dt \over  \sqrt{t^2-1}}  \bigg ( C_2 (tr+ix) + C_2 (tr+ix)^* \bigg )
\eea
It is straightforward to show directly that this expression is a solution
to the original differential equation (\ref{Psieq}) for any holomorphic
functions $C_1$ and $C_2$,  by using integration by parts. But, using
the steps we have taken, we now know that this is the most general solution.
One may also directly work out the expression for $G$,
\bea
G  (x,r)& = & r \int _{-1} ^1 { dt \over \sqrt{1-t^2}} \bigg (
(1-t) C_1 ' (tr+ix) + (1+t) C_1'(tr+ix)^* \bigg )
\no \\ && +
r \int _1 ^\infty { dt \over \sqrt{t^2-1}} \bigg (
(1-t) C_2 ' (tr+ix) + (1+t) C_2'(tr+ix)^* \bigg )
\eea
This result gives the complete local solution to the BPS equations
in exact form.  The above expression is indeterminate in the limit 
$r \rightarrow 0$ due to the divergence of the integrals and the overall 
factor of $r$.  It will therefore be useful to have an asymptotic form for 
$G(r,x)$ near $r = 0$.  First we expand $\Psi$ in a series about $r = 0$ 
using the expression (\ref{psi5})
\bea
\Psi_k(r) \approx \psi_2(k) \bigg( {1 \over r} + \half r \ln r \bigg) + \cO(r)
\eea
Computing first $\Psi(x,r)$ and then $G(x,r)$ using (\ref{Psidef}) we obtain
\bea
G(x,r) = \int_0^\infty {dk \over 2\pi} \bigg(k \, \psi_2(k) e^{-i k x} + c.c. \bigg) + \cO(r)
\eea
which is just the statement that $G(x,0)$ is an arbitrary function of $x$.

\subsection{Equations for $\Psi $ and $G$ in general coordinates}

From  equation (\ref{Psidef}), we deduce that
\bea
G  = G _r + i G _x =  2 \p_{\bar u}  ( h \Psi )
\hskip 1in u = r+ i x = h + i \tilde h
\eea
This expression is cast in the form of the special coordinates
$u$ adapted to the arbitrary harmonic function $h$. We can, however,
also express $G$ in terms of $\Psi$ with general coordinates $w$,
by changing conformal coordinates from $u$ to $w$. We use,
$\p_w u = \p_w (h + i \tilde h) = 2 \p_w h$, and its inverse,
to give the following form for $G$ in arbitrary conformal coordinates $w$,
\bea
G  = { \p_{\bar w} (h  \Psi )\over \p_{\bar w} h}
\eea
The first order differential equation for $G$ of (\ref{bps3})
may be recast in terms of a second order differential equation for $h \Psi$,
for a general coordinate system $w, \bar w$,
by eliminating $G$ in terms of $h \Psi$. This equation is given by,
\bea
2 \p_w \p_{\bar w} \Psi
+ \p_w  \Psi \p_{\bar w} \ln h + \p_{\bar w}  \Psi \p_w \ln h
+ 2  \Psi \p_w \p_{\bar w} \ln h =0
\eea
valid for arbitrary conformal coordinates $w$ and $\bar w$. The general
solution to this equation is simply given by changing variables in the
result for $\Psi$ expressed in terms of coordinates $x,y$ in (\ref{Psi3}),
and we find, in general coordinates $w,\bar w$,
\bea
\Psi (\tilde h, h)& = &
\int _{-1} ^1 { t \, dt \over  \sqrt{1-t^2}}
\bigg ( C_1 (th+i\tilde h ) + C_1 (th +i \tilde h)^* \bigg )
\no \\ &&
+ \int _1 ^\infty { t \, dt \over  \sqrt{t^2-1}}
\bigg ( C_2 (th +iy \tilde h ) + C_2 (th+iy \tilde h)^* \bigg )
\eea
Similarly, $G$ in general coordinates takes the form,
\bea
\label{G7}
G  (\tilde h, h) & = & h \int _{-1} ^1 { dt \over \sqrt{1-t^2}} \bigg (
(1-t) C_1 ' (t h+i \tilde h) + (1+t) C_1'(t h+i \tilde h)^* \bigg )
\no \\ && +
h \int _1 ^\infty { dt \over \sqrt{t^2-1}} \bigg (
(1-t) C_2 ' (t h+i \tilde h) + (1+t) C_2'(t h+i \tilde h)^* \bigg )
\eea
In both equations, it is understood that $h$ and $\tilde h$ are
to be expressed as functions of $w, \bar w$.

\subsection{The inverse problem of determining  $C_1, C_2$ from $G$}

In the case of solutions $G$ such that
\bea
G_\p (\tilde h) = \lim _{h \to 0} \left ( { G \over h} \right )
\eea
is finite, there is a simple method of recovering the functions $C_{1,2}$
from $G$. Assuming that $G_\p$ is a finite function of $\tilde h$,
we take the corresponding limit in the integral of (\ref{G7}).
Both integrals involving $C_2$ on the second line of (\ref{G7}) diverge
in the limit $h \to 0$. Canceling the leading linear divergence
requires  $C_2 ' (\tilde h) = C_2 ' (\tilde h)^*$ for all $\tilde h$. Canceling also 
the remaining logarithmic divergence requires $C_2 ' (\tilde h) = 0$ for all $\tilde h$.
Since $C_2(th + i \tilde h)$ is a holomorphic function of $th + i \tilde h$
which vanishes on the line $h=0$, it must be identically zero,
$C_2 (th + i \tilde h)=0$.
The remaining $t$-integrals multiplying $C_1'(\tilde h)$ and $C_1'(\tilde h)^*$
may be evaluated and equal  $\pi$. We thus obtain
\bea
C_1 ' (i \tilde h) + C_1 ' (i \tilde h)^* = {G_\p  (\tilde h) \over \pi}
\eea
This is enough information to determine the real part of the $C_1'$.
In addition, it shows that $G $ is real on the boundary defined by $h=0$
and the imaginary part of the $C_1'$ only affects the value of $G$ away
from the boundary defined by $h = 0$.

\subsection{The functions $\Psi$, $C_1, C_2$ for the $AdS_7 \times S^4$ case}

We now proceed to evaluate the functions $\Psi$, $C_1$ and $C_2$ for the 
$AdS_7 \times S^4$ case.  First we quote expressions for $G$, $h$, and $\tilde h$,
\bea
\label{Ghh}
G(w, \bar w) &=& -i {\sh(w - \bar w) \over \sh(2 \bar w)}
\no\\
r= h(w, \bar w) &=& -i (\ch(2w) - \ch(2 \bar w))
\no\\
x= \tilde h(w, \bar w) &=& - (\ch(2w)+ \ch(2 \bar w))
\eea
The expression for $h \Psi$ is readily computed and we find,
\bea
h \Psi (\tilde h, h) = - 2 \ch (w-\bar w)
\eea
To obtain $\Psi (\tilde h, h)$ as a function of $\tilde h$ and $h$ is 
surprisingly complicated, as $\Psi$ is found to satisfy a 4-th degree 
polynomial equation, 
\bea
{ \tilde h ^2 \over h^2 \Psi^2 } + { h^2 \over h^2 \Psi ^2 -4} =1
\eea
In other words, the equipotential lines for $h \Psi (\tilde h,h)$ are ellipses,
when $|h \Psi (\tilde h ,h)| \geq 2$, and hyperbolas, when  $|h \Psi (\tilde h, h)| \leq 2$.
It would be difficult to carry out the Fourier integrals needed to 
derive the functions $C_1$ and $C_2$. Fortunately, we may use 
the more indirect methods discussed above to extract $C_1$ and $C_2$
from the limiting behavior as $h \to 0$.
As a result,  we begin by first calculating the ratio $G/h$.
The harmonic function  $h$ vanishes whenever $w-\bar w = i n \pi$ for any
integer $n$, so that $w = s + i n \pi/2$, for any real $s$. For simplicity, consider the 
case $n=0$, the other cases being completely analogous. For these values,
we have $\tilde h = - 2 \ch (2s)$, and 
\bea
G_\p = { 1  \over 2 \, \sh (2s)^2 } = -  { 2  \over  (i\tilde h)^2 + 4}
\eea
The limit $h \to 0$ of $G/h$ is a finite function $G_\p$ of $\tilde h$, and we
conclude that we must have $C_2=0$. The result further determines the real
part of the holomorphic function $C_1' (th + i \tilde h)$ on the boundary $h=0$.
Assuming that the imaginary part of this function
vanishes for $h=0$, we find that
\bea
\label{C1}
C_1 ' (v) = -  {1 \over \pi} \, { 1  \over  v^2 + 4} 
= { i \over 4 \pi} \left ( { 1 \over v - 2 i}  - { 1 \over v + 2i} \right )
\eea
for all complex $v=th+i\tilde h$. In turn, substituting this expression
and $C_2=0$ into (\ref{G7}), and performing the $t$-integrations
yields back expression (\ref{Ghh}) for $G$ in terms of $w, \bar w$, 
as we shall show explicitly in section 8.7 below.

\subsection{The function $\Psi$ for the $AdS_4 \times S^7$ case}

Next, we proceed to evaluating the functions $\Psi$, $C_1$ and $C_2$ for the 
$AdS_4 \times S^7$ case.  First we quote expressions for $G$, $h$, and $\tilde h$,
\bea
\label{Ghh4}
G(w, \bar w) &=& i {\ch(w + \bar w) \over \ch(2 \bar w)}
\no\\
r= h(w, \bar w) &=& 4i (\sh(2w) - \sh(2 \bar w))
\no\\
x= \tilde h(w, \bar w) &=& 4 (\sh(2w) + \sh(2 \bar w))
\eea
The expression for $h \Psi$ is readily computed and we find,
\bea
h \Psi (\tilde h, h) = 8 \, \sh (w + \bar w)
\eea
To obtain $\Psi (\tilde h, h)$ as a function of $\tilde h$ and $h$ is 
surprisingly complicated, as $\Psi$ is found to satisfy a 4-th degree 
polynomial equation, 
\bea
{ \tilde h ^2 \over h^2 \Psi^2 } + { h^2 \over h^2 \Psi ^2 + 64} =1
\eea
The equipotential lines for $h \Psi (\tilde h,h)$ are always ellipses.

\subsection{Simple poles in $C_1'$ and $C_2'$}

To derive the functions $C_1$ and $C_2$ of (\ref{Psi3})
for the case $AdS_4 \times S^7$, it would be difficult 
to carry out the Fourier integrals on $\Psi$, and its is also not possible
to use the methods of section 8.4, since the quantity $G/h$ does 
not have a finite limit as $h \to 0$. Having the result for the $AdS_7 \times S^4$
case in terms of simple poles for $C_1$, it is natural to see a result
for $AdS_4 \times S^7$ also in terms of simple poles for $C_1$ and $C_2$.
This will allow us to check, by direct calculation, that both cases
correspond to simple poles for $C_1'$ and $C_2'$.

\sm

The key ingredients are the following integrals, for $z \in \bC$, 
\bea
\int _{-1} ^1 { dt \over \sqrt{ 1 - t^2} } \, { 1 \over t + z} & = & 
{ \pi \over \sqrt{ z^2 -1} }
\no \\
\int _1 ^\infty { dt \over \sqrt{ t^2 -1 } } \, { 1 \over t + z} & = & 
{ 1 \over \sqrt{ z^2 -1} } \, \ln \Big ( z + \sqrt{z^2-1} \Big ) 
\eea
Here, the branches have been defined so that for $z \in \bR$ and $z >1$
both integrals are real and positive, with a real and positive branch 
is chosen for the square rots and for the logarithm. Analytic continuations 
$z \to -z$ must be carried out with care, and we find, 
\bea
\int _{-1} ^1 { dt \over \sqrt{ 1 - t^2} } \, { 1 \over t - z} & = & 
- { \pi \over \sqrt{ z^2 -1} }
\no \\
\int _1 ^\infty { dt \over \sqrt{ t^2 -1 } } \, { 1 \over t - z} & = & 
- { 1 \over \sqrt{ z^2 -1} } \, \ln \Big ( - z - \sqrt{z^2-1} \Big ) 
\eea
In particular, by putting together the following combinations, 
\bea
\int _1 ^\infty { dt \over \sqrt{ t^2 -1 } } \, \left ( { 1 \over t + z} + { 1 \over t -z} \right )
& = &  { \pi \over \sqrt{ 1- z^2} } 
 \eea 
Using these integrals we may now check that $C_1'$ of (\ref{C1})
together with $C_2'=0$ indeed yields $G$ as given by (\ref{Ghh}).
We shall also evaluate the contribution to $G$ from a simple pole 
in $C_2'$. Both calculations result from considering either $C_1'$ or
$C_2'$ to be of the form,
\bea
\label{pole}
C'(v) = { a \over v +b}
\eea
for $a,b \in \bC$. The integrands for both calculations may be simplified as follows,
\bea
r (1-t) C'(tr + i x ) + r (1+t)  C'(tr+ix)^*
= - (a - \bar a) + a { z+1 \over t+z} - \bar a { \bar z -1 \over t + \bar z}
\eea
where we have used the abbreviation
\bea 
\label{zz}
z = ( i x + b )/ r
\eea

\subsubsection{Contributions from poles in $C_1'$}

The contribution $G_1(x,r)$ to $G(x,r)$ of a simple pole (\ref{pole})  in $C_1'$
is given by, 
\bea
G_1 (x,r) 
& = & 
- \pi (a - \bar a) + \pi a \sqrt{ { z+1 \over z-1} } 
- \pi \bar a \sqrt{ { \bar z-1 \over \bar z+1} }
\no \\
& = & 
- \pi (a - \bar a) + \pi a \sqrt{{ r + i x + b \over - r + i x +b}} 
-   \pi \bar a  \sqrt{{ - r - i x + \bar b \over r - i x + \bar b}}
\eea
For purely imaginary $b$, the two square roots are equal to one another.
To recover (\ref{C1}), we add the contribution from $b=-2i$ and the opposite 
of that from $b=2i$, 
with the same  value of $a$ cancels the $- \pi (a-\bar a)$ term, and gives, 
\bea
G(x,r) =  \pi  ( a - \bar a) \left (
 \sqrt{{ r + i x - 2i \over -r + i x - 2i}} -   \sqrt{{ r + i x + 2i \over -r + i x + 2i}} \right )
\eea
Using now the combinations 
\bea
r + i x & = & - 2 i \ch (2w)
\no \\
r -i x & = &  + 2 i \ch (2 \bar w)
\eea
gives
\bea
G(x,r) = \pi ( a - \bar a) \left ( { \ch w \over \ch \bar w} - { \sh w \over \sh \bar w} \right )
= -2 \pi (a - \bar a) { \sh (w - \bar w) \over \sh (2 \bar w)}
\eea
Clearly, we must choose $ a = i /( 4 \pi)$ to recover (\ref{Ghh}),
and this value agrees with the one found in (\ref{C1}).

\sm

\subsubsection{Contributions from poles in $C_2'$}

The contribution $G_2(x,r)$ to $G(x,r)$ of a simple pole (\ref{pole}) in $C_2'$,
 is given by
\bea
G_2 (x,r) 
& = & - (a - \bar a) \ln \Lambda 
+  a \sqrt{ { z+1 \over z-1} } \ln \Big ( z + \sqrt{ z^2 -1} \Big )
\no \\ && \hskip 1in 
-  \bar a \sqrt{ { \bar z-1 \over \bar z+1} } \ln \Big ( \bar z + \sqrt{ \bar z^2 -1} \Big )
\eea
where we have again used the definition (\ref{zz}) for $z$,
and $\Lambda$ is a constant cutoff for the divergent integral over $t$.
We now add the contributions for $(a,b)$ and the opposite of $(\bar a, - \bar b)$, 
which  is a combination chosen so that the logarithms will only contribute
through their discontinuities, namely though $\ln (z) - \ln (-z) = - i \pi$. 
The result for $C_2'$ is given by,
\bea
C_2 '(v) = { a \over v +b} - { \bar a \over v - \bar b}
\eea
and the resulting contribution to $G$ is given by
\bea
G(x,r) =
-i \pi  a \sqrt{{ r + i x + b \over - r + i x + b }} 
- i \pi \bar a \sqrt{{ -r - i x + \bar b \over r - i x + \bar b}}
\eea
Using now the combinations 
\bea
r + i x & = & +8 i \sh (2w)
\no \\
r -i x & = &  -8 i \sh (2 \bar w)
\eea
and the choice $b = -8$, we find, 
\bea
G(x,r) =
-i \pi  a \sqrt{{  i \, \sh (2w) -1 \over i \, \sh (2 \bar w) -1}} 
- i \pi \bar a \sqrt{{ i \, \sh (2 w) +1  \over i \, \sh (2 \bar w) +1 }}
\eea
Using now the identities,
\bea
i \, \sh (2 w) \pm  1  =  \pm  2  \ch \left ( w \pm i  \pi /4 \right )^2
\eea
and their complex conjugate relations, we readily show that
\bea
G(x,r) = -i \pi a { \ch (w - i \pi /4) \over \ch (\bar w - i \pi /4)}
- i  \pi \bar a  { \ch (w + i \pi /4) \over \ch (\bar w + i \pi /4)}
\eea
To recover the expression for $G$ of (\ref{Ghh4}), it suffices to take $C_1'=0$, 
$a$ real and equal to $a = -1/( 4 \pi)$, and to use the following identities,
\bea
\ch (2 \bar w)
& = & 
2 \ch (\bar w + i \pi /4) \ch (\bar w - i \pi /4)
\no \\
\ch (w+\bar w)
& = & 
\ch (w - i \pi /4) \ch (\bar w + i \pi /4) +  \ch (w + i \pi /4) \ch (\bar w - i \pi /4)
\eea
This concludes our calculation of $C_1'$ and $C_2'$ for the case $AdS_4 \times S^7$.

\newpage

%%%%%%%%%%%%%%%%%%%%%%%%%%%%%%%%%%%%%%%%%%%%%%%
%%%%%%%%%%%%%%%%%%%%%%%%%%%%%%%%%%%%%%%%%%%%%%%
\section{Bianchi Identities and Field equations}
\setcounter{equation}{0}
%%%%%%%%%%%%%%%%%%%%%%%%%%%%%%%%%%%%%%%%%%%%%%%
%%%%%%%%%%%%%%%%%%%%%%%%%%%%%%%%%%%%%%%%%%%%%%%

We have now completely solved the BPS equations for cases I, II, and III,
by reducing the problem to a set of linear equations.
We shall now show that the Bianchi identities and the field equations  are
automatically satisfied, as soon as the BPS equations hold.  

\sm

We shall present explicitly only the case III, for which $c_1 = c_2$.  The fact that
cases I and II are related to case III by analytic continuation then automatically
guarantees that the Bianchi identities and field equations follow from the BPS
equations also for cases I and II.

\subsection{Bianchi identities}

The reduced Bianchi identities (\ref{bianchipre}) may be expressed as,
\bea
\p_w b_i = - \rho f_i^3 g_{zi} \hskip 1in i=1,2,3
\eea
where the $b_i$ are the real gauge potentials.  Using equation (\ref{rads1}) 
for the metric factors and equation (\ref{gs}) for the fluxes in terms of $\a$, $\b$ 
and $\psi$, integrability of the Bianchi identities is
equivalent to conservation of the following currents
\bea
\p_w b_1 \sim j_w^1 &=& \rho (\ba \a + \bb \b)^3 (\a^2 + \b^2) \psi
\no\\
\p_w b_2 \sim j_w^2 &=& - \rho (\ba \a - \bb \b)^3 (\a^2 - \b^2) \psi
\no\\
\p_w b_3\sim j^3 _w &=&  \left ({\bb \over \b} - {\ba \over \a} \right )^3
\left [ - 3  \rho \a^2 \b^2 + \rho (\a^4 - \b^4) \a^2 \b^2 \psi \right ]
\eea
where the  $\sim$ sign stands for equality up to an overall constant factor.
Current conservation is expressed here as the closure of differential forms,
and takes the form,
\bea
\p_{\bar w} j_w^i - \p_w j_{\bar w}^i = 0 \hskip 1in  i=1,2,3
\eea
Instead of working with $j_w^1$ and $j_w^2$, it will be convenient to work 
with $j^{\pm}_w$ defined as
\bea
j^{\pm}_w = {1 \over 2} (j_w^1 \pm j_2^2)
\eea
To confirm that the BPS equations imply the Bianchi identities, we  first
re-write the currents in terms of $G$ and $\bar G$ and then use the remaining 
BPS equations (\ref{Psidef}) and (\ref{Psieq}) to show that the Bianchi identities 
are automatically satisfied when the BPS equations are.  

\sm

As a first step we express the Bianchi identities in terms of $\f$ and $\tilde \rho$ 
using equations (\ref{phich}) to eliminate $\a$ and $\b$, equation (\ref{rhotildedef}) 
to eliminate $\rho$, and equation (\ref{psi4}) as well as the fact 
$\p_w \bar \sigma = 2 \bar \kappa \sh(2 \bar \f) \p_w \bar \f$ 
to eliminate $\psi$,
\bea
j^+ _w & = &
{\sqrt{2} \p_w \bar \f \over   \tilde \rho ^{3/2}\,  \sh (2 \bar \f) \sh (\f + \bar \f)}
\bigg  (  2 \ch (\f + \bar \f)  \ch (2 \bar \f) - \ch (\f - \bar \f)  \bigg )
\no \\
j^- _w & = &
- {\p_w \bar \f \over \sqrt{2}  \tilde \rho ^{3/2}\,  |\sh (2 \f)  | \sh (\f + \bar \f)}
\bigg (  \sh ( 2 \bar \f) \ch (2 \f)  + 3  \sh (2 \f)  \ch (2 \bar \f)   \bigg )
\no\\
j^3 _w & = &   { \sh (\f - \bar \f)^3 \over \sh ( 2 \bar \f) |\sh (2 \f)|}
\left ( -12 \kappa
+ 8 \sqrt{2} { \p_w \bar \f \,   \over \tilde \rho ^{3/2} \,  } ~
{  \sh \f (\ch \bar \f)^3  -  \ch \f (\sh \bar \f)^3  \over |\sh (2 \f) |^2 \, \sh (\f + \bar \f) }
\right )
\eea
Next we recast the conserved currents in terms of the variables $h$, $\tet$ and $\mu$.  
This is done by using the definition of $\hat \rho$ (\ref{rhohatdef}) and the fact 
$\hat \rho^{-3/2} = h$ as well as the definitions for $\tet$ and $\mu$ (\ref{ab1}) 
and the identities (\ref{ab2}) and (\ref{ab3}).  Finally, we may use the definition of 
$G$ (\ref{Gdef}) to write the conserved currents in terms of $G$ and $\bar G$ as follows
\bea
j_w^+ &=&
2  i  h \,
\bigg( \bar G (G - 3 \bar G + 4 G \bar G^2) \p_w G + G (G + \bar G) \p_w \bar G \bigg)
\no\\&&\hskip0.5in \times
\bigg((G - \bar G)^2 - 4 G^3 \bar G\bigg)  (G + \bar G)^{-1} W^{-4}
%\no\\&&
\no\\
j_w^- &=& 2 h \, G
\bigg( \bar G (G - 3 \bar G + 4 G \bar G^2) \p_w G
+ G(G + \bar G) \p_w \bar G \bigg)
\no\\ && \hskip0.5in \times
\bigg(-2 G \bar G + 3 \bar G^2 - G^2 + 4 G^2 \bar G^2 \bigg)
(G+\bar G)^{-1} W^{-4}
%\no\\&&
\no\\
j_w^3 &=& 
3  \p_w h \, {W^2 \over G (1 - G \bar G)}
-2 h \, {(1 + G^2) \over G (G + \bar G) (1 - G \bar G)^2} 
\no\\ && \hskip 1in 
\times \bigg( \bar G (G - 3 \bar G + 4 G \bar G^2) \p_w G +
G( G+ \bar G) \p_w \bar G \bigg) 
\eea
Integrability may now be checked as follows.  First we choose  conformal coordinates 
such that $h = r$ as in section (\ref{solvng BPS eqns}).  Next we eliminate $G$ and 
$\bar G$ in terms of $\Psi$ using $(\ref{Psidef})$.  Next we compute 
$\p_{\bar w} j_w^i - \p_w j_{\bar w}^i$ where $i = +,-,3$.  Upon using the fact $h$ is 
harmonic and the differential equation for $\Psi$ (\ref{Psieq}) to eliminate the 
terms with second order derivatives and higher in $r$ the resulting expressions 
will automatically vanish.  This shows that the BPS equations imply the Bianchi identities 
are automatically satisfied. Finally, we give the expressions for the $\p_w b_i$ with the 
correct normalization factors
\bea
\label{bi}
\p_w b_1 &=& -2 (j_w^+ + j_w^-)
\no\\
\p_w b_2 &=& 2 (j_w^+ - j_w^-)
\no\\
\p_w b_3 &=& - {1 \over 8} \, j_w^3
\eea

\subsection{Field  equations for the 4-form $F$}

The field equation for the 4-form $F$ is given by (\ref{Maxwell})
\footnote{Our convention for the Hodge dual is given as follows.  
Let $F_{(p)}$ be a p-form $F_{(p)} = {1 \over p!} F_{a_1 a_2 ... a_p} e^{a_1 a_2 ... a_p}$.  
The Hodge dual is defined by 
$* F_{(p)} = {1 \over p! (11-p)!} \epsilon_{a_1 a_2 ... a_p b_{p+1} b_{p+2} ... b_{11}} 
F^{a_1 a_2 ... a_p} e^{b_{p+1} b_{p+2} ... b_{11}}$, and $\ep_{a_1 a_2 ... a_{11}}$ 
is the anti-symmetric tensor with $\ep_{0 1 2 3 ... 9 \natural} = +1$.  
In particular, we have the following results needed in the calculations of the present
paper, $* e^{012a} = - \epsilon^a {}_b e^{345678b}$, 
$* e^{345a} = - \epsilon^a {}_b e^{012678b}$ and 
$* e^{678a} = + \epsilon^a {}_b e^{012678b}$.}
\bea
d * F + {1 \over 2} F \wedge F = 0
\eea
Writing out in components we have
\bea
\label{maxwelleqns}
0 &=& 
\p_{\bar w} \p_w b_1 
+ \half \left ( \p_{\bar w} b_1 \p_w \ln \bigg ( {f_2 f_3 \over f_1} \bigg ) ^3 + c.c. \right )
+ {i \over 4} \bigg( {f_1 \over f_2 f_3} \bigg)^3 (\p_w b_2 \p_{\bar w} b_3 - c.c.)
\no\\
0 &=& 
\p_{\bar w} \p_w b_2 
+ \half \left ( \p_{\bar w} b_2 \p_w \ln \bigg ( {f_1 f_3 \over f_2} \bigg ) ^3 + c.c. \right )
+ {i \over 4} \bigg( {f_2 \over f_1 f_3} \bigg)^3 (\p_w b_1 \p_{\bar w} b_3 - c.c.)
\no\\
0 &=& 
\p_{\bar w} \p_w b_3 
+ \half \left ( \p_{\bar w} b_3 \p_w \ln \bigg ( {f_1 f_2 \over f_3} \bigg ) ^3 + c.c. \right )
- {i \over 4} \bigg( {f_3 \over f_1 f_2} \bigg)^3 (\p_w b_1 \p_{\bar w} b_2 - c.c.)
\eea
where we have made use of the relation (\ref{bianchipre}) to write the equations in 
terms of the $b_i$.

\sm

Using the BPS equations, the 4-form equations (\ref{maxwelleqns})  may be shown 
to be satisfied automatically. This is done as follows. First we express the equations 
in terms of $G$, $\bar G$ and $h$ using  $(\ref{bi})$ for the fluxes.  Next, we choose a conformal gauge such that $h = r$ as in section (\ref{solvng BPS eqns}).  Next, we 
re-express $G$ in terms of $\Psi$ using (\ref{Psidef}).  Finally we use the differential 
equation (\ref{Psieq}) for $\Psi$  to eliminate terms with second order derivatives 
and higher in $r$.  The resulting equations will then be satisfied automatically.  
The resulting calculations are lengthy and will not be repeated here; they were 
confirmed using MATHEMATICA.

\subsection{Einstein's equations}

Einstein's equations are given by (\ref{Einstein})
\bea
R_{MN} - {1 \over 12} F_{MPQR} F_N {}^{PQR} 
+ {1 \over 144} g_{MN} F_{PQRS} F^{PQRS} =0
\eea
To obtain Einstein's equations, we will need to calculate the Ricci tensor $R_{MN}$ this is most easily done by first computing the curvature two-form $\Omega^A {}_B$ and then the Ricci tensor using
\bea
\Omega^A {}_B &=& d \omega^A {}_B + \omega^A {}_C \wedge \omega^C {}_B
\no\\
i_{e^A} \Omega^A {}_B &=& R_{BD} e^D
\eea
The expressions for the curvature two-forms on the symmetric spaces $AdS_3$,
$S_2^3$ and $S_3^3$ as well as the two-dimensional base space $\Sigma$ are given by
\bea
\hat \Omega ^m {}_n = - \hat e^m \wedge \hat e_n
&\qquad& \qquad
\hat \Omega ^{i_1} {}_{j_1} = + \hat e^{i_1} \wedge \hat e_{j_1}
\no \\
\hat \Omega ^{i_2} {}_{j_2} = + \hat e^{i_2} \wedge \hat e_{j_2}
&\qquad& \qquad
\Omega ^a {}_b  =  R^{(2)} e^a \wedge e^b
\eea
The curvature two-form form can then computed using the above equations 
and the expressions for the spin-connection (\ref{spincon}). The {\sl block
diagonal entries} are given as follows, 
\bea
\Omega ^m {}_n & = &
\left ( - {1 \over f_1^2} - | D_a \ln f_1 |^2 \right ) e^m \wedge e_n
\no \\
\Omega ^{i_1} {}_{j_1} & = &
\left ( + {1 \over f_2^2} - | D_a \ln f_2 |^2 \right ) e^{i_1} \wedge e_{j_1}
\no \\
\Omega ^{i_2} {}_{j_2} & = &
\left ( + {1 \over f_3^2} - | D_a \ln f_3 |^2 \right ) e^{i_2} \wedge e_{j_2}
\no \\
\Omega ^a {}_b & = & R^{(2)} e^a \wedge e^b
\eea
where we use the notation $|D_a f |^2 \equiv D^a f D_a f$.
The {\sl block off-diagonal entries} between two of the factor spaces $AdS_3$,
$S_2^3$ and $S_3^3$ are given by,
\bea
\Omega ^m {}_{i_1} & = & - (D^a \ln f_2) (D_a \ln f_1)  e^m \wedge e_{i_1}
\no \\ 
\Omega ^m {}_{i_2} & = &  - (D^a \ln f_3) (D_a \ln f_1)  e^m \wedge e_{i_2}
\no \\ 
\Omega ^{i_1} {}_{i_2} & = & - (D^a \ln f_2) (D_a \ln f_3)  e^{i_1} \wedge e_{i_2}
\eea
The {\sl block off-diagonal entries} between one of the factor spaces $AdS_3$,
$S_2^3$ and $S_3^3$ and the surface $\Sigma$  are given by, 
\bea
&& 
\Omega ^m {}_a  = {1 \over f_1} (D_b D_a f_1) + \epsilon^c {}_a {D_c f_1 \over f_1} e^m \wedge \hat \omega
\no \\ &&
\Omega ^{i_1} {}_a = {1 \over f_2} (D_b D_a f_2) e^b \wedge e^{i_1} + \epsilon^c {}_a {D_c f_2 \over f_2} e^{i_1} \wedge \hat \omega
\no \\ &&
\Omega ^{i_2} {}_a = {1 \over f_3} (D_b D_a f_3) e^b \wedge e^{i_2} + \epsilon^c {}_a {D_c f_3 \over f_3} e^{i_2} \wedge \hat \omega
\eea
The resulting components of the Ricci tensor (in frame index convention) are given by
\bea
R_{mn} & = & \eta _{mn} \left (
- {2 \over f_1^2} - 2 |D_a \ln f_1 |^2 - 3 (D^a \ln f_1) (D_a \ln f_2 f_3) - {D^a D_a f_1 \over f_1} - \hat \omega_a (D^b \ln f_1) \epsilon_b {}^a \right )
\no \\
R_{i_1 j_1} & = & \delta _{i_1 j_1} \left (
+ {2 \over f_2^2} - 2 |D_a \ln f_2 |^2 - 3 (D^a \ln f_2) (D_a \ln f_1 f_3) - {D^a D_a f_2 \over f_2} - \hat \omega_a (D^b \ln f_2) \epsilon_b {}^a  \right )
\no \\
R_{i_2 j_2} & = & \delta _{i_2 j_2} \left (
+ {2 \over f_3^2} - 2 |D_a \ln f_3 |^2 - 3 (D^a \ln f_3) (D_a \ln f_1 f_2) - {D^a D_a f_3 \over f_3} - \hat \omega_a (D^b \ln f_3) \epsilon_b {}^a  \right )
\no \\
R_{ab} & = &
- 3 { D_b D_a f_1 \over f_1}  - 3 { D_b D_a f_2 \over f_2}
- 3 { D_b D_a f_3 \over f_3}  - 3 \, \hat \omega_b (D_c \ln f_1) \epsilon^c {}_a
\no\\&&
- 3 \, \hat \omega_b (D_c \ln f_2) \epsilon^c{}_a
- 3 \, \hat \omega_b (D_c \ln f_3) \epsilon^c {}_a
+ R^{(2)} \delta _{ab}
\eea
while all other components vanish.  $R^{(2)}$ is the two-dimensional curvature of $\Sigma$, 
and may be computed using the conventions in \S {\ref{compconv}.  It is given by
\bea
R^{(2)} = - {1 \over \rho^2} \p_w \p_{\bar w} \ln \rho
\eea
The  contributions from the 4-form $F$ are as follows,
\bea\
&&
- {1 \over 12} F_{MPQR} F_N {}^{PQR} + {1 \over 144} g_{MN} F_{PQRS} F^{PQRS} =
\hskip2.7in \no\\ &&\hskip1.3in
\left \{ \matrix{
\eta _{mn} \left( + {1 \over 3} g_{1 a}^2 + {1 \over 6} g_{2 a}^2 + {1 \over 6} g_{3 a}^2 \right)
&& (M,N) = (m,n) \cr
\delta _{i_1 j_1} \left( - {1 \over 6} g_{1 a}^2 - {1 \over 3} g_{2 a}^2 + {1 \over 6} g_{3 a}^2 \right)
&& (M,N) = (i_1,j_1) \cr
\delta_{i_2 j_2} \left( - {1 \over 6} g_{1 a}^2 + {1 \over 6} g_{2 a}^2 - {1 \over 3} g_{3 a}^2 \right)
&& (M,N) = (i_2,j_2) \cr
\half g_{1 a} g_{1 b} - \half g_{2 a} g_{2 b} - \half g_{3 a} g_{3 b}
&& (M,N) = (a,b) \cr
\hskip 0.4in + \delta _{ab} \left( - {1 \over 6} g_{1 a}^2 + {1 \over 6} g_{2 a}^2 + {1 \over 6} g_{3 a}^2 \right) && \cr
0 && \hbox{otherwise} \cr }
\right.
\eea
The equations along the symmetric spaces $AdS_3$, $S^3_1$, and $S^3_2$ are
\bea
\label{einstein2}
0 &=& - {2 \rho^2 \over f_1^2} - 2 |\p_w \ln f_1 |^2 - {3 \over 2}
\bigg( (\p_w \ln f_1) (\p_{\bar w} \ln f_2 f_3) + c.c. \bigg) - {\p_w \p_{\bar w} f_1 \over f_1}
\no\\ &&
+ {1 \over 3} {|\p_w b_1|^2 \over f_1^6} + {1 \over 6} {|\p_w b_2|^2 \over f_2^6} + {1 \over 6} {|\p_w b_3|^2 \over f_3^6}
\no\\
0 &=& + {2 \rho^2 \over f_2^2} - 2 |\p_w \ln f_2 |^2 - {3 \over 2}
\bigg( (\p_w \ln f_2) (\p_{\bar w} \ln f_1 f_3) + c.c. \bigg) - {\p_w \p_{\bar w} f_2 \over f_2}
\no\\ &&
- {1 \over 6} {|\p_w b_1|^2 \over f_1^6} - {1 \over 3} {|\p_w b_2|^2 \over f_2^6} + {1 \over 6} {|\p_w b_3|^2 \over f_3^6}
\no\\
0 &=& + {2 \rho^2 \over f_3^2} - 2 |\p_w \ln f_3 |^2 - {3 \over 2}
\bigg( (\p_w \ln f_3) (\p_{\bar w} \ln f_1 f_2) + c.c. \bigg) - {\p_w \p_{\bar w} f_3 \over f_3}
\no\\ &&
- {1 \over 6} {|\p_w b_1|^2 \over f_1^6} + {1 \over 6} {|\p_w b_2|^2 \over f_2^6} - {1 \over 3} {|\p_w b_3|^2 \over f_3^6}
\eea
With respect to the frame rotation group $SO(2)$ of $\Sigma$, all three equations are of weight $(0,0)$.
The equations along $\Sigma$ contain both a weight $(0,0)$ part and a weight $(2,0)$ part, and it will be useful to separate them.  They are correspondingly
\bea
\label{einstein3}
0 &=& -3 \sum_{i = 1}^3 {\p_w \p_{\bar w} f_i \over f_i} + 2 \rho^2 R^{(2)}
+ {1 \over 6} \bigg( {|\p_w b_1| \over f_1^6} - {|\p_w b_2|^2 \over f_2^6} - {|\p_w b_3|^2 \over f_3^6} \bigg)
\no\\
0 &=& -3 \sum_{i = 1}^3 \left( {\p_w^2 f_i \over f_i} - 2 \, (\p_w \ln \rho) \, (\p_w \ln f_i) \right) + {1 \over 2} \bigg( {|\p_w b_1| \over f_1^6} - {|\p_w b_2|^2 \over f_2^6} - {|\p_w b_3|^2 \over f_3^6} \bigg)
\eea
Using the BPS equations,  the Einstein equations (\ref{einstein2}) and (\ref{einstein3}) 
may be shown to hold automatically in a manner similar to the arguments 
given for the field equations of the 4-form $F$.  First we express the reduced Einstein 
equations in terms of $G$, $\bar G$ and $h$ using $(\ref{metricsolution})$, 
$(\ref{metricsolution2})$, and $(\ref{rhosolution})$ for the metric factors.  We choose 
conformal coordinates such that $h = r$ and  re-express $G$ in terms of $\Psi$ 
using (\ref{Psidef}).  Finally we use the differential equation for $\Psi$ $(\ref{Psieq})$ 
to eliminate terms with second order derivatives and higher in $r$.  
Again, these calculations were checked using MATHEMATICA.

\newpage

\appendix

%%%%%%%%%%%%%%%%%%%%%%%%%%%%%%%%%%%%%%%%%%%
%%%%%%%%%%%%%%%%%%%%%%%%%%%%%%%%%%%%%%%%%%%
\section{Clifford algebra basis adapted to the Ansatz}
\setcounter{equation}{0}
\label{appA}
%%%%%%%%%%%%%%%%%%%%%%%%%%%%%%%%%%%%%%%%%%%
%%%%%%%%%%%%%%%%%%%%%%%%%%%%%%%%%%%%%%%%%%%

The Clifford algebra of 11-dimensional $\G$-matrices is defined by
\bea
\{ \G^A, \G^B \} = 2 \eta ^{AB}
\eea
where $A,B$ are 11-dimensional Lorentz frame indices, $A,B=0,1,\cdots ,9,\natural=10$,
and $\eta ^{AB}$ is the Minkowski metric  $\eta^{AB} = {\rm diag} (- + \cdots ++)$.
A choice of $\G$-matrices which is well adapted to the product structure
$AdS_3 \times S_2^3 \times S_3^3 \times \Sigma$ of 11-dimensional space is as follows,
\bea
i_1=0,1,2 & \hskip 1in &
\G ^{i_1} = \g^{i_1} \otimes I_2 \otimes \, I_2 \otimes \s^1 \otimes \s^3
\no \\
i_2=3,4,5 &&
\G ^{i_2} = \, I_2 \otimes \g^{i_2} \otimes \, I_2 \otimes \s^2 \otimes \s^3
\no \\
i_3=6,7,8 &&
\G ^{i_3} = \, I_2 \otimes \, I_2 \otimes \g^{i_3} \otimes \s^3 \otimes \s^3
\no \\
a=9,\natural \hskip 0.2in &&
\G ^{a} = \, I_2 \otimes \, I_2 \otimes \, I_2 \otimes \, I_2 \otimes \s^a
\eea
where we have introduced the following $\g$-matrices associated with each of the
3-dimensional symmetric spaces $AdS_3$, $S_2^3$, and $S_3^3$,
\bea
-i \gamma^0 = \gamma^3 = \gamma^6 = \s^\natural & =  & \sigma^2
\no \\
\gamma^1 = \gamma^4 = \gamma^7  = \s^9 & = &  \sigma^1
\no \\
\gamma^2 = \gamma^5 = \gamma^8 & = & \sigma^3
\eea
Using the defining property of  complex conjugation,
\bea
B \Gamma^M B^{-1} =  (\Gamma^M)^*
\eea
the  complex conjugation matrix $B$ is given by
\bea
B = 1\; \otimes\sigma^2\otimes\sigma^2\otimes\sigma^3\otimes\sigma^1
\eea
The Majorana condition on the spinors $\zeta^*= B\zeta$ can be solved in terms of the eigenspinors of $\sigma^1$ and $\sigma^2$ respectively.

\newpage

%%%%%%%%%%%%%%%%%%%%%%%%%%%%%%%%%%%%%%%%%%%%%%%
%%%%%%%%%%%%%%%%%%%%%%%%%%%%%%%%%%%%%%%%%%%%%%%
\section{Geometry of Killing spinors in odd dimensions}
\label{Killing}
\setcounter{equation}{0}
%%%%%%%%%%%%%%%%%%%%%%%%%%%%%%%%%%%%%%%%%%%%%%%
%%%%%%%%%%%%%%%%%%%%%%%%%%%%%%%%%%%%%%%%%%%%%%%

In this appendix, we review the geometry of Killing spinors on spheres
and Minkowski signature hyperbolic spaces of odd dimensions $d=2n+1$.

\subsection{Killing spinors on $S^d$}

The $2^n$-dimensional Clifford algebra generators of $SO(d)$
$\g^i$ obey $\{ \g^i, \g^j \} = 2 \delta ^{ij}$, where $i,j=1,2,\cdots, d$.
The $2^{n+1}$-dimensional $SO(d+1)$ Clifford algebra generators $\G^I$
and the associated  chirality matrix $\bar \G$ may be constructed out of the
generators $\g^i$ by
\bea
\G^i = \g^i \otimes \s^1 \hskip 1in
\G^{d+1} = I \otimes \s^2 \hskip 1in
\bar \G = I \otimes \s^3
\eea
The sphere may be represented as the coset space $S^d=SO(d+1)/SO(d)$,
which is maximally symmetric. The Maurer-Cartan 1-form $\o^{(t)}$ of $SO(d+1)$
provides a flat connection on $SO(d+1)$ with torsion, and obeys,
\bea
d \o ^{(t)} + \o ^{(t)} \wedge \o^{(t)}=0
\hskip 1in
\o^{(t)} = U^\dagger dU
\eea
where $U$ parametrizes $SO(d+1)$ in the spinor representation.
Under the subgroup $SO(d)$, the Maurer-Cartan form $\o^{(t)}_{IJ} $
(with $I,J=1,2,\cdots, d, d+1$) decomposes into the canonical orthonormal
frame 1-form $e_i$  and the associated torsion-free orthonormal connection $\o_{ij}$,
$\o ^{(t)} _{i\, d+1} = e_i$, and $\o^{(t)} _{ij} = \o_{ij}$, so that
\bea
\o^{(t)} = {1 \over 4} \o^{(t)} _{IJ} \G^{IJ} = \half e_i \G^i \G^{d+1} + {1 \over 4} \o_{ij} \G^{ij}
\eea
The spin connection $\o^{(t)}$ may be consistently restricted to the
$\bar \G$-chirality $\eta= \pm 1$ eigenspace. We shall denote this restriction by
$\o ^{(t)} _\eta$, and express it  in the basis of $\G^I$ adapted to $\g^i$,
\bea
\o^{(t)}_\eta  = {i \eta \over 2}  e_i \g^i  + {1 \over 4} \o_{ij} \g^{ij}
\eea
By construction, this connection also satisfies the Maurer-Cartan equation.
The equation for $SO(d+1)$-covariantly constant spinors $\ep_\eta $ on $S^d$ then
becomes identical to the Killing spinor equation on $S^d$, and is given by,
\bea
\Big ( d + \o ^{(t)}_\eta  \Big ) \ep_\eta
= \Big ( d  + {1 \over 4} \o_{ij} \g^{ij}  +i  { \eta \over 2}  e_i \g^i \Big ) \ep_\eta  =0
\eea
Its solution space is of maximal rank, and given by
\bea
\ep_\eta  = \half (I +\eta \bar \G) U \ep_0
\eea
where $\ep_0$ is an arbitrary constant Dirac spinor, satisfying $d \ep_0=0$.

\sm

Under complex conjugation, the $d+1$-dimensional Euclidean signature
Dirac matrices behave as follows, (recall that we set $d=2n+1$),
\bea
(\g^i)^* & = &  (-)^n B_E \g^i B_E ^{-1}
\no \\
B_E B_E ^* & = &  - (-)^m I
\hskip 1in
m = \left [ { n \over 2} \right ]
\eea
where $[~]$ denotes the integer part of the argument. Under
complex conjugation, and the use of the complex conjugation matrix $B_E$,
the Killing spinor equation becomes,
\bea
\Big ( d  + {1 \over 4} \o_{ij} \g^{ij}
- i (-)^n { \eta \over 2}  e_i \g^i \Big ) B_E ^{-1} \ep_\eta ^* =0
\eea
Therefore, the combination $B_E ^{-1} \ep_\eta ^*$ is again a
Killing spinor, but for $ \eta \to \eta ' \equiv -(-)^n \eta$,
and we may identify those two spinors as follows,
\bea
B_E ^{-1} \ep_\eta ^* = \tilde \ep _{ \eta '}
\eea
Whenever $n$ is odd, such as in the case of the spheres $S^3$ in
the present paper, we have  $\eta ' = \eta$.

\subsection{Killing spinors on Minkowski signature $AdS_d$}

The $2^n$-dimensional Clifford algebra generators of $SO(1,2n)$
$\g^\mu$ obey $\{ \g^\mu, \g^\nu \} = 2 \delta ^{\mu \nu}$, where
$\mu, \nu=0,1,\cdots, 2n$. The $2^{n+1}$-dimensional $SO(2,2n)$
Clifford algebra generators  $\G^{\bar \mu}$ and the associated
chirality matrix $\bar \G$ may be constructed out of the
generators $\g^\mu $ by
\bea
\G^\mu  = \g^\mu \otimes \s^1 \hskip 1in
\G^{2n+1} = - i I \otimes \s^2 \hskip 1in
\bar \G = I \otimes \s^3
\eea
Minkowski signature anti-de Sitter space  may be represented as the
coset space $AdS_d=SO(2,2n)/SO(1,2n)$,  which is maximally symmetric.
The Maurer-Cartan 1-form $\o^{(t)}$ of $SO(2,2n)$
provides a flat connection on $SO(2,2n)$ with torsion, and obeys,
\bea
d \o ^{(t)} + \o ^{(t)} \wedge \o^{(t)}=0
\hskip 1in
\o^{(t)} = U^\dagger dU
\eea
where $U$ parametrizes $SO(2,2n)$ in the spinor representation.
Under the subgroup $SO(1,2n)$, the Maurer-Cartan form
$\o^{(t)}_{\bar \mu \bar \nu} $  (with $\bar \mu ,\bar \nu =0,1,2,\cdots, 2n, 2n+1$) decomposes into the canonical orthonormal frame 1-form $e_i$
and the associated torsion-free orthonormal connection $\o_{\mu \nu}$,
$\o ^{(t)} _{\mu\, d} = e_\mu$, and $\o^{(t)} _{\mu \nu } = \o_{\mu \nu}$,
so that
\bea
\o^{(t)} = {1 \over 4} \o^{(t)} _{\bar \mu \bar \nu} \G^{\bar \mu \bar \nu}
= \half e_\mu \G^\mu \G^d + {1 \over 4} \o_{\mu \nu} \G^{\mu \nu}
\eea
The spin connection $\o^{(t)}$ may be consistently restricted to the
$\bar \G$-chirality $\eta= \pm 1$ eigenspace. We shall denote this
restriction by  $\o ^{(t)} _\eta$, and express it  in the basis of
$\G^{\bar \mu}$ adapted to $\g^\mu$,
\bea
\o^{(t)}_\eta  = { \eta \over 2}  e_\mu \g^\mu
+ {1 \over 4} \o_{\mu \nu } \g^{\mu \nu}
\eea
By construction, this connection also satisfies the Maurer-Cartan equation.
The equation for $SO(2,2n)$-covariantly constant spinors $\ep_\eta $
on $AdS_d$ then
becomes identical to the Killing spinor equation on $AdS_d$, and is given by,
\bea
\Big ( d + \o ^{(t)}_\eta  \Big ) \ep_\eta
= \Big ( d  + {1 \over 4} \o_{\mu \nu } \g^{\mu \nu }
+  { \eta \over 2}  e_\mu \g^\mu \Big ) \ep_\eta  =0
\eea
Its solution space is of maximal rank, and given by
\bea
\ep_\eta  = \half (I +\eta \bar \G) U \ep_0
\eea
where $\ep_0$ is an arbitrary constant Dirac spinor, satisfying $d \ep_0=0$.

\sm

Under complex conjugation, the $d+1$-dimensional Minkowski signature
Dirac matrices behave as follows,
\bea
(\g^\mu )^* & = &  -(-)^n B_M \g^\mu B_M ^{-1}
\no \\
B_M B_M ^* & = &   (-)^m I
\hskip 1in
m = \left [ { n \over 2} \right ]
\eea
where $[~]$ denotes the integer part of the argument. Under
complex conjugation, and the use of the complex conjugation matrix $B_M$,
the Killing spinor equation becomes,
\bea
\Big ( d  + {1 \over 4} \o_{\mu \nu} \g^{\mu \nu }
-(-)^n { \eta \over 2}  e_\mu  \g^\mu  \Big ) B_M ^{-1} \ep_\eta ^* =0
\eea
Therefore, the combination $B_M ^{-1} \ep_\eta ^*$ is again a
Killing spinor, but for $\eta \to \eta ' = -(-)^n \eta$,
and we may identify those two spinors as follows,
\bea
B_M ^{-1} \ep_\eta ^* = \tilde \ep _{\eta '}
\eea
Whenever $n$ is odd, as is the case of the $AdS_3$ in the present paper,
we have $\eta' = \eta$.


\newpage

\begin{thebibliography}{99}

{\small



%\cite{Maldacena:1997re}
\bibitem{Maldacena:1997re}
  J.~M.~Maldacena,
   ``The large N limit of superconformal field theories and supergravity,''
  %
  Adv.\ Theor.\ Math.\ Phys.\  {\bf 2}, 231 (1998)
  [Int.\ J.\ Theor.\ Phys.\  {\bf 38}, 1113 (1999)]
  [arXiv:hep-th/9711200].
  %%CITATION = HEP-TH 9711200;%%

%\cite{Gubser:1998bc}
\bibitem{Gubser:1998bc}
  S.~S.~Gubser, I.~R.~Klebanov and A.~M.~Polyakov,
   ``Gauge theory correlators from non-critical string theory,''
  %
  Phys.\ Lett.\ B {\bf 428}, 105 (1998)
  [arXiv:hep-th/9802109].
  %%CITATION = HEP-TH 9802109;%%

%\cite{Witten:1998qj}
\bibitem{Witten:1998qj}
  E.~Witten,
   ``Anti-de Sitter space and holography,''
  %
  Adv.\ Theor.\ Math.\ Phys.\  {\bf 2}, 253 (1998)
  [arXiv:hep-th/9802150].
  %%CITATION = HEP-TH 9802150;%%

%\cite{Aharony:1999ti}
\bibitem{Aharony:1999ti}
  O.~Aharony, S.~S.~Gubser, J.~M.~Maldacena, H.~Ooguri and Y.~Oz,
  ``Large N field theories, string theory and gravity,''
  Phys.\ Rept.\  {\bf 323} (2000) 183
  [arXiv:hep-th/9905111].
  %%CITATION = HEP-TH 9905111;%%

%\cite{D'Hoker:2002aw}
\bibitem{D'Hoker:2002aw}
  E.~D'Hoker and D.~Z.~Freedman,
  ``Supersymmetric gauge theories and the AdS/CFT correspondence,''
   in {\sl Strings, Branes, and Extra Dimensions},
   S.S. Gubser, J.D. Lykken, Eds,
   World Scientific (2004),
  arXiv:hep-th/0201253.
  %%CITATION = HEP-TH 0201253;%%

%\cite{Lin:2004nb}
\bibitem{Lin:2004nb}
  H.~Lin, O.~Lunin and J.~M.~Maldacena,
  ``Bubbling AdS space and 1/2 BPS geometries,''
  JHEP {\bf 0410}, 025 (2004)
  [arXiv:hep-th/0409174].
  %%CITATION = HEP-TH 0409174;%%

%\cite{Berenstein:2004kk}
\bibitem{Berenstein:2004kk}
  D.~Berenstein,
  ``A toy model for the AdS/CFT correspondence,''
  JHEP {\bf 0407}, 018 (2004)
  [arXiv:hep-th/0403110].
  %%CITATION = JHEPA,0407,018;%%

%\cite{D'Hoker:2007xy}
\bibitem{D'Hoker:2007xy}
  E.~D'Hoker, J.~Estes and M.~Gutperle,
  ``Exact half-BPS Type IIB interface solutions I: Local solution and
  supersymmetric Janus,''
  JHEP {\bf 0706} (2007) 021
  [arXiv:0705.0022 [hep-th]].
  %%CITATION = JHEPA,0706,021;%%

%\cite{D'Hoker:2007xz}
\bibitem{D'Hoker:2007xz}
  E.~D'Hoker, J.~Estes and M.~Gutperle,
  ``Exact half-BPS type IIB interface solutions. II: Flux solutions and
  multi-janus,''
  JHEP {\bf 0706} (2007) 022
  [arXiv:0705.0024 [hep-th]].
  %%CITATION = JHEPA,0706,022;%%

%\cite{Bak:2003jk}
\bibitem{Bak:2003jk}
  D.~Bak, M.~Gutperle and S.~Hirano,
   ``A dilatonic deformation of AdS(5) and its field theory dual,''
  JHEP {\bf 0305}, 072 (2003)
  [arXiv:hep-th/0304129].
  %%CITATION = HEP-TH 0304129;%%

%\cite{Clark:2005te}
\bibitem{Clark:2005te}
  A.~Clark and A.~Karch,
  ``Super Janus,''
  JHEP {\bf 0510} (2005) 094
  [arXiv:hep-th/0506265].
  %%CITATION = HEP-TH 0506265;%%

  %\cite{D'Hoker:2006uu}
\bibitem{D'Hoker:2006uu}
  E.~D'Hoker, J.~Estes and M.~Gutperle,
  ``Ten-dimensional supersymmetric Janus solutions,''
  Nucl.\ Phys.\  B {\bf 757} (2006) 79
  [arXiv:hep-th/0603012].
  %%CITATION = NUPHA,B757,79;%%

%\cite{DeWolfe:2001pq}
\bibitem{DeWolfe:2001pq}
  O.~DeWolfe, D.~Z.~Freedman and H.~Ooguri,
   ``Holography and defect conformal field theories,''
  Phys.\ Rev.\ D {\bf 66}, 025009 (2002)
  [arXiv:hep-th/0111135].
  %%CITATION = HEP-TH 0111135;%%




%\cite{Clark:2004sb}
\bibitem{Clark:2004sb}
  A.~B.~Clark, D.~Z.~Freedman, A.~Karch and M.~Schnabl,
  ``The dual of Janus ((<:) <--> (:>)) an interface CFT,''
  Phys.\ Rev.\  D {\bf 71}, 066003 (2005)
  [arXiv:hep-th/0407073].
  %%CITATION = PHRVA,D71,066003;%%

%\cite{D'Hoker:2006uv}
\bibitem{D'Hoker:2006uv}
  E.~D'Hoker, J.~Estes and M.~Gutperle,
  ``Interface Yang-Mills, supersymmetry, and Janus,''
  Nucl.\ Phys.\  B {\bf 753} (2006) 16
  [arXiv:hep-th/0603013].
  %%CITATION = NUPHA,B753,16;%%
  
  
%\cite{Gomis:2006cu}
\bibitem{Gomis:2006cu}
  J.~Gomis and C.~Romelsberger,
  ``Bubbling defect CFT's,''
  JHEP {\bf 0608}, 050 (2006)
  [arXiv:hep-th/0604155].
  %%CITATION = HEP-TH 0604155;%%


%\cite{Gaiotto:2008sd}
\bibitem{Gaiotto:2008sd}
  D.~Gaiotto and E.~Witten,
  ``Janus Configurations, Chern-Simons Couplings, 
  And The Theta-Angle in N=4
  Super Yang-Mills Theory,''
  arXiv:0804.2907 [hep-th].
  %%CITATION = ARXIV:0804.2907;%%
  
  
%\cite{Gaiotto:2008sa}
\bibitem{Gaiotto:2008sa}
  D.~Gaiotto and E.~Witten,
  ``Supersymmetric Boundary Conditions in N=4 Super Yang-Mills Theory,''
  arXiv:0804.2902 [hep-th].
  %%CITATION = ARXIV:0804.2902;%%

%\cite{D'Hoker:2007fq}
\bibitem{D'Hoker:2007fq}
  E.~D'Hoker, J.~Estes and M.~Gutperle,
  ``Gravity duals of half-BPS Wilson loops,''
  JHEP {\bf 0706} (2007) 063
  [arXiv:0705.1004 [hep-th]].
  %%CITATION = JHEPA,0706,063;%%

%\cite{Drukker:1999zq}
\bibitem{Drukker:1999zq}
  N.~Drukker, D.~J.~Gross and H.~Ooguri,
  ``Wilson loops and minimal surfaces,''
  Phys.\ Rev.\  D {\bf 60} (1999) 125006
  [arXiv:hep-th/9904191].
  %%CITATION = PHRVA,D60,125006;%%


%\cite{Yamaguchi:2006te}
\bibitem{Yamaguchi:2006te}
  S.~Yamaguchi,
  ``Bubbling geometries for half BPS Wilson lines,''
  arXiv:hep-th/0601089.
  %%CITATION = HEP-TH/0601089;%%


 %\cite{Gomis:2006sb}
\bibitem{Gomis:2006sb}
  J.~Gomis and F.~Passerini,
  ``Holographic Wilson loops,''
  JHEP {\bf 0608} (2006) 074
  [arXiv:hep-th/0604007].
  %%CITATION = JHEPA,0608,074;%%

%\cite{Lunin:2006xr}
\bibitem{Lunin:2006xr}
  O.~Lunin,
  ``On gravitational description of Wilson lines,''
  JHEP {\bf 0606} (2006) 026
  [arXiv:hep-th/0604133].
  %%CITATION = JHEPA,0606,026;%%

 
  %\cite{Bagger:2007jr}
\bibitem{Bagger:2007jr}
  J.~Bagger and N.~Lambert,
  ``Gauge Symmetry and Supersymmetry of Multiple M2-Branes,''
  Phys.\ Rev.\  D {\bf 77} (2008) 065008
  [arXiv:0711.0955 [hep-th]].
  %%CITATION = PHRVA,D77,065008;%%

 %\cite{Bagger:2007vi}
\bibitem{Bagger:2007vi}
  J.~Bagger and N.~Lambert,
  ``Comments On Multiple M2-branes,''
  JHEP {\bf 0802} (2008) 105
  [arXiv:0712.3738 [hep-th]].
  %%CITATION = JHEPA,0802,105;%%
 
 
 %\cite{Gustavsson:2008dy}
\bibitem{Gustavsson:2008dy}
  A.~Gustavsson,
  ``Selfdual strings and loop space Nahm equations,''
  JHEP {\bf 0804} (2008) 083
  [arXiv:0802.3456 [hep-th]].
  %%CITATION = JHEPA,0804,083;%%

%\cite{Boonstra:1998yu}
\bibitem{Boonstra:1998yu}
  H.~J.~Boonstra, B.~Peeters and K.~Skenderis,
  ``Brane intersections, anti-de Sitter spacetimes and dual superconformal
  theories,''
  Nucl.\ Phys.\  B {\bf 533}, 127 (1998)
  [arXiv:hep-th/9803231].
  %%CITATION = NUPHA,B533,127;%%

%\cite{Gauntlett:1998kc}
\bibitem{Gauntlett:1998kc}
  J.~P.~Gauntlett, R.~C.~Myers and P.~K.~Townsend,
  ``Supersymmetry of rotating branes,''
  Phys.\ Rev.\  D {\bf 59}, 025001 (1999)
  [arXiv:hep-th/9809065].
  %%CITATION = PHRVA,D59,025001;%%
  
%\cite{de Boer:1999rh}
\bibitem{de Boer:1999rh}
  J.~de Boer, A.~Pasquinucci and K.~Skenderis,
  ``AdS/CFT dualities involving large 2d N = 4 superconformal symmetry,''
  Adv.\ Theor.\ Math.\ Phys.\  {\bf 3}, 577 (1999)
  [arXiv:hep-th/9904073].
  %%CITATION = 00203,3,577;%%


%\cite{Gauntlett:2006ns}
\bibitem{Gauntlett:2006ns}
  J.~P.~Gauntlett, N.~Kim and D.~Waldram,
  ``Supersymmetric AdS(3), AdS(2) and bubble solutions,''
  JHEP {\bf 0704} (2007) 005
  [arXiv:hep-th/0612253].
  %%CITATION = JHEPA,0704,005;%%
  
  
 %\cite{Gauntlett:2006qw}
\bibitem{Gauntlett:2006qw}
  J.~P.~Gauntlett, O.~A.~P.~Mac Conamhna, T.~Mateos and D.~Waldram,
  ``New supersymmetric AdS(3) solutions,''
  Phys.\ Rev.\  D {\bf 74} (2006) 106007
  [arXiv:hep-th/0608055].
  %%CITATION = PHRVA,D74,106007;%%

  
%\cite{Gauntlett:2007ts}
\bibitem{Gauntlett:2007ts}
  J.~P.~Gauntlett and N.~Kim,
  ``Geometries with Killing Spinors and Supersymmetric AdS Solutions,''
  arXiv:0710.2590 [hep-th].
  %%CITATION = ARXIV:0710.2590;%%



%\cite{Lunin:2007ab}
\bibitem{Lunin:2007ab}
  O.~Lunin,
  ``1/2-BPS states in M theory and defects in the dual CFTs,''
  JHEP {\bf 0710}, 014 (2007)
  [arXiv:0704.3442 [hep-th]].
  %%CITATION = JHEPA,0710,014;%%

\bibitem{DEGKS} E. D'Hoker, J. Estes, M. Gutperle, D. Krym, and P. Sorba,
``Half-BPS supergravity solutions and supergroups", in preparation.

 
\bibitem{Liouville}
E.~D'Hoker and R.~Jackiw,
``Liouville Field Theory,''
Phys.\ Rev.\  D {\bf 26}, 3517 (1982);\\
E.~D'Hoker and R.~Jackiw,
 ``Space Translation Breaking And Compactification In The Liouville Theory,''
Phys.\ Rev.\ Lett.\  {\bf 50}, 1719 (1983);\\
E.~D'Hoker, D.~Z.~Freedman and R.~Jackiw,
``SO(2,1) Invariant Quantization Of The Liouville Theory,''
Phys.\ Rev.\  D {\bf 28}, 2583 (1983).

%\cite{Cremmer:1978km}
\bibitem{Cremmer:1978km}
  E.~Cremmer, B.~Julia and J.~Scherk,
  ``Supergravity theory in 11 dimensions,''
  Phys.\ Lett.\  B {\bf 76}, 409 (1978).
  %%CITATION = PHLTA,B76,409;%%



 
}

\end{thebibliography}
\end{document}